# A theory of quantum electrodynamics with nonlocal interaction

T. Mei

(Department of Journal, Central China Normal University, Wuhan, Hubei PRO, People's Republic of China

E-Mail:   meitao@mail.ccnu.edu.cn     meitaowh@public.wh.hb.cn )

**Abstract:** In this paper, we present a theory of quantum electrodynamics with nonlocal interaction, a main characteristic of the theory is that a charged particle situated $x^\mu$ interacts with electromagnetic field situated $y^\mu$, where $x^\mu = y^\mu + aA^\mu(y)$, $A^\mu(y)$ reads electromagnetic 4-potential, $a$ is a constant. All the action, the equations of motion of charged particle and electromagnetic field are given. For the case of free fields, charged particle and electromagnetic field obey the Dirac equation and the Maxwell equation of free fields, respectively; for the case with interaction, both the equations of motion of charged particle and electromagnetic field lead to current conservation $j^\mu{}_{,\mu}(x) = 0$ naturally. The theory is Lorentz invariant and gauge invariant under a generalized gauge transformation, the generalized gauge transformation can guarantee that the temporal gauge condition $A^0(x) = 0$ holds or the Lorentz gauge condition $A^\mu{}_{,\mu}(y) = 0$ holds, respectively. The theory returns to the current QED when $a = 0$. Finally, taking advantage of the Lehmann-Symanzik-Zimmermann approach, we establish the corresponding quantum theory.

In this paper, we set $c = \hbar = 1$, the Greek alphabet is raised or lowered by $\eta^{\alpha\beta} = \mathrm{diag}\,(+1, -1, -1, -1)$ or $\eta_{\alpha\beta}$, respectively; the value of the Greek and Latin alphabet are 0, 1, 2, 3 and 1, 2, 3, respectively.

## 1  The current QED

As well-known, the action of the current QED is

$$S = \int \mathrm{d}^4 x L(x) = \int \mathrm{d}^4 x \big(L_\mathrm{D}(x) + L_\mathrm{EM}(x) + L_\mathrm{I}(x)\big), \qquad (1\text{-}1)$$

$$L_\mathrm{D}(x) = \overline{\psi}(x)\left(\mathrm{i}\gamma^\mu \frac{\partial}{\partial x^\mu} - m\right)\psi(x), \qquad (1\text{-}2)$$

$$L_{EM}(x) = -\frac{1}{4} F_{\mu\nu}(x) F^{\mu\nu}(x), \tag{1-3}$$

$$L_I(x) = -ej^\mu(x) A_\mu(x), \tag{1-4}$$

$$F_{\mu\nu}(x) = \frac{\partial A_\nu(x)}{\partial x^\mu} - \frac{\partial A_\mu(x)}{\partial x^\nu} = A_{\nu,\mu}(x) - A_{\mu,\nu}(x), \tag{1-5}$$

$$j^\mu(x) = \bar{\psi}(x) \gamma^\mu \psi(x). \tag{1-6}$$

The equation of motion of charged particle and electromagnetic field read

$$\left( i\gamma^\mu \frac{\partial}{\partial x^\mu} - m \right) \psi(x) = e\gamma^\mu A_\mu(x) \psi(x), \tag{1-7}$$

$$-\frac{\partial F^{\mu\nu}(x)}{\partial x^\nu} = \frac{D^2 A^\mu(x)}{Dx^2} = ej^\mu(x). \tag{1-8}$$

For arbitrary function $W^\mu(x)$, the operator $\frac{D^2 W^\mu(x)}{Dx^2}$ is defined as

$$\frac{D^2 W^\mu(x)}{Dx^2} \equiv \eta^{\alpha\beta} \frac{\partial^2 W^\mu(x)}{\partial x^\alpha \partial x^\beta} - \eta^{\mu\alpha} \frac{\partial^2 W^\beta(x)}{\partial x^\alpha \partial x^\beta} = W^{\mu,\alpha}{}_{,\alpha}(x) - W^{\alpha,\mu}{}_{,\alpha}(x). \tag{1-9}$$

Both the equations of motion (1-7) and (1-8) lead to the current conservation equation

$$\frac{\partial j^\mu(x)}{\partial x^\mu} = 0. \tag{1-10}$$

All the action (1-1), the equations of motion (1-7) and (1-8) are invariable under the following gauge transformation

$$A'_\mu(x) = A_\mu(x) + \frac{\partial \chi(x)}{\partial x^\mu}, \tag{1-11}$$

$$\psi'(x) = e^{-ie\chi(x)} \psi(x). \tag{1-12}$$

There are other equivalent forms of the standard form (1-1) of the action of the current QED. For example, we can separate formally $A^\mu(x)$ into two parts

$$\begin{aligned}
A^\mu(x) &= \int d^4 y \delta^4(x-y) A^\mu(y) = \int d^4 y \frac{d^4 k}{(2\pi)^4} e^{ik\cdot(x-y)} \delta^\mu_\nu A^\nu(y) \\
&= \int \frac{d^4 k}{(2\pi)^4} \frac{k^2 \delta^\mu_\nu - k^\mu k_\nu}{k^2} e^{ik\cdot(x-y)} A^\nu(y) d^4 y + \int \frac{d^4 k}{(2\pi)^4} \frac{k^\mu k_\nu}{k^2} e^{ik\cdot(x-y)} A^\nu(y) d^4 y,
\end{aligned} \tag{1-13}$$

Via integration by parts we have

$$\int \frac{\mathrm{d}^4 k}{(2\pi)^4} \frac{k^2 \delta_\nu^\mu - k^\mu k_\nu}{k^2} \mathrm{e}^{\mathrm{i}k\cdot(x-y)} A^\nu(y) \mathrm{d}^4 y = \int \frac{\mathrm{d}^4 k}{(2\pi)^4} \frac{\mathrm{i}k_\nu}{k^2}\left(A^\mu(y)\frac{\partial}{\partial y_\nu} - A^\nu(y)\frac{\partial}{\partial y_\mu}\right)\mathrm{e}^{\mathrm{i}k\cdot(x-y)}\mathrm{d}^4 y$$

$$= \int \frac{\mathrm{d}^4 k}{(2\pi)^4} \frac{-\mathrm{i}k_\nu}{k^2} \mathrm{e}^{\mathrm{i}k\cdot(x-y)}\left(A^{\mu,\nu}(y) - A^{\nu,\mu}(y)\right)\mathrm{d}^4 y$$

$$+ \int \frac{\mathrm{d}^4 k}{(2\pi)^4} \frac{\mathrm{i}k_\nu}{k^2}\left(\frac{\partial(A^\mu(y)\mathrm{e}^{\mathrm{i}k\cdot(x-y)})}{\partial y_\nu} - \frac{\partial(A^\nu(y)\mathrm{e}^{\mathrm{i}k\cdot(x-y)})}{\partial y_\mu}\right)\mathrm{d}^4 y \quad (1\text{-}14)$$

$$= \int \mathrm{d}^4 y F^{\mu\nu}(y) \frac{\mathrm{d}^4 k}{(2\pi)^4} \frac{\mathrm{i}k_\nu}{k^2} \mathrm{e}^{\mathrm{i}k\cdot(x-y)} + \int \frac{\mathrm{d}^4 k}{(2\pi)^4} \frac{\mathrm{i}k_\nu}{k^2}\left(\frac{\partial(A^\mu(y)\mathrm{e}^{\mathrm{i}k\cdot(x-y)})}{\partial y_\nu} - \frac{\partial(A^\nu(y)\mathrm{e}^{\mathrm{i}k\cdot(x-y)})}{\partial y_\mu}\right)\mathrm{d}^4 y$$

$$= \int \mathrm{d}^4 y F^{\mu\nu}(y) \frac{\mathrm{d}^4 k}{(2\pi)^4} \frac{-\mathrm{i}k_\nu}{k^2} \mathrm{e}^{-\mathrm{i}k\cdot(x-y)} + \int \frac{\mathrm{d}^4 k}{(2\pi)^4} \frac{\mathrm{i}k_\nu}{k^2}\left(\frac{\partial(A^\mu(y)\mathrm{e}^{\mathrm{i}k\cdot(x-y)})}{\partial y_\nu} - \frac{\partial(A^\nu(y)\mathrm{e}^{\mathrm{i}k\cdot(x-y)})}{\partial y_\mu}\right)\mathrm{d}^4 y,$$

Ignoring the surface terms in (1-14), we have

$$A^\mu(x) = A^\mu_\perp(x) + A^\mu_{//}(x), \quad (1\text{-}15)$$

$$A^\mu_\perp(x) = \int \mathrm{d}^4 y F^{\mu\nu}(y) \frac{\mathrm{d}^4 k}{(2\pi)^4} \frac{-\mathrm{i}k_\nu}{k^2} \mathrm{e}^{-\mathrm{i}k\cdot(x-y)}, \quad (1\text{-}16)$$

$$A^\mu_{//}(x) = \int \frac{\mathrm{d}^4 k}{(2\pi)^4} \frac{k^\mu k_\nu}{k^2} \mathrm{e}^{\mathrm{i}k\cdot(x-y)} A^\nu(y) \mathrm{d}^4 y = \int \frac{\mathrm{d}^4 k}{(2\pi)^4} \frac{k^\mu k_\nu}{k^2} \mathrm{e}^{-\mathrm{i}k\cdot(x-y)} A^\nu(y) \mathrm{d}^4 y. \quad (1\text{-}17)$$

According to (1-16) and (1-17) we have

$$S_\mathrm{I}^{(0)} \equiv \int \mathrm{d}^4 x L_\mathrm{I}(x) = -e \int \mathrm{d}^4 x j^\mu(x) A_\mu(x) = S_{\mathrm{I}\perp}^{(0)} + S_{\mathrm{I}//}^{(0)}, \quad (1\text{-}18)$$

$$S_{\mathrm{I}\perp}^{(0)} = -e \int \mathrm{d}^4 x j_\mu(x) A^\mu_\perp(x) = -e \int \mathrm{d}^4 x j_\mu(x) \mathrm{d}^4 z F^{\mu\nu}(z) \frac{\mathrm{d}^4 k}{(2\pi)^4} \frac{-\mathrm{i}k_\nu}{k^2} \mathrm{e}^{-\mathrm{i}k\cdot(x-z)}, \quad (1\text{-}19)$$

$$S_{\mathrm{I}//}^{(0)} = -e \int \mathrm{d}^4 x j_\mu(x) A^\mu_{//}(x) = -e \int \mathrm{d}^4 x j_\mu(x) \mathrm{d}^4 z A^\nu(z) \frac{\mathrm{d}^4 k}{(2\pi)^4} \frac{k^\mu k_\nu}{k^2} \mathrm{e}^{-\mathrm{i}k\cdot(x-z)}. \quad (1\text{-}20)$$

And, further, using integration by parts and considering the characteristic $\eta_{\alpha\beta} k^\alpha k_\lambda F^{\beta\lambda}(x) = k_\beta k_\lambda F^{\beta\lambda}(x) = 0$ due to $F^{\beta\lambda}(x) = -F^{\lambda\beta}(x)$, we can prove

$$\int_{t_a}^{t_b} \mathrm{d}t \int \mathrm{d}^3 x L_\mathrm{EM}(A^\gamma(x)) = -\frac{1}{4}\eta_{\alpha\beta}\eta_{\lambda\tau} \int_{t_a}^{t_b} \mathrm{d}t \int \mathrm{d}^3 x F^{\alpha\tau}(x) F^{\beta\lambda}(x)$$

$$= \int_{t_a}^{t_b} \mathrm{d}x^0 \int \mathrm{d}^3 x \widetilde{L}_\mathrm{EM}(t_a, t_b; \boldsymbol{x}) + \frac{1}{2}\eta_{\alpha\beta}\int_{t_a}^{t_b} \mathrm{d}x^0 \int \mathrm{d}^3 x \frac{\partial}{\partial x_\lambda}\left(A^\alpha(x) F^{\beta\lambda}(x)\right)$$

$$+ \frac{1}{2}\eta_{\alpha\beta}\int_{t_a}^{t_b} \mathrm{d}x^0 \int \mathrm{d}^3 x \frac{\partial}{\partial x_\alpha}\left(A^\tau(x)\int_{t_a}^{t_b}\mathrm{d}z^0\int \mathrm{d}^3 z \int \frac{\mathrm{d}^4 k}{(2\pi)^4} \frac{k_\tau k_\lambda}{k^2} \mathrm{e}^{\mathrm{i}k\cdot(x-z)} F^{\beta\lambda}(z)\right) \quad (1\text{-}21)$$

$$- \frac{1}{2}\eta_{\alpha\beta}\int_{t_a}^{t_b} \mathrm{d}x^0 \int \mathrm{d}^3 x \frac{\partial}{\partial x_\tau}\left(A^\alpha(x)\int_{t_a}^{t_b}\mathrm{d}z^0\int \mathrm{d}^3 z \int \frac{\mathrm{d}^4 k}{(2\pi)^4} \frac{k_\tau k_\lambda}{k^2} \mathrm{e}^{\mathrm{i}k\cdot(x-z)} F^{\beta\lambda}(z)\right),$$

where

$$\widetilde{L}_{\text{EM}}(t_a, t_b; \boldsymbol{x}) = -\frac{1}{2}\eta_{\alpha\beta}F^{\alpha\tau}(x)\int_{t_a}^{t_b}dz^0\int d^3z\int\frac{d^4k}{(2\pi)^4}\frac{k_\tau k_\lambda}{k^2}e^{ik\cdot(x-z)}F^{\beta\lambda}(z). \tag{1-22}$$

Hence, dropping the total derivative terms in (1-21), we can employ

$$S_{\text{EM}}^{(0)} = \int d^4x \widetilde{L}_{\text{EM}}(x) = -\frac{1}{2}\eta_{\alpha\beta}\int d^4z F^{\alpha\tau}(z)\frac{d^4k}{(2\pi)^4}\frac{k_\tau k_\lambda}{k^2}e^{ik\cdot(z-x)}d^4x F^{\beta\lambda}(x) \tag{1-23}$$

as action of electromagnetic field, and the action (1-1) can be written to the form

$$S = \int d^4x L_D(x) + S_{\text{EM}}^{(0)} + S_{\text{I}\perp}^{(0)} + S_{\text{I}//}^{(0)}. \tag{1-24}$$

We see that all the Lagrangians in $S_{\text{EM}}^{(0)}$, $S_{\text{I}\perp}^{(0)}$ and $S_{\text{I}//}^{(0)}$ are nonlocal formally.

We can prove that the variational equation $\dfrac{\delta S}{\delta A^\mu(y)} = 0$ leads to the equation of motion of electromagnetic field (1-8). At first, we have

$$\begin{aligned}
\frac{\delta S_{\text{EM}}^{(0)}}{\delta A^\mu(y)} &= -\frac{1}{2}\eta_{\alpha\beta}\int d^4z \frac{\partial F^{\alpha\tau}(z)}{\partial A^\mu_{,\nu}(z)}\frac{\partial \delta^4(z-y)}{\partial z^\nu}\frac{d^4k}{(2\pi)^4}\frac{k_\tau k_\lambda}{k^2}e^{ik\cdot(z-x)}d^4x F^{\beta\lambda}(x) \\
&\quad -\frac{1}{2}\eta_{\alpha\beta}\int d^4z F^{\alpha\tau}(z)\frac{d^4k}{(2\pi)^4}\frac{k_\tau k_\lambda}{k^2}e^{ik\cdot(z-x)}d^4x \frac{\partial F^{\beta\lambda}(x)}{\partial A^\mu_{,\nu}(x)}\frac{\partial \delta^4(x-y)}{\partial z^\nu} \\
&= \eta_{\alpha\beta}\left(\delta^\tau_\mu\eta^{\nu\alpha} - \delta^\alpha_\mu\eta^{\nu\tau}\right)\int d^4z F^{\beta\lambda}(z)\frac{d^4k}{(2\pi)^4}\frac{-ik_\tau k_\lambda k_\nu}{k^2}e^{-ik\cdot(y-z)} \\
&= \int d^4z F^{\nu\lambda}(z)\frac{d^4k}{(2\pi)^4}\frac{-ik_\mu k_\lambda k_\nu}{k^2}e^{-ik\cdot(y-z)} - \eta_{\mu\beta}\int d^4z F^{\beta\lambda}(z)\frac{d^4k}{(2\pi)^4}\frac{-ik^\nu k_\lambda k_\nu}{k^2}e^{-ik\cdot(y-z)} \\
&= \eta_{\mu\beta}\int d^4z F^{\beta\lambda}(z)\frac{d^4k}{(2\pi)^4}ik_\lambda e^{-ik\cdot(y-z)} = \eta_{\mu\beta}\int d^4z F^{\beta\lambda}(z)\frac{d^4k}{(2\pi)^4}\frac{\partial}{\partial z^\lambda}e^{-ik\cdot(y-z)} \\
&= -\eta_{\mu\beta}\int d^4z \frac{\partial F^{\beta\lambda}(z)}{\partial z^\lambda}\frac{d^4k}{(2\pi)^4}e^{-ik\cdot(y-z)} = -\eta_{\mu\beta}\int d^4z \frac{\partial F^{\beta\lambda}(z)}{\partial z^\lambda}\delta^4(y-z) = -\eta_{\mu\beta}\frac{\partial F^{\beta\lambda}(y)}{\partial y^\lambda},
\end{aligned} \tag{1-25}$$

$$\begin{aligned}
\frac{\delta S_{\text{I}\perp}^{(0)}}{\delta A^\mu(y)} &= -e\int d^4x j_\alpha(x)d^4z \frac{\partial F^{\alpha\tau}(z)}{\partial A^\mu_{,\nu}(z)}\frac{\partial \delta^4(z-y)}{\partial z^\nu}\frac{d^4k}{(2\pi)^4}\frac{-ik_\tau}{k^2}e^{-ik\cdot(x-z)} \\
&= e\left(\delta^\tau_\mu\eta^{\nu\alpha} - \delta^\alpha_\mu\eta^{\nu\tau}\right)\int d^4x j_\alpha(x)\frac{d^4k}{(2\pi)^4}\frac{k_\tau k_\nu}{k^2}e^{-ik\cdot(x-y)} \\
&= e\int d^4x j_\alpha(x)\frac{d^4k}{(2\pi)^4}\frac{k_\mu k^\alpha}{k^2}e^{-ik\cdot(x-y)} - e\int d^4x j_\mu(x)\frac{d^4k}{(2\pi)^4}\frac{k^\nu k_\nu}{k^2}e^{-ik\cdot(x-y)} \\
&= e\int d^4x j_\alpha(x)\frac{d^4k}{(2\pi)^4}\frac{k_\mu k^\alpha}{k^2}e^{-ik\cdot(x-y)} - ej_\mu(y),
\end{aligned} \tag{1-26}$$

$$\begin{aligned}
\frac{\delta S_{\text{I}//}^{(0)}}{\delta A^\mu(y)} &= -e\int d^4x j_\alpha(x)d^4z \frac{\partial A^\nu(z)}{\partial A^\mu(z)}\delta^4(z-y)\frac{d^4k}{(2\pi)^4}\frac{k^\alpha k_\nu}{k^2}e^{-ik\cdot(x-z)} \\
&= -e\int d^4x j_\alpha(x)\frac{d^4k}{(2\pi)^4}\frac{k^\alpha k_\mu}{k^2}e^{-ik\cdot(x-z)}.
\end{aligned} \tag{1-27}$$

According to (1-25) ~ (1-27) we now obtain

$$\frac{\delta S}{\delta A^\mu(y)} = \frac{\delta S_{\text{EM}}^{(0)}}{\delta A^\mu(y)} + \frac{\delta S_{\text{I}\perp}^{(0)}}{\delta A^\mu(y)} + \frac{\delta S_{\text{I}//}^{(0)}}{\delta A^\mu(y)} = -\eta_{\mu\beta}\frac{\partial F^{\beta\lambda}(y)}{\partial y^\lambda} - ej_\mu(y) = 0,$$

This is just (1-8).

Although one has obtained wonderful results in accord with experiments[1] from the current QED based on the renormalization method, other models, e.g., nonlocal interaction theories, had also been studied. (There are a number of literatures on nonlocal interaction, see, for example, Refs [2~5].) In this paper, we present a theory of quantum electrodynamics with nonlocal interaction, a main characteristic of the theory is that a charged particle situated $x^\mu$ interacts with electromagnetic field situated $y^\mu$, where $x^\mu = y^\mu + aA^\mu(y)$, $a$ is a constant.

## 2  Some quantities and their characteristics

At first, we introduce a constant $a$, the dimension of $aA^\mu(x)$ is length, e.g., meter in the SI units.

### 2.1  The functions $J(x)$, $J_\beta^\alpha(x)$ and $\Omega_\nu^\mu(x)$

We define

$$J_\beta^\alpha(x) = \frac{\partial(x^\alpha + aA^\alpha(x))}{\partial x^\beta} = \delta_\beta^\alpha + aA^\alpha{}_{,\beta}(x),\tag{2-1}$$

the determinant $J(x)$ of $J_\beta^\alpha(x)$ is

$$J(x) \equiv \left\| \delta_\beta^\alpha + a\frac{\partial A^\alpha(x)}{\partial x^\beta} \right\| = \begin{vmatrix} 1+aA^0{}_{,0}(x) & aA^0{}_{,1}(x) & aA^0{}_{,2}(x) & aA^0{}_{,3}(x) \\ aA^1{}_{,0}(x) & 1+aA^1{}_{,1}(x) & aA^1{}_{,2}(x) & aA^1{}_{,3}(x) \\ aA^2{}_{,0}(x) & aA^2{}_{,1}(x) & 1+aA^2{}_{,2}(x) & aA^2{}_{,3}(x) \\ aA^3{}_{,0}(x) & aA^3{}_{,1}(x) & aA^3{}_{,2}(x) & 1+aA^3{}_{,3}(x) \end{vmatrix}$$

$$\equiv 1 + \sum_{n=1}^{4} \frac{a^n}{n!} J_{(n)}(x);\tag{2-2}$$

$$J_{(1)}(x) = A^\lambda{}_{,\lambda}(x),\tag{2-3}$$

$$J_{(2)}(x) = A^\mu{}_{,\mu}(x)A^\nu{}_{,\nu}(x) - A^\mu{}_{,\nu}(x)A^\nu{}_{,\mu}(x),\tag{2-4}$$

$$J_{(3)}(x) = A^\alpha{}_{,\alpha}(x)A^\beta{}_{,\beta}(x)A^\gamma{}_{,\gamma}(x) + 2A^\alpha{}_{,\beta}(x)A^\beta{}_{,\gamma}(x)A^\gamma{}_{,\alpha}(x) \\ - 3A^\lambda{}_{,\lambda}(x)A^\mu{}_{,\nu}(x)A^\nu{}_{,\mu}(x),\tag{2-5}$$

$$J_{(4)}(x) = A^\mu{}_{,\mu}(x)A^\nu{}_{,\nu}(x)A^\rho{}_{,\rho}(x)A^\sigma{}_{,\sigma}(x) + 3A^\mu{}_{,\nu}(x)A^\nu{}_{,\mu}(x)A^\rho{}_{,\sigma}(x)A^\sigma{}_{,\rho}(x) \\ + 8A^\lambda{}_{,\lambda}(x)A^\alpha{}_{,\beta}(x)A^\beta{}_{,\gamma}(x)A^\gamma{}_{,\alpha}(x) - 6A^\mu{}_{,\mu}(x)A^\nu{}_{,\nu}(x)A^\rho{}_{,\sigma}(x)A^\sigma{}_{,\rho}(x) \\ - 6A^\mu{}_{,\nu}(x)A^\nu{}_{,\rho}(x)A^\rho{}_{,\sigma}(x)A^\sigma{}_{,\mu}(x).\tag{2-6}$$

The inverse function $\Omega^\mu_\nu(x)$ of $J^\mu_\nu(x)$ satisfies

$$J^\mu_\lambda(x)\Omega^\lambda_\nu(x) = \delta^\mu_\nu , \quad \Omega^\mu_\lambda(x)J^\lambda_\nu(x) = \delta^\mu_\nu. \tag{2-7}$$

Concretely, we have

$$\Omega^\mu_\nu(x) = \frac{1}{J(x)}\left(\delta^\mu_\nu + \sum_{n=1}^{3}\frac{a^n}{n!}\Omega^\mu_{\nu(n)}(x)\right), \tag{2-8}$$

$$\Omega^\mu_{\nu(1)}(x) = \delta^\mu_\nu J_{(1)}(x) - A^\mu{}_{,\nu}(x), \tag{2-9}$$

$$\Omega^\mu_{\nu(2)}(x) = \delta^\mu_\nu J_{(2)}(x) - 2A^\mu{}_{,\nu}(x)J_{(1)}(x) + 2A^\mu{}_{,\lambda}(x)A^\lambda{}_{,\nu}(x), \tag{2-10}$$

$$\Omega^\mu_{\nu(3)}(x) = \delta^\mu_\nu J_{(3)}(x) - 3A^\mu{}_{,\nu}(x)J_{(2)}(x) + 6A^\mu{}_{,\lambda}(x)A^\lambda{}_{,\nu}(x)J_{(1)}(x) \\ - 6A^\mu{}_{,\alpha}(x)A^\alpha{}_{,\beta}(x)A^\beta{}_{,\nu}(x). \tag{2-11}$$

And we can prove the following result by direct calculation:

$$\Omega^\mu_{\lambda(3)}(x)A^\lambda{}_{,\nu}(x) = \frac{1}{4}\delta^\mu_\nu J_{(4)}(x). \tag{2-12}$$

According to the above formulas (2-8) ~ (2-12), we can obtain another form of $\Omega^\mu_\nu(x)$:

$$\Omega^\mu_\nu(x) = \delta^\mu_\nu - \frac{a}{J}\sum_{n=0}^{3}\frac{a^n}{n!}\omega^\mu_{\nu(n)}(x), \tag{2-13}$$

$$\omega^\mu_{\nu(0)}(x) = A^\mu{}_{,\nu}(x), \tag{2-14}$$

$$\omega^\mu_{\nu(1)}(x) = A^\mu{}_{,\nu}(x)J_{(1)}(x) - A^\mu{}_{,\lambda}(x)A^\lambda{}_{,\nu}(x), \tag{2-15}$$

$$\omega^\mu_{\nu(2)}(x) = A^\mu{}_{,\nu}(x)J_{(2)}(x) - 2A^\mu{}_{,\lambda}(x)A^\lambda{}_{,\nu}(x)J_{(1)}(x) + 2A^\mu{}_{,\alpha}(x)A^\alpha{}_{,\beta}(x)A^\beta{}_{,\nu}(x), \tag{2-16}$$

$$\omega^\mu_{\nu(3)}(x) = A^\mu{}_{,\nu}(x)J_{(3)}(x) - 3A^\mu{}_{,\lambda}(x)A^\lambda{}_{,\nu}(x)J_{(2)}(x) \\ + 6A^\mu{}_{,\alpha}(x)A^\alpha{}_{,\beta}(x)A^\beta{}_{,\nu}(x)J_{(1)}(x) - 6A^\mu{}_{,\alpha}(x)A^\alpha{}_{,\beta}(x)A^\beta{}_{,\gamma}(x)A^\gamma{}_{,\nu}(x). \tag{2-17}$$

If we set

$$\Omega^\mu_\nu(x) = \sum_{n=0}^{\infty}a^n\widetilde{\Omega}^\mu_{\nu(n)}(x), \tag{2-18}$$

substituting the above form to (2-7), and comparing the coefficients of the power of $a$ one by one, we obtain

$$\widetilde{\Omega}^\mu_{\nu(0)}(x) = \delta^\mu_\nu, \widetilde{\Omega}^\mu_{\nu(1)}(x) = -A^\mu{}_{,\nu}(x), \widetilde{\Omega}^\mu_{\nu(2)}(x) = A^\mu{}_{,\lambda}(x)A^\lambda{}_{,\nu}(x), \cdots, \\ \widetilde{\Omega}^\mu_{\nu(n)}(x) = -A^\mu{}_{,\lambda}(x)\widetilde{\Omega}^\lambda_{\nu(n-1)}(x) = (-1)^n A^\mu{}_{,\lambda_1}(x)A^{\lambda_1}{}_{,\lambda_2}(x)\cdots A^{\lambda_{n-1}}{}_{,\nu}(x), \cdots. \tag{2-19}$$

Both $J(x)$ and $\Omega^\mu_\nu(x)$ are functions of the first derivative of $A^\mu(x)$:

$$J = J(A^\rho{}_{,\sigma}(x)), \quad \Omega^\mu_\mu = \Omega^\mu_\nu(A^\rho{}_{,\sigma}(x)). \tag{2-20}$$

According to the characteristics of determinant we have

$$dJ(x) = J(x)\Omega_\mu^\nu(x) \cdot dJ_\nu^\mu(x),  \qquad (2\text{-}21)$$

based on the above formula we obtain

$$\frac{\partial J(x)}{\partial x^\rho} = J(x)\Omega_\mu^\nu(x) \cdot \frac{\partial J_\nu^\mu(x)}{\partial x^\rho} = J(x)\Omega_\mu^\nu(x) \cdot \frac{\partial(\delta_\nu^\mu + aA^\mu{}_{,\nu}(x))}{\partial x^\rho} = aJ(x)\Omega_\mu^\nu(x)A^\mu{}_{,\nu,\rho}(x). \qquad (2\text{-}22)$$

$$\begin{aligned}\frac{\partial J(x)}{\partial A^\rho{}_{,\sigma}(x)} &= J(x)\Omega_\mu^\nu(x)\frac{\partial J_\nu^\mu(x)}{\partial A^\rho{}_{,\sigma}(x)} = J(x)\Omega_\mu^\nu(x)\frac{\partial(\delta_\nu^\mu + aA^\mu{}_{,\nu}(x))}{\partial A^\rho{}_{,\sigma}(x)} \\ &= J(x)\Omega_\mu^\nu(x) \cdot a\delta_\rho^\mu \delta_\nu^\sigma = aJ(x)\Omega_\rho^\sigma(x).\end{aligned} \qquad (2\text{-}23)$$

According to (2-7) we have

$$0 = \frac{\partial\bigl(J_\lambda^\mu(x)\Omega_\nu^\lambda(x)\bigr)}{\partial x^\rho} = \frac{\partial(\delta_\lambda^\mu + aA^\mu{}_{,\lambda}(x))}{\partial x^\rho}\Omega_\nu^\lambda(x) + J_\lambda^\mu(x)\frac{\partial \Omega_\nu^\lambda(x)}{\partial x^\rho} = aA^\mu{}_{,\lambda,\rho}(x)\Omega_\nu^\lambda(x) + J_\lambda^\mu(x)\frac{\partial \Omega_\nu^\lambda(x)}{\partial x^\rho},$$

$$\begin{aligned}0 &= \frac{\partial}{\partial A^\rho{}_{,\sigma}(x)}\bigl(J_\lambda^\mu(x)\Omega_\nu^\lambda(x)\bigr) = \frac{\partial(\delta_\lambda^\mu + aA^\mu{}_{,\lambda}(x))}{\partial A^\rho{}_{,\sigma}(x)}\Omega_\nu^\lambda(x) + J_\lambda^\mu(x)\frac{\partial \Omega_\nu^\lambda(x)}{\partial A^\rho{}_{,\sigma}(x)} \\ &= a\delta_\rho^\mu\delta_\lambda^\sigma\Omega_\nu^\lambda(x) + J_\lambda^\mu(x)\frac{\partial\Omega_\nu^\lambda(x)}{\partial A^\rho{}_{,\sigma}(x)} = a\delta_\rho^\mu\Omega_\nu^\sigma(x) + J_\lambda^\mu(x)\frac{\partial\Omega_\nu^\lambda(x)}{\partial A^\rho{}_{,\sigma}(x)},\end{aligned}$$

multiplying $J_\mu^\kappa(x)$ on the above two formulas, respectively, according to (2-7) we obtain

$$\frac{\partial\Omega_\nu^\mu(x)}{\partial x^\lambda} = -a\Omega_\rho^\mu(x)\Omega_\nu^\sigma(x)A^\rho{}_{,\sigma,\lambda}(x), \qquad (2\text{-}24)$$

$$\frac{\partial\Omega_\nu^\mu(x)}{\partial A^\rho{}_{,\sigma}(x)} = -a\Omega_\rho^\mu(x)\Omega_\nu^\sigma(x). \qquad (2\text{-}25)$$

Using the (2-22) and (2-24), we can prove

$$\frac{\partial\bigl(J(x)\Omega_\sigma^\rho(x)\bigr)}{\partial x^\rho} = 0. \qquad (2\text{-}26)$$

## 2.2 The function $\delta^4(x - (y + aA(y)))$

At first, we have

$$\frac{\partial}{\partial y^\sigma}e^{ik_\lambda(y^\lambda + aA^\lambda(y))} = ik_\rho\frac{\partial(y^\rho + aA^\rho(y))}{\partial y^\sigma}e^{ik_\lambda(y^\lambda + aA^\lambda(y))} = ik_\rho J_\sigma^\rho(y)e^{ik_\lambda(y^\lambda + aA^\lambda(y))}, \qquad (2\text{-}27)$$

multiplying $\Omega_\zeta^\sigma(y)$ on the above formula, according to (2-7) we have

$$ik_\sigma e^{ik_\lambda(y^\lambda + aA^\lambda(y))} = \Omega_\sigma^\rho(y)\frac{\partial}{\partial y^\rho}e^{ik_\lambda(y^\lambda + aA^\lambda(y))}; \qquad (2\text{-}28)$$

And we have

$$\frac{\partial}{\partial A^\mu(y)}e^{ik_\lambda(y^\lambda + aA^\lambda(y))} = aik_\mu e^{ik_\lambda(y^\lambda + aA^\lambda(y))}. \qquad (2\text{-}29)$$

According to (2-27) ~ (2-29) we obtain

$$\frac{\partial}{\partial y^{\sigma}} e^{ik_{\lambda}(x^{\lambda}-(y^{\lambda}+aA^{\lambda}(y)))} = -ik_{\rho}J_{\sigma}^{\rho}(y)e^{ik_{\lambda}(x^{\lambda}-(y^{\lambda}+aA^{\lambda}(y)))} = -J_{\sigma}^{\rho}(y)\frac{\partial}{\partial x^{\rho}}e^{ik_{\lambda}(x^{\lambda}-(y^{\lambda}+aA^{\lambda}(y)))}, \quad (2\text{-}30)$$

$$\frac{\partial}{\partial x^{\sigma}} e^{ik_{\lambda}(x^{\lambda}-(y^{\lambda}+aA^{\lambda}(y)))} = ik_{\sigma}e^{ik_{\lambda}(x^{\lambda}-(y^{\lambda}+aA^{\lambda}(y)))} = -\Omega_{\sigma}^{\rho}(y)\frac{\partial}{\partial y^{\rho}}e^{ik_{\lambda}(x^{\lambda}-(y^{\lambda}+aA^{\lambda}(y)))}, \quad (2\text{-}31)$$

$$\frac{\partial}{\partial A^{\mu}(y)} e^{ik_{\lambda}(x^{\lambda}-(y^{\lambda}+aA^{\lambda}(y)))} = -a\frac{\partial}{\partial x^{\mu}}e^{ik_{\lambda}(x^{\lambda}-(y^{\lambda}+aA^{\lambda}(y)))} = a\Omega_{\mu}^{\nu}(y)\frac{\partial}{\partial y^{\nu}}e^{ik_{\lambda}(x^{\lambda}-(y^{\lambda}+aA^{\lambda}(y)))}. \quad (2\text{-}32)$$

Hence, for the function

$$\delta^{4}(x-(y+aA(y))) = \int \frac{d^{4}k}{(2\pi)^{4}} e^{ik_{\lambda}(x^{\lambda}-(y^{\lambda}+aA^{\lambda}(y)))}, \quad (2\text{-}33)$$

we have

$$\frac{\partial \delta^{4}(x-(y+aA(y)))}{\partial y^{\nu}} = -J_{\nu}^{\mu}(y)\frac{\partial \delta^{4}(x-(y+aA(y)))}{\partial x^{\mu}}, \quad (2\text{-}34)$$

$$\frac{\partial \delta^{4}(x-(y+aA(y)))}{\partial x^{\nu}} = -\Omega_{\nu}^{\mu}(y)\frac{\partial \delta^{4}(x-(y+aA(y)))}{\partial y^{\mu}}; \quad (2\text{-}35)$$

$$\frac{\partial \delta^{4}(x-(y+aA(y)))}{\partial A^{\mu}(y)} = -a\frac{\partial \delta^{4}(x-(y+aA(y)))}{\partial x^{\mu}} = a\Omega_{\mu}^{\nu}(y)\frac{\partial \delta^{4}(x-(y+aA(y)))}{\partial y^{\nu}}. \quad (2\text{-}36)$$

Generally speaking, for arbitrary function $W(x, y) \equiv W(x-(y+aA(y)))$, we can prove easily

$$\frac{\partial W(x-(y+aA(y)))}{\partial x^{\mu}} + \Omega_{\mu}^{\nu}(y)\frac{\partial W(x-(y+aA(y)))}{\partial y^{\nu}} = 0. \quad (2\text{-}37)$$

If $K^{\mu}(x)$ is infinitesimal, then ignoring all the qualities of higher than first infinitesimal, we obtain

$$\begin{aligned}
\delta^{4}(x-(y+aA(y)+aK(y))) &= \int \frac{d^{4}k}{(2\pi)^{4}} e^{ik_{\lambda}(x^{\lambda}-(y^{\lambda}+aA^{\lambda}(y)+aK^{\lambda}(y)))} \\
&= \int \frac{d^{4}k}{(2\pi)^{4}} e^{-aik_{\rho}K^{\rho}(y)} e^{ik_{\lambda}(x^{\lambda}-(y^{\lambda}+aA^{\lambda}(y)))} \approx \int \frac{d^{4}k}{(2\pi)^{4}}(1-aik_{\rho}K^{\rho}(y))e^{ik_{\lambda}(x^{\lambda}-(y^{\lambda}+aA^{\lambda}(y)))} \\
&= \delta^{4}(x-(y+aA(y))) - aK^{\rho}(y)\frac{\partial \delta^{4}(x-(y+aA(y)))}{\partial x^{\rho}} \\
&= \delta^{4}(x-(y+aA(y))) + aK^{\rho}(y)\Omega_{\rho}^{\sigma}(y)\frac{\partial \delta^{4}(x-(y+aA(y)))}{\partial y^{\sigma}}.
\end{aligned} \quad (2\text{-}38)$$

For arbitrary functions $f(y)$ and $g(x)$, using the above formulas and integration by parts, we have

$$\begin{aligned}
\frac{\partial}{\partial x^{\nu}} \int d^{4}y \delta^{4}(x-(y+aA(y)))f(y) &= \int d^{4}y \frac{\partial \delta^{4}(x-(y+aA(y)))}{\partial x^{\nu}} f(y) \\
&= \int d^{4}y \frac{\partial \delta^{4}(x-(y+aA(y)))}{\partial y^{\mu}}(-\Omega_{\nu}^{\mu}(y))f(y) = \int d^{4}y \delta^{4}(x-(y+aA(y)))\frac{\partial(\Omega_{\nu}^{\mu}(y)f(y))}{\partial y^{\mu}},
\end{aligned} \quad (2\text{-}39)$$

$$\frac{\partial}{\partial y^\nu}\int d^4x \delta^4(x-(y+aA(y)))g(x) = \int d^4x \frac{\partial \delta^4(x-(y+aA(y)))}{\partial y^\nu}g(x)$$
$$= \int d^4x \left(-J_\nu^\mu(y)\right)\frac{\partial \delta^4(x-(y+aA(y)))}{\partial x^\mu}g(x) = J_\nu^\mu(y)\int d^4x \delta^4(x-(y+aA(y)))\frac{\partial g(x)}{\partial x^\mu}.$$
(2-40)

For the formula
$$\int d^4\xi \delta^4(\xi-x)f(\xi) = f(x),\tag{2-41}$$

setting
$$\xi^\mu = y^\mu + aA^\mu(y),\tag{2-42}$$

we have
$$d^4\xi = \left\|\frac{\partial(y^\alpha + aA^\alpha(y))}{\partial y^\beta}\right\| d^4y = \left\|J_\beta^\alpha(y)\right\| d^4y = J(y)d^4y;\tag{2-43}$$

Substituting (2-42) and (2-43) to (2-41), we obtain
$$\int J(y)d^4y \delta^4(x-(y+aA(y)))f(y^\mu + aA^\mu(y)) = f(x).\tag{2-44}$$

We see that $J(y)$ is as the Jacobian determinant in the above formula.

Especially, if $f(x) = \delta^4(x-z)$ in (2-41), then from (2-44) we have
$$\int J(y)d^4y \delta^4(x-(y+aA(y)))\delta^4(y+aA(y)-z) = \delta^4(x-z).\tag{2-45}$$

We now consider the integral $\int d^4y \delta^4((y+aA(y))-(z+aA(z)))f(y)$. If we set (2-43) and denote $\xi_0^\mu = z^\mu + aA^\mu(z)$, then the corresponding inverse functions are $y^\mu = y^\mu(\xi)$ and $y^\mu(\xi_0) = z^\mu$, respectively; hence, considering (2-44), we obtain
$$\int d^4y \delta^4((y+aA(y))-(z+aA(z)))f(y) = \int J(y)d^4y \delta^4((y+aA(y))-(z+aA(z)))\frac{f(y)}{J(y)}$$
$$= \int d^4\xi \delta^4(\xi-\xi_0)\frac{f(y(\xi))}{J(y(\xi))} = \frac{f(y(\xi_0))}{J(y(\xi_0))} = \frac{f(z)}{J(z)}.$$
(2-46)

If there is a relation between functions $f(y)$ and $g(x)$:
$$\int d^4y \delta^4(x-(y+aA(y)))f(y) = g(x),\tag{2-47}$$

then by calculating the integral $\int d^4x \delta^4(x-(z+aA(z)))$ and using (2-48) and (2-46), we obtain
$$\int d^4x \delta^4(x-(z+aA(z)))g(x) = \int d^4y \left(\int d^4x \delta^4(x-(z+aA(z)))\delta^4(x-(y+aA(y)))\right)f(y)$$
$$= \int d^4y \delta^4((y+aA(y))-(z+aA(z)))f(y) = \frac{f(z)}{J(z)}.$$
(2-48)

## 2.3 Transverse and longitudinal four-vector, a Green function

At first, we define transverse four-vector and longitudinal four-vector:

$$\text{If } \begin{matrix} W^{\mu}{}_{,\mu}(x) = 0, \\ W^{[\mu,\nu]}(x) = 0, \end{matrix} \text{ then } W^{\mu}(x) \text{ is a } \begin{matrix} \text{transverse} \\ \text{longitudinal} \end{matrix} \text{ four-vector}, \quad (2\text{-}49)$$

where the operator $W^{[\mu,\nu]}(x)$ is defined as

$$W^{[\mu,\nu]}(x) \equiv W^{\mu,\nu}(x) - W^{\nu,\mu}(x) = \eta^{\nu\lambda}\frac{\partial W^{\mu}(x)}{\partial x^{\lambda}} - \eta^{\mu\lambda}\frac{\partial W^{\nu}(x)}{\partial x^{\lambda}}. \quad (2\text{-}50)$$

In this paper, we use $W_{\perp}^{\mu}(x)$ and $W_{//}^{\mu}(x)$ to denote transverse and longitudinal four-vector, respectively.

Formally, arbitrary four-vector $W^{\mu}(x)$ can be separated into transverse part $W_{\perp}^{\mu}(x)$ and longitudinal part $W_{//}^{\mu}(x)$:

$$W^{\mu}(x) = W_{\perp}^{\mu}(x) + W_{//}^{\mu}(x), \quad (2\text{-}51)$$

because

$$W^{\mu}(x) = \int d^4y \delta^4(x-y) W^{\mu}(y) = \int d^4y \frac{d^4k}{(2\pi)^4} e^{ik\cdot(x-y)} \delta^{\mu}_{\nu} W^{\nu}(y)$$
$$= \int \frac{d^4k}{(2\pi)^4} \frac{k^2 \delta^{\mu}_{\nu} - k^{\mu}k_{\nu}}{k^2} e^{ik\cdot(x-y)} W^{\nu}(y) d^4y + \int \frac{d^4k}{(2\pi)^4} \frac{k^{\mu}k_{\nu}}{k^2} e^{ik\cdot(x-y)} W^{\nu}(y) d^4y, \quad (2\text{-}52)$$

we therefore can introduce

$$W_{\perp}^{\mu}(x) = \int \frac{d^4k}{(2\pi)^4} \frac{k^2 \delta^{\mu}_{\nu} - k^{\mu}k_{\nu}}{k^2} e^{ik\cdot(x-y)} W^{\nu}(y) d^4y, \quad (2\text{-}53)$$

$$W_{//}^{\mu}(x) = \int \frac{d^4k}{(2\pi)^4} \frac{k^{\mu}k_{\nu}}{k^2} e^{ik\cdot(x-y)} W^{\nu}(y) d^4y, \quad (2\text{-}54)$$

and, (2-51) thus holds. According to the definition (2-49), we can verify easily that $W_{\perp}^{\mu}(x)$ and $W_{//}^{\mu}(x)$ given by (2-53) and (2-54) are transverse four-vector and longitudinal four-vector indeed, respectively.

As well-known, for arbitrary functions $U^{\mu}(x)$ and $V^{\mu}(x)$, if the equation

$$\frac{D^2 U^{\mu}(x)}{Dx^2} = V^{\mu}(x) \quad (2\text{-}55)$$

has solution, where the definition of the operator $\frac{D^2 U^{\mu}(x)}{Dx^2}$ is given by (1-9), then $\frac{\partial V^{\mu}(x)}{\partial x^{\mu}} = 0$.

And, further, if $\frac{\partial V^{\mu}(x)}{\partial x^{\mu}} = 0$, then the solution of (2-55) can be written to the form:

$$U^{\mu}(x) = \overline{U}^{\mu}(x) + U_{\perp}^{\mu}(x) + U_{//}^{\mu}(x), \quad (2\text{-}56)$$

where $\overline{U}^{\mu}(x)$ is a solution of the equation $\dfrac{D^2 \overline{U}^{\mu}(x)}{Dx^2}=0$,

$$U_{\perp}^{\mu}(x)=\int d^4 y D(x-y) V^{\mu}(y), \qquad (2\text{-}57)$$

where the Green function $D(x-y)$ is

$$D(x)=\int \frac{d^4 k}{(2\pi)^4}\frac{-1}{k^2}e^{-ik\cdot x}=\int \frac{d^3 k}{(2\pi)^3}e^{i\boldsymbol{k}\cdot\boldsymbol{x}}\int \frac{dk_0}{2\pi}\frac{-1}{k_0^2-\boldsymbol{k}\cdot\boldsymbol{k}}e^{-ik_0\cdot x^0}, \qquad (2\text{-}58)$$

which satisfies

$$\eta^{\mu\nu}\frac{\partial^2 D(x-y)}{\partial x^{\mu}\partial x^{\nu}}=\delta^4(x-y). \qquad (2\text{-}59)$$

According to the definition (2-49), we can verify easily that $U_{\perp}^{\mu}(x)$ given by (2-57) is a transverse four-vector; $U_{//}^{\mu}(x)$ in (2-56) is an arbitrary longitudinal four-vector.

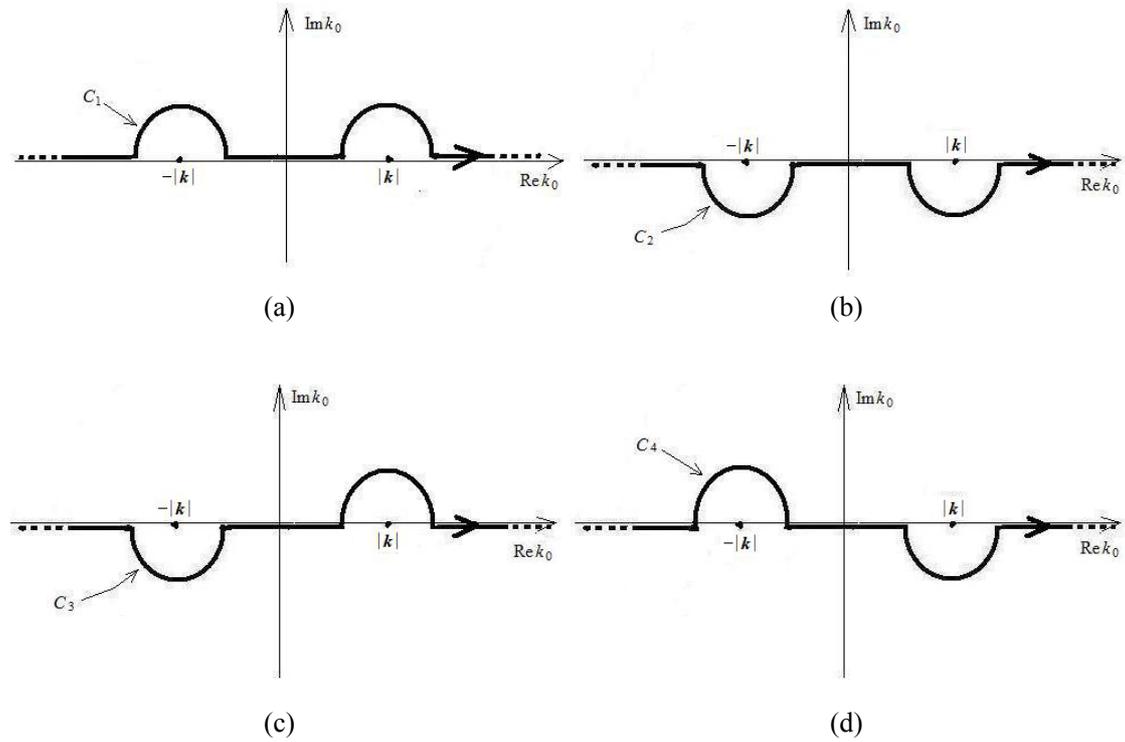

(a)    (b)

(c)    (d)

**Fig.1**

As is well-known, generally speaking, integral $\int\dfrac{d^4 k}{(2\pi)^4}\dfrac{1}{k^2}\Sigma(k)=\int\dfrac{d^4 k}{(2\pi)^4}\dfrac{1}{k_0^2-\boldsymbol{k}\cdot\boldsymbol{k}}\Sigma(k)$

appearing in (2-53), (2-54) and (2-58) is significative only when we designate an appropriate path for the integral of the variable $k_0$. For example, for any one of the four paths $C_1$, $C_2$, $C_3$ and $C_4$ in Fig1. (a), (b), (c) and (d), (2-58) is significative. Especially, (2-58) are so called the retarded, advanced Green functions and the Feynman propagator for the paths $C_1$, $C_2$, $C_3$ in

Fig1. (a), (b), (c), respectively. For the paths $C_1$ and $C_2$, concretely, we have[1, 6]

$$D_{(1) \atop (2)}(x) = D_{\text{ret} \atop \text{adv}}(x) = \theta(\pm x^0)\frac{1}{4\pi|\mathbf{x}|}\delta(x^0 \mp |\mathbf{x}|); \quad \theta(t) = \begin{cases} 1, & t > 0; \\ \frac{1}{2}, & t = 0; \\ 0, & t < 0. \end{cases} \quad (2\text{-}60)$$

In this paper, once integral of the form $\int \frac{d^4k}{(2\pi)^4}\frac{1}{k^2}\Sigma(k)$ appears, we do not show a special path and always assume that an appropriate path has been already implied in it. However, we assume that all paths are the same in different integrals $\int \frac{d^4k}{(2\pi)^4}\frac{1}{k^2}\Sigma_1(k)$, $\int \frac{d^4k}{(2\pi)^4}\frac{1}{k^2}\Sigma_2(k)$, ……

In the discussion below, generally, we no longer point out what formula in this section is used one by one.

## 3  A theory of quantum electrodynamics with nonlocal interaction

### 3.1  An accessorial strength $\widetilde{F}^{\mu\nu}(x)$ and an accessorial 4-potential $\Phi^\mu(x)$

For the electromagnetic field strength $F^{\mu\nu}(x)$ given by (1-5), we define an *accessorial strength* $\widetilde{F}^{\mu\nu}(x)$:

$$\widetilde{F}^{\mu\nu}(x) = \int d^4y\, \delta^4(x-(y+aA(y))) J^\mu_\rho(y) J^\nu_\sigma(y) F^{\rho\sigma}(y), \quad (3\text{-}1)$$

it is obvious that $\widetilde{F}^{\mu\nu}(x)$ is antisymmetric in the pair of indices due to $F^{\rho\sigma}(y) = -F^{\sigma\rho}(y)$:

$$\widetilde{F}^{\mu\nu}(x) = -\widetilde{F}^{\nu\mu}(x). \quad (3\text{-}2)$$

According to the formulas given by §2, we have

$$\varepsilon_{\mu\alpha\beta\gamma}\partial^\alpha \widetilde{F}^{\beta\gamma}(x) = \varepsilon_{\mu\alpha\beta\gamma}\eta^{\alpha\nu}\frac{\partial \widetilde{F}^{\beta\gamma}(x)}{\partial x^\nu} = \varepsilon_{\mu\alpha\beta\gamma}\eta^{\alpha\nu}\frac{\partial}{\partial x^\nu}\int d^4y\,\delta^4(x-(y+aA(y))) J^\beta_\rho(y) J^\gamma_\sigma(y) F^{\rho\sigma}(y)$$

$$= \varepsilon_{\mu\alpha\beta\gamma}\eta^{\alpha\nu}\int d^4y\,\delta^4(x-(y+aA(y)))\frac{\partial}{\partial y^\lambda}\left(\Omega^\lambda_\nu(y) J^\beta_\rho(y) J^\gamma_\sigma(y) F^{\rho\sigma}(y)\right)$$

$$= \int d^4y\,\delta^4(x-(y+aA(y)))\frac{1}{J^2(y)}\left(2a\varepsilon_{\mu\alpha\beta\gamma}\left(A^{\alpha,\beta}(y)A^{\lambda,\gamma}_{,\lambda}(y) + 2A^{\lambda,\alpha}(y)A^{\beta,\gamma}_{,\lambda}(y)\right) + O(a^2)\right),$$

we see $\varepsilon_{\mu\alpha\beta\gamma}\partial^\alpha \widetilde{F}^{\beta\gamma}(x) \neq 0$, this means that there is not any function $\phi^\mu(x)$ such that

$$\widetilde{F}_{\mu\nu}(x) = \frac{\partial \phi_\nu(x)}{\partial x^\mu} - \frac{\partial \phi_\mu(x)}{\partial x^\nu}.$$

If we define

$$\bar{\bar{F}}^{\mu\nu}(x) = \tilde{F}^{\mu\nu}(x) + \int d^4 y \delta^4(x - (y + aA(y))) l^{\mu\nu}(y)$$

$$= \int d^4 y \delta^4(x - (y + aA(y))) \left( J^\mu_\rho(y) J^\nu_\sigma(y) F^{\rho\sigma}(y) + l^{\mu\nu}(y) \right),$$

and ask that the following three conditions hold:

① $\bar{\bar{F}}^{\mu\nu}(x) = -\bar{\bar{F}}^{\nu\mu}(x)$, which leads to $l^{\mu\nu}(y) = l^{\nu\mu}(y)$;

② $\dfrac{\partial \bar{\bar{F}}^{\mu\nu}(x)}{\partial x^\nu} = \dfrac{\partial \tilde{F}^{\mu\nu}(x)}{\partial x^\nu}$, which leads to $\dfrac{\partial (\Omega^\lambda_\nu(y) l^{\mu\nu}(y))}{\partial y^\lambda} = 0$;

③ $\varepsilon_{\mu\alpha\beta\gamma} \partial^\alpha \bar{\bar{F}}^{\beta\gamma}(x) = 0$, which leads to

$$\varepsilon_{\mu\alpha\beta\gamma} \eta^{\alpha\nu} \dfrac{\partial}{\partial y^\lambda} \left( \Omega^\lambda_\nu(y) \left( J^\beta_\rho(y) J^\gamma_\sigma(y) F^{\rho\sigma}(y) + l^{\beta\gamma}(y) \right) \right) = 0,$$

then we can prove that such $l^{\mu\nu}(y)$, and, further, such $\bar{\bar{F}}^{\mu\nu}(x)$, do not exist.

On the other hand, according to (3-2) we have

$$\frac{\partial^2 \tilde{F}^{\mu\nu}(x)}{\partial x^\mu \partial x^\nu} = \frac{\partial}{\partial x^\mu} \left( \frac{\partial \tilde{F}^{\mu\nu}(x)}{\partial x^\nu} \right) = 0, \tag{3-3}$$

we therefore can define an *accessorial 4-potential* $\Phi^\mu(x)$ which satisfies:

$$\frac{D^2 \Phi^\mu(x)}{Dx^2} = -\frac{\partial \tilde{F}^{\mu\nu}(x)}{\partial x^\nu}. \tag{3-4}$$

The solution of (3-4) is

$$\Phi^\mu(x) = \Phi^\mu_\perp(x) + \Phi^\mu_{//}(x), \tag{3-5}$$

$$\Phi^\mu_\perp(x) = \int d^4 y D(x - y) \left( -\frac{\partial \tilde{F}^{\mu\nu}(y)}{\partial y^\nu} \right). \tag{3-6}$$

Notice that as the definition of $\Phi^\mu_\perp(x)$, we ignore the part $\bar{\Phi}^\mu(x)$ that satisfies $\dfrac{D^2 \bar{\Phi}^\mu(x)}{Dx^2} = 0$ in the solution of the equation (3-4), and, on the other hand, we can choose arbitrary one of the two paths $C_1$ and $C_2$ in Fig. 1, even some combination of different paths, but not be localized a special path for the function $D(x - y)$ in (3-6). For example, if we do not consider the paths $C_3$ and $C_4$ in classical case, then we can define

$$\Phi^\mu_\perp(x) = \sum_{i=1}^{2} d_i \Phi^\mu_{(i)\perp}(x), \quad \Phi^\mu_{(i)\perp}(x) = \int d^4 y D_{(i)}(x - y) \left( -\frac{\partial \tilde{F}^{\mu\nu}(y)}{\partial y^\nu} \right) \tag{3-7}$$

where both $d_1$ and $d_2$ are constants, $D_{(i)}(x)$ ($i = 1, 2$) are given by (2-60). In this paper, we only discuss the form of (3-6), the method analyzing (3-6) can be generated to the form of (3-7) easily.

If $\Phi_\perp^\mu(x)$ given by (3-6) could be written to the form of

$$W_\perp^\mu(x) = \int \frac{\mathrm{d}^4 k}{(2\pi)^4} \frac{k^2 \delta_\nu^\mu - k^\mu k_\nu}{k^2} \mathrm{e}^{ik\cdot(x-y)} \widetilde{W}^\nu(y) \mathrm{d}^4 y ,$$

then we could find out the corresponding longitudinal part $W_{//}^\mu(x) = \int \frac{\mathrm{d}^4 k}{(2\pi)^4} \frac{k^\mu k_\nu}{k^2} \mathrm{e}^{ik\cdot(x-y)} \widetilde{W}^\nu(y) \mathrm{d}^4 y$ naturally. But I cannot write $\Phi_\perp^\mu(x)$ to such form (see the Appendix A of this paper), hence, determining the form of $W_{//}^\mu(x)$, and, further, that of $W^\mu(x)$, is arbitrary to a certain extent.

For example, if we define

$$\Phi_{(0)}^\mu(x) \equiv \int \mathrm{d}^4 y \, \delta^4(x - (y + aA(y))) A^\mu(y), \tag{3-8}$$

then we can try to choose $\Phi_{//}^\mu(x) = \int \frac{\mathrm{d}^4 k}{(2\pi)^4} \frac{k^\mu k_\nu}{k^2} \mathrm{e}^{ik\cdot(x-y)} \Phi_{(0)}^\mu(y) \mathrm{d}^4 y$. However, in this paper, we choose

$$\begin{aligned}
\Phi_{//}^\mu(x) &= \int \frac{\mathrm{d}^4 k}{(2\pi)^4} \frac{k^\mu k_\nu}{k^2} \mathrm{e}^{ik\cdot(x-y)} \int \mathrm{d}^4 y \left( \int \mathrm{d}^4 z \, \delta^4(y - (z + aA(z))) \left( b_1 J_\lambda^\nu(z) A^\lambda(z) + b_2 J(z) A^\nu(z) \right) \right) \\
&= \int \frac{\mathrm{d}^4 k}{(2\pi)^4} \frac{k^\mu k_\nu}{k^2} \mathrm{e}^{ik\cdot(x - (z + aA(z)))} \left( b_1 J_\lambda^\nu(z) A^\lambda(z) + b_2 J(z) A^\nu(z) \right) \\
&= \frac{\partial^2}{\partial x_\mu \partial x^\nu} \int \mathrm{d}^4 y \, D(x - (y + aA(y))) \left( b_1 J_\lambda^\nu(y) A^\lambda(y) + b_2 J(y) A^\nu(y) \right) \\
&\equiv \eta^{\mu\nu} \frac{\partial \Theta(x)}{\partial x^\nu} = \partial^\mu \Theta(x),
\end{aligned} \tag{3-9}$$

where

$$\begin{aligned}
\Theta(x) &= \frac{\partial}{\partial x^\nu} \int \mathrm{d}^4 y \, D(x - (y + aA(y))) \left( b_1 J_\lambda^\nu(y) A^\lambda(y) + b_2 J(y) A^\nu(y) \right) \\
&= -\int \mathrm{d}^4 y \, \frac{D(x - (y + aA(y)))}{\partial y^\mu} \Omega_\nu^\mu(y) \left( b_1 J_\lambda^\nu(y) A^\lambda(y) + b_2 J(y) A^\nu(y) \right) \\
&= \int \mathrm{d}^4 y \, D(x - (y + aA(y))) \frac{\partial}{\partial y^\mu} \left( b_1 A^\mu(y) + b_2 J(y) \Omega_\nu^\mu(y) A^\nu(y) \right) \\
&= \int \mathrm{d}^4 y \, D(x - (y + aA(y))) \left( b_1 \frac{\partial A^\lambda(y)}{\partial y^\lambda} + b_2 J(y) \Omega_\rho^\sigma(y) \frac{\partial A^\rho(y)}{\partial y^\sigma} \right);
\end{aligned} \tag{3-10}$$

Both $b_1$ and $b_2$ in (3-9) are constants, and

$$b_1 + b_2 = 1. \tag{3-11}$$

If we use (3-7), then we can choose

$$\Phi_{//}^{\mu}(x) = \sum_{i=1}^{2} d_i \Phi_{(i)//}^{\mu}(x), \quad \Phi_{(i)//}^{\mu}(x) = \partial^{\mu}\Theta_{(i)}(x),$$

$$\Theta_{(i)}(x) = \int d^4 y D_{(i)}(x-(y+aA(y)))\left(b_1 \frac{\partial A^{\lambda}(y)}{\partial y^{\lambda}} + b_2 J(y)\Omega_{\nu}^{\sigma}(y)\frac{\partial A^{\nu}(y)}{\partial y^{\sigma}}\right) \quad (3\text{-}12)$$

as the corresponding longitudinal four-vector. However, we only discuss the form of (3-9).

The function $\Phi^{\mu}(x) = \Phi_{\perp}^{\mu}(x) + \Phi_{//}^{\mu}(x)$ is thus fully determined. Some other expressions of $\Phi_{\perp}^{\mu}(x)$ and $\Phi^{\mu}(x)$ will be found in the Appendix A of this paper.

### 3.2 The action and the equations of motion of charged particle and electromagnetic field

The action of the system reads

$$S = \int d^4 x L(x) = \int d^4 x (L_D(x) + L_{EM}(x) + L_I(x)), \quad (3\text{-}13)$$

$$L_{EM}(x) = -\frac{1}{4}(\Phi_{\mu,\nu}(x) - \Phi_{\nu,\mu}(x))(\Phi^{\mu,\nu}(x) - \Phi^{\nu,\mu}(x))$$

$$= -\frac{1}{4}\eta_{\mu\rho}\eta_{\nu\sigma}(\Phi_{\perp}^{\mu,\nu}(x) - \Phi_{\perp}^{\nu,\mu}(x))(\Phi_{\perp}^{\rho,\sigma}(x) - \Phi_{\perp}^{\sigma,\rho}(x)), \quad (3\text{-}14)$$

$$L_I(x) = -ej^{\mu}(x)\Phi_{\mu}(x) = -ej^{\mu}(x)\Phi_{\mu\perp}(x) - ej^{\mu}(x)\Phi_{\mu//}(x), \quad (3\text{-}15)$$

where $L_D(x)$ and $j^{\mu}(x)$ are still given by (1-2) and (1-6), respectively; $\Phi_{\mu\perp}(x)$ and $\Phi_{\mu//}(x)$ are given by (3-6) and (3-9), respectively.

If we use (3-7) and (3-12), then instead of (3-14), we have

$$L_{EM}(x) = \sum_{i=1}^{2} d_i L_{EM(i)}(x),$$

$$L_{EM(i)}(x) = -\frac{1}{4}(\Phi_{(i)\mu,\nu}(x) - \Phi_{(i)\nu,\mu}(x))(\Phi_{(i)}^{\mu,\nu}(x) - \Phi_{(i)}^{\nu,\mu}(x)) \quad (3\text{-}16)$$

$$= -\frac{1}{4}\eta_{\mu\rho}\eta_{\nu\sigma}(\Phi_{(i)\perp}^{\mu,\nu}(x) - \Phi_{(i)\perp}^{\nu,\mu}(x))(\Phi_{(i)\perp}^{\rho,\sigma}(x) - \Phi_{(i)\perp}^{\sigma,\rho}(x));$$

For this case, the forms of $\Phi_{\mu\perp}(x)$ and $\Phi_{\mu//}(x)$ in (3-15) are (3-7) and (3-12), respectively. However, we only discuss the form of (3-14).

According to the Euler-Langrange equation $\frac{\partial L(x)}{\partial \overline{\psi}(x)} - \partial_{\mu}\frac{\partial L(x)}{\partial(\partial_{\mu}\overline{\psi}(x))} = 0$ we obtain the equation of motion of charged particle field immediately

$$\left(i\gamma^{\mu}\frac{\partial}{\partial x^{\mu}} - m\right)\psi(x) = e\gamma^{\mu}\Phi_{\mu}(x)\psi(x) = e\gamma^{\mu}\Phi_{\mu\perp}(x)\psi(x) + e\gamma^{\mu}\Phi_{\mu//}(x)\psi(x). \quad (3\text{-}17)$$

According to (3-17), the equation of motion of free particle field is the usual form:

$$\left(i\gamma^{\mu}\frac{\partial}{\partial x^{\mu}} - m\right)\psi(x) = 0. \quad (3\text{-}18)$$

We can verify easily that the equation (3-17) leads to the current conservation equation (1-10).

The classical limit of the equation (3-17) is

$$m\frac{dU^{\mu}}{d\tau} = e\left(\Phi^{\mu,\nu}(x) - \Phi^{\nu,\mu}(x)\right)U_{\nu} = e\left(\Phi_{\perp}^{\mu,\nu}(x) - \Phi_{\perp}^{\nu,\mu}(x)\right)U_{\nu},$$

where $U^{\mu} = \dfrac{dx^{\mu}}{d\tau}$, this is a generalization of the Lorentz force.

In the Appendix B of this paper, we shall prove that the variational equation $\dfrac{\delta S}{\delta A^{\mu}(y)} = 0$ leads to the equation of motion of electromagnetic field

$$-\frac{\partial F^{\mu\nu}(y)}{\partial y^{\nu}} = \frac{D^2}{Dy^2}A^{\mu}(y) = eJ(y)\Omega_{\nu}^{\mu}(y)\int d^4x j^{\nu}(x)\delta^4(x-(y+aA(y))) \qquad (3\text{-}19)$$
$$= eJ(y)\Omega_{\nu}^{\mu}(y)j^{\nu}(y+aA(y)).$$

Among the four equations of (3-19), there is not any second time derivative term in the equation corresponding to $\mu = 0$:

$$\frac{\partial}{\partial y^i}\left(A^{0,i}(y) - A^{i,0}(y)\right) = eJ(y)\Omega_{\nu}^{0}(y)\int d^4x j^{\nu}(x)\delta^4(x-(y+aA(y))); \qquad (3\text{-}20)$$

and there is a second time derivative term in the rest three equation corresponding to $\mu = i = 1, 2, 3$, respectively:

$$\frac{\partial}{\partial y^0}\left(A^{i,0}(y) - A^{0,i}(y)\right) + \frac{\partial}{\partial y^j}\left(A^{i,j}(y) - A^{j,i}(y)\right)$$
$$= eJ(y)\Omega_{\nu}^{i}(y)\int d^4x j^{\nu}(x)\delta^4(x-(y+aA(y))). \qquad (3\text{-}21)$$

It is important that according to (3-19) the equation of motion of free electromagnetic field is the usual form:

$$\frac{D^2}{Dy^2}A^{\mu}(y) = -\frac{\partial F^{\mu\nu}(y)}{\partial y^{\nu}} = 0. \qquad (3\text{-}22)$$

From (3-19) we have

$$0 = -\frac{\partial^2 F^{\mu\nu}(y)}{\partial y^{\mu}\partial y^{\nu}} = \frac{\partial}{\partial y^{\mu}}\left(\frac{D^2}{Dy^2}A^{\mu}(y)\right) = e\frac{\partial}{\partial y^{\mu}}\left(J(y)\Omega_{\nu}^{\mu}(y)\int d^4x j^{\nu}(x)\delta^4(x-(y+aA(y)))\right)$$
$$= e\frac{\partial(J(y)\Omega_{\nu}^{\mu}(y))}{\partial y^{\mu}}\int d^4x j^{\nu}(x)\delta^4(x-(y+aA(y))) + eJ(y)\Omega_{\nu}^{\mu}(y)\int d^4x j^{\nu}(x)\frac{\partial \delta^4(x-(y+aA(y)))}{\partial y^{\mu}}$$
$$= -eJ(y)\Omega_{\nu}^{\mu}(y)J_{\mu}^{\lambda}(y)\int d^4x j^{\nu}(x)\frac{\partial \delta^4(x-(y+aA(y)))}{\partial x^{\lambda}} = eJ(y)\delta_{\nu}^{\lambda}\int d^4x \frac{\partial j^{\nu}(x)}{\partial x^{\lambda}}\delta^4(x-(y+aA(y)))$$
$$= eJ(y)\int d^4x \frac{\partial j^{\lambda}(x)}{\partial x^{\lambda}}\delta^4(x-(y+aA(y))), \qquad (3\text{-}23)$$

we therefore see that (3-19) leads to the current conservation equation (1-10).

If we calculate $\int d^4y \delta^4(x-(y+aA(y)))J_{\mu}^{\rho}(y)$ for (3-19), then we obtain:

$$ej^{\mu}(x) = -\int d^4 y \delta^4(x-(y+aA(y)))J_{\rho}^{\mu}(y)\frac{\partial F^{\rho\sigma}(y)}{\partial y^{\sigma}}$$
$$= \int d^4 y \delta^4(x-(y+aA(y)))J_{\nu}^{\mu}(y)\frac{D^2}{Dy^2}A^{\nu}(y);$$
(3-24)

This is an equivalent form of (3-19); and from (3-24) we have

$$e\frac{\partial j^{\mu}(x)}{\partial x^{\mu}} = -\int d^4 y \frac{\partial \delta^4(x-(y+aA(y)))}{\partial x^{\mu}}J_{\rho}^{\mu}(y)\frac{\partial F^{\rho\sigma}(y)}{\partial y^{\sigma}}$$
$$= \int d^4 y \frac{\partial \delta^4(x-(y+aA(y)))}{\partial y^{\nu}}\Omega_{\mu}^{\nu}(y)J_{\rho}^{\mu}(y)\frac{\partial F^{\rho\sigma}(y)}{\partial y^{\sigma}}$$
$$= -\int d^4 y \delta^4(x-(y+aA(y)))\frac{\partial}{\partial y^{\nu}}\left(\delta_{\rho}^{\nu}\frac{\partial F^{\rho\sigma}(y)}{\partial y^{\sigma}}\right)$$
$$= -\int d^4 y \delta^4(x-(y+aA(y)))\frac{\partial^2 F^{\rho\sigma}(y)}{\partial y^{\rho}\partial y^{\sigma}} = 0,$$

we therefore obtain the current conservation equation (1-10) again from (3-24).

Notice $A^{\mu}{}_{,\rho,\sigma}F^{\rho\sigma}(y) = 0$ due to $F^{\rho\sigma}(y) = -F^{\sigma\rho}(y)$, we have,

$$\frac{\partial (J_{\rho}^{\mu}(y)F^{\rho\sigma}(y))}{\partial y^{\sigma}} = \frac{\partial J_{\rho}^{\mu}(y)}{\partial y^{\sigma}}F^{\rho\sigma}(y) + J_{\rho}^{\mu}(y)\frac{\partial F^{\rho\sigma}(y)}{\partial y^{\sigma}}$$
$$= aA^{\mu}{}_{,\rho,\sigma}F^{\rho\sigma}(y) + J_{\rho}^{\mu}(y)\frac{\partial F^{\rho\sigma}(y)}{\partial y^{\sigma}} = J_{\rho}^{\mu}(y)\frac{\partial F^{\rho\sigma}(y)}{\partial y^{\sigma}},$$

And, further

$$\int d^4 y \delta^4(x-(y+aA(y)))J_{\rho}^{\mu}(y)\frac{\partial F^{\rho\sigma}(y)}{\partial y^{\sigma}}$$
$$= \int d^4 y \delta^4(x-(y+aA(y)))\frac{\partial (J_{\rho}^{\mu}(y)F^{\rho\sigma}(y))}{\partial y^{\sigma}}$$
$$= -\int d^4 y \frac{\partial \delta^4(x-(y+aA(y)))}{\partial y^{\sigma}}J_{\rho}^{\mu}(y)F^{\rho\sigma}(y)$$
$$= \int d^4 y \frac{\partial \delta^4(x-(y+aA(y)))}{\partial x^{\nu}}J_{\sigma}^{\nu}(y)J_{\rho}^{\mu}(y)F^{\rho\sigma}(y)$$
$$= \frac{\partial}{\partial x^{\nu}}\int d^4 y \delta^4(x-(y+aA(y)))J_{\rho}^{\mu}(y)J_{\sigma}^{\nu}(y)F^{\rho\sigma}(y)$$
$$= \frac{\partial \widetilde{F}^{\mu\nu}(x)}{\partial x^{\nu}} = -\frac{D^2\Phi^{\mu}(x)}{Dx^2};$$
(3-25)

Taking advantage of (3-25), from (3-24) we obtain the third form of the equation of motion of electromagnetic field:

$$\frac{D^2\Phi^{\mu}(x)}{Dx^2} = -\frac{\partial \widetilde{F}^{\mu\nu}(x)}{\partial x^{\nu}} = ej^{\mu}(x).$$
(3-26)

Considering (3-3), from (3-26) we can obtain the current conservation equation (1-10) immediately.

## 4  Gauge invariance

We investigate the transformational relations between two groups of fields $\psi'(x), A'^{\mu}(x)$ and $\psi(x), A^{\mu}(x)$. We assume that the transformational relation between $\psi'(x)$ and $\psi(x)$ is still (1-12), however, instead of (1-11), we assume that the transformational relation between $A'^{\mu}(x)$ and $A^{\mu}(x)$ is

$$A'^{\mu}(x) = A^{\mu}(x) + K^{\mu}(x). \tag{4-1}$$

At first, we ask that the equation of motion of electromagnetic field (3-26) is invariable under the transformation (1-12) and (4-1). Due to

$$j'^{\mu}(x) = \overline{\psi'}(x)\gamma^{\mu}\psi'(x) = \overline{\psi}(x)e^{ie\chi(x)}\gamma^{\mu}e^{-ie\chi(x)}\psi(x) = \overline{\psi}(x)\gamma^{\mu}\psi(x) = j^{\mu}(x),$$

we see that the equation of motion of electromagnetic field (3-26) is invariable under the transformation (1-12) and (4-1) if and only if

$$\frac{\partial \widetilde{F}'^{\mu\nu}(x)}{\partial x^{\nu}} = \frac{\partial \widetilde{F}^{\mu\nu}(x)}{\partial x^{\nu}}. \tag{4-2}$$

under the transformation (4-1). For guaranteeing that (4-2) holds, what condition must be satisfied for the function $K^{\mu}(x)$? We discuss this question for the cases that $K^{\mu}(x)$ is infinitesimal and finite, respectively.

① The case that $K^{\mu}(x)$ is infinitesimal

According to (4-1), we have

$$\begin{aligned}
J'^{\mu}_{\rho}(y) &= \frac{\partial(y^{\mu} + aA'^{\mu}(y))}{\partial y^{\rho}} = \frac{\partial(y^{\mu} + aA^{\mu}(y))}{\partial y^{\rho}} + \frac{\partial(aK^{\mu}(y))}{\partial y^{\rho}} = J^{\mu}_{\rho}(y) + aK^{\mu}{}_{,\rho}(y), \\
F'^{\rho\sigma}(y) &= A'^{\sigma,\rho}(y) - A'^{\rho,\sigma}(y) = A^{\sigma,\rho}(y) - A^{\rho,\sigma}(y) + K^{\sigma,\rho}(y) - K^{\rho,\sigma}(y) \\
&= F^{\rho\sigma}(y) - (K^{\rho,\sigma}(y) - K^{\sigma,\rho}(y)),
\end{aligned} \tag{4-3}$$

when $K^{\mu}(x)$ is infinitesimal, from (3-25) for $A'^{\mu}(x)$ and (4-1) we have

$$\begin{aligned}
\frac{\partial \widetilde{F}'^{\mu\nu}(x)}{\partial x^{\nu}} &= \int d^4 y\, \delta^4(x-(y+aA'(y))) J'^{\mu}_{\rho}(y) \frac{\partial F'^{\rho\sigma}(y)}{\partial y^{\sigma}} \\
&= \int d^4 y\, \delta^4(x-(y+aA(y)+aK(y)))\left(J^{\mu}_{\rho}(y)+aK^{\mu}{}_{,\rho}(y)\right)\left[\frac{\partial F^{\rho\sigma}(y)}{\partial y^{\sigma}} - \frac{\partial(K^{\rho,\sigma}(y)-K^{\sigma,\rho}(y))}{\partial y^{\sigma}}\right] \\
&= \int d^4 y \left[\delta^4(x-(y+aA(y))) + a\Omega^{\tau}_{\lambda}(y)K^{\lambda}(y)\frac{\partial \delta^4(y+aA(y)-x)}{\partial y^{\tau}}\right] \\
&\quad \times \left(J^{\mu}_{\rho}(y)+aK^{\mu}{}_{,\rho}(y)\right)\left(\frac{\partial F^{\rho\sigma}(y)}{\partial y^{\sigma}} - \frac{D^2 K^{\rho}(y)}{Dy^2}\right) \\
&= \int d^4 y\, \delta^4(x-(y+aA(y)))\left[J^{\mu}_{\rho}(y)\frac{\partial F^{\rho\sigma}(y)}{\partial y^{\sigma}} - J^{\mu}_{\rho}(y)\frac{D^2 K^{\rho}(y)}{Dy^2} + aK^{\mu}{}_{,\rho}(y)\frac{\partial F^{\rho\sigma}(y)}{\partial y^{\sigma}}\right. \\
&\quad \left. - a\frac{\partial}{\partial y^{\tau}}\left(\Omega^{\tau}_{\lambda}(y)K^{\lambda}(y)J^{\mu}_{\rho}(y)\frac{\partial F^{\rho\sigma}(y)}{\partial y^{\sigma}}\right)\right].
\end{aligned} \tag{4-4}$$

In the above expression, we have ignored all the qualities of higher than first infinitesimal, such as

$$K^\mu_{\ ,\rho}(y)\frac{D^2 K^\rho(y)}{Dy^2},\ K^\lambda(y)K^\mu_{\ ,\rho}(y)\frac{\partial F^{\rho\sigma}(y)}{\partial y^\sigma}\text{ , etc.}$$

Due to (4-2), we compare (4-4) and $\dfrac{\partial \widetilde{F}^{\mu\nu}(x)}{\partial x^\nu}$ given by (3-25) and have

$$-J^\mu_\rho(y)\frac{D^2 K^\rho(y)}{Dy^2}+aK^\mu_{\ ,\rho}(y)\frac{\partial F^{\rho\sigma}(y)}{\partial y^\sigma}-a\frac{\partial}{\partial y^\tau}\left(\Omega^\tau_\lambda(y)K^\lambda(y)J^\mu_\rho(y)\frac{\partial F^{\rho\sigma}(y)}{\partial y^\sigma}\right)=0,$$

multiplying $\Omega^\nu_\mu(y)$, we obtain

$$\frac{D^2 K^\mu(y)}{Dy^2}=a\Omega^\mu_\nu(y)\left[K^\nu_{\ ,\rho}(y)\frac{\partial F^{\rho\sigma}(y)}{\partial y^\sigma}-\frac{\partial}{\partial y^\tau}\left(\Omega^\tau_\lambda(y)K^\lambda(y)J^\nu_\rho(y)\frac{\partial F^{\rho\sigma}(y)}{\partial y^\sigma}\right)\right], \qquad (4\text{-}5)$$

the infinitesimal $K^\mu(x)$ thus can be determined by the equation (4-5).

② The case that $K^\mu(x)$ is finite

If we want to use the method similar to that of discussing the case that $K^\mu(x)$ is infinitesimal to discuss the case that $K^\mu(x)$ is finite, then at first we have

$$\delta^4(x-(y+aA(y)+aK(y)))=\int\frac{\mathrm{d}^4 k}{(2\pi)^4}\mathrm{e}^{ik_\lambda(x^\lambda-(y^\lambda+aA^\lambda(y)+aK^\lambda(y)))}$$

$$=\int\frac{\mathrm{d}^4 k}{(2\pi)^4}\mathrm{e}^{-aik_\tau K^\tau(y)}\mathrm{e}^{ik_\lambda(x^\lambda-(y^\lambda+aA^\lambda(y)))}$$

$$=\int\frac{\mathrm{d}^4 k}{(2\pi)^4}\left[1+\sum_{n=1}^\infty\frac{(-a\mathrm{i})^n}{n!}k_{\lambda_1}K^{\lambda_1}(y)k_{\lambda_2}K^{\lambda_2}(y)\cdots k_{\lambda_n}K^{\lambda_n}(y)\right]\mathrm{e}^{ik_\lambda(x^\lambda-(y^\lambda+aA^\lambda(y)))},$$

we see that this method is very complicated.

Another method that discusses the case that $K^\mu(x)$ is finite can be found in the Appendix C of this paper.

Here we consider the Taylor's expansion of $\delta^4(x-(y+aA(y)))$ about the power of $a$

$$\delta^4(x-(y+aA(y)))=\delta^4(y+aA(y)-x)=\delta^4(y-x)+aA^\lambda(y)\frac{\partial\delta^4(y-x)}{\partial y^\lambda}$$
$$+\frac{1}{2!}a^2 A^\rho(y)A^\sigma(y)\frac{\partial^2\delta^4(y-x)}{\partial y^\rho\partial y^\sigma}+\cdots+\frac{1}{n!}a^n A^n(y)\frac{\partial^n\delta^4(y-x)}{\partial y^n}+\cdots, \qquad (4\text{-}6)$$

where $A^n(y)\dfrac{\partial^n\delta^4(y-x)}{\partial y^n}$ is the abbreviation for $A^{\lambda_1}(y)A^{\lambda_2}(y)\cdots A^{\lambda_n}(y)\dfrac{\partial^n\delta^4(y-x)}{\partial y^{\lambda_1}\partial y^{\lambda_2}\cdots\partial y^{\lambda_n}}$.

According to the above formula and integration by parts, (3-1) becomes

$$\widetilde{F}^{\mu\nu}(x) = J^{\mu}_{\rho}(x)J^{\nu}_{\sigma}(x)F^{\rho\sigma}(x) - a\frac{\partial}{\partial x^{\lambda}}\left(A^{\lambda}(x)J^{\mu}_{\rho}(x)J^{\nu}_{\sigma}(x)F^{\rho\sigma}(x)\right)$$
$$+\frac{1}{2!}a^{2}\frac{\partial^{2}}{\partial x^{\alpha}\partial x^{\beta}}\left(A^{\alpha}(x)A^{\beta}(x)J^{\mu}_{\rho}(x)J^{\nu}_{\sigma}(x)F^{\rho\sigma}(x)\right)+\cdots$$
$$+(-1)^{n}\frac{1}{n!}a^{n}\frac{\partial^{n}}{\partial x^{n}}\left(A^{n}(x)J^{\mu}_{\rho}(x)J^{\nu}_{\sigma}(x)F^{\rho\sigma}(x)\right)+\cdots \tag{4-7}$$
$$= F^{\mu\nu}(x) + \sum_{n=1}^{\infty}\frac{(-1)^{n}a^{n}}{n!}F^{\mu\nu}_{(n)}(x),$$

$$F^{\mu\nu}_{(1)}(x) = \frac{\partial}{\partial x^{\lambda}}\left(A^{\lambda}(x)F^{\mu\nu}(x)\right) - \left(A^{\lambda,\mu}(x)A^{\nu}_{,\lambda}(x) - A^{\lambda,\nu}(x)A^{\mu}_{,\lambda}(x)\right),$$

$$F^{\mu\nu}_{(2)}(x) = \frac{\partial^{2}}{\partial x^{\alpha}\partial x^{\beta}}\left(A^{\alpha}(x)A^{\beta}(x)F^{\mu\nu}(x)\right) - 2\frac{\partial}{\partial x^{\rho}}\left(A^{\rho}(x)\left(A^{\sigma,\mu}(x)A^{\nu}_{,\sigma}(x) - A^{\sigma,\nu}(x)A^{\mu}_{,\sigma}(x)\right)\right)$$
$$+ 2A^{\mu}_{,\rho}(x)A^{\nu}_{,\sigma}(x)F^{\rho\sigma}(x), \tag{4-8}$$

$$F^{\mu\nu}_{(n)}(x) = \frac{\partial^{n}}{\partial x^{n}}\left(A^{n}(x)F^{\mu\nu}(x)\right) - n\frac{\partial}{\partial x^{n-1}}\left(A^{n-1}(x)\left(A^{\lambda,\mu}(x)A^{\nu}_{,\lambda}(x) - A^{\lambda,\nu}(x)A^{\mu}_{,\lambda}(x)\right)\right)$$
$$+ n(n-1)\frac{\partial}{\partial x^{n-2}}\left(A^{n-2}(x)A^{\mu}_{,\rho}(x)A^{\nu}_{,\sigma}(x)F^{\rho\sigma}(x)\right), \quad n=3,4,\cdots.$$

We see that $F^{\mu\nu}_{(n)}(x)$ is a function of $A^{\rho}(x)$ and its derivative $\frac{\partial^{m}A^{\rho}(x)}{\partial x^{m}} \equiv \frac{\partial^{m}A^{\rho}(x)}{\partial x^{\sigma_{1}}\partial x^{\sigma_{2}}\cdots\partial x^{\sigma_{m}}}$, and we have

$$\frac{\partial \widetilde{F}^{\mu\nu}(x)}{\partial x^{\nu}} = -\frac{D^{2}A^{\mu}(x)}{Dx^{2}} + \sum_{n=1}^{\infty}\frac{(-1)^{n}a^{n}}{n!}\frac{\partial}{\partial x^{\nu}}F^{\mu\nu}_{(n)}\left(A^{\rho}(x),\frac{\partial^{m}A^{\rho}(x)}{\partial x^{m}}\right); \tag{4-9}$$

For $A'^{\mu}(x)$, (4-9) becomes

$$\frac{\partial \widetilde{F}'^{\mu\nu}(x)}{\partial x^{\nu}} = -\frac{D^{2}A'^{\mu}(x)}{Dx^{2}} + \sum_{n=1}^{\infty}\frac{(-1)^{n}a^{n}}{n!}\frac{\partial}{\partial x^{\nu}}F^{\mu\nu}_{(n)}\left(A'^{\rho}(x),\frac{\partial^{m}A'^{\rho}(x)}{\partial x^{m}}\right)$$
$$= -\frac{D^{2}A^{\mu}(x)}{Dx^{2}} - \frac{D^{2}K^{\mu}(x)}{Dx^{2}} + \sum_{n=1}^{\infty}\frac{(-1)^{n}a^{n}}{n!}\frac{\partial}{\partial x^{\nu}}F^{\mu\nu}_{(n)}\left(A^{\rho}(x)+K^{\rho}(x),\frac{\partial^{m}\left(A^{\rho}(x)+K^{\rho}(x)\right)}{\partial x^{m}}\right). \tag{4-10}$$

According to (4-2) and, thus, comparing (4-9) with (4-10), we see that (4-2) holds if and only if $K^{\mu}(x)$ satisfies

$$\frac{D^{2}K^{\mu}(x)}{Dx^{2}} = \sum_{n=1}^{\infty}\frac{(-1)^{n}a^{n}}{n!}\frac{\partial}{\partial x^{\nu}}\left[F^{\mu\nu}_{(n)}\left(A^{\rho}(x)+K^{\rho}(x),\frac{\partial^{m}\left(A^{\rho}(x)+K^{\rho}(x)\right)}{\partial x^{m}}\right)\right.$$
$$\left. - F^{\mu\nu}_{(n)}\left(A^{\rho}(x),\frac{\partial^{m}A^{\rho}(x)}{\partial x^{m}}\right)\right]. \tag{4-11}$$

Assuming

$$K^{\mu}(x) = \sum_{n=0}^{\infty}a^{n}K^{\mu}_{(n)}(x), \tag{4-12}$$

substituting the above expression to (4-11) and comparing the power of $a$ one by one, we first have

$$\frac{D^2 K_{(0)}^{\mu}(x)}{Dx^2} = 0, \tag{4-13}$$

the solution of the above equation is

$$K_{(0)}^{\mu}(x) = \partial^{\mu}\theta_{(0)}(x) = \eta^{\mu\nu}\frac{\partial\theta_{(0)}(x)}{\partial x^{\nu}}. \tag{4-14}$$

where $\theta_{(0)}(x)$ is an arbitrary scalar function. And, further,

$$\frac{D^2 K_{(n)}^{\mu}(x)}{Dx^2} = R_{(n)}^{\mu}\left(x; A^{\rho}, A^{\lambda}_{,\tau}, \cdots, A^{\alpha}_{,\beta,\gamma,\cdots,\chi}; K_{(1)}^{\mu}, K_{(2)}^{\mu}, \cdots, K_{(n-1)}^{\mu}\right) \quad (n=1,2,3,\cdots), \tag{4-15}$$

where $R_{(n)}^{\mu}$, of which the independent variable is $x$, is a function of $A^{\rho}, A^{\alpha}_{,\beta,\gamma,\cdots,\chi}; K_{(1)}^{\mu}, K_{(2)}^{\mu}, \cdots, K_{(n-1)}^{\mu}$ and satisfies

$$\frac{\partial R_{(n)}^{\mu}\left(x; A^{\rho}, A^{\lambda}_{,\tau}, \cdots, A^{\alpha}_{,\beta,\gamma,\cdots,\chi}; K_{(1)}^{\mu}, K_{(2)}^{\mu}, \cdots, K_{(n-1)}^{\mu}\right)}{\partial x^{\mu}} = 0. \tag{4-16}$$

And, further, if we assume

$$K_{(n)}^{\mu}(x) = K_{\perp(n)}^{\mu}(x) + K_{//(n)}^{\mu}(x) \quad (n=1,2,3,\cdots), \tag{4-17}$$

then (4-15) cannot determine $K_{//(n)}^{\mu}(x)$ duo to $\frac{D^2 K_{(n)}^{\mu}(x)}{Dx^2} = \frac{D^2 K_{\perp(n)}^{\mu}(x)}{Dx^2}$, we therefore have

$$K_{//(n)}^{\mu}(x) = \partial^{\mu}\theta_{(n)}(x) = \eta^{\mu\nu}\frac{\partial\theta_{(n)}(x)}{\partial x^{\nu}} \quad (n=1,2,3,\cdots), \tag{4-18}$$

where $\theta_{(n)}(x)$ ($n=1,2,3,\cdots$) are arbitrary scalar functions; the form of the function $R_{(n)}^{\mu}$ in (4-15) becomes

$$\begin{aligned}&R_{(n)}^{\mu}\left(x; A^{\rho}, A^{\lambda}_{,\tau}, \cdots, A^{\alpha}_{,\beta,\gamma,\cdots,\chi}; K_{(1)}^{\mu}, K_{(2)}^{\mu}, \cdots, K_{(n-1)}^{\mu}\right) \\ &= R_{(n)}^{\mu}\left(x; A^{\rho}, A^{\lambda}_{,\tau}, \cdots, A^{\alpha}_{,\beta,\gamma,\cdots,\chi}; K_{\perp(1)}^{\mu}, K_{\perp(2)}^{\mu}, \cdots, K_{\perp(n-1)}^{\mu}; \theta_{(0)}, \theta_{(1)}, \cdots, \theta_{(n-1)}\right).\end{aligned} \tag{4-19}$$

According to (4-18), (4-15) and (4-19), we can obtain $K_{\perp(n)}^{\mu}(x)$ ($n=1,2,3,\cdots$) one by one.

We therefore have determined the transformational relation (4-1) under the condition (4-2). Contrarily, it is obvious that if $K^{\mu}(x)$ satisfies (4-12) ~ (4-19), then under the transformational relation (4-1), (4-2) holds and the equation of motion of electromagnetic field (3-26) is invariable.

As a consequence, according to the above discussion, $\Phi_{\perp}^{\mu}(x)$ defined by (3-6) is also invariable under the transformational relation (4-1):

$$\Phi_{\perp}^{\prime\mu}(x) = \int d^4 y D(x-y)\left(-\frac{\partial \widetilde{F}^{\prime\mu\nu}(y)}{\partial y^{\nu}}\right) = \int d^4 y D(x-y)\left(-\frac{\partial \widetilde{F}^{\mu\nu}(y)}{\partial y^{\nu}}\right) = \Phi_{\perp}^{\mu}(x), \tag{4-20}$$

where (4-2) is used. $L_{\text{EM}}(x)$ given by (3-14) is thus invariable under the transformational relation (4-1).

Under the transformation (1-12) and (4-1), the transformation manner of the equation of motion of charged particle field (3-17) reads

$$0 = \left(i\gamma^\mu \frac{\partial}{\partial x^\mu} - m\right)\psi'(x) - e\gamma^\mu \Phi'_{\mu\perp}(x)\psi'(x) - e\gamma^\mu \Phi'_{\mu//}(x)\psi'(x)$$

$$= \left(i\gamma^\mu \frac{\partial}{\partial x^\mu} - m\right)\left(e^{-ie\chi(x)}\psi(x)\right) - e\gamma^\mu \Phi_{\mu\perp}(x)e^{-ie\chi(x)}\psi(x) - e\gamma^\mu \Phi_{\mu//}(x)e^{-ie\chi(x)}\psi(x)$$

$$- e\gamma^\mu \left(\Phi'_{\mu//}(x) - \Phi_{\mu//}(x)\right)e^{-ie\chi(x)}\psi(x) \qquad (4\text{-}21)$$

$$= e^{-ie\chi(x)}\left[\left(i\gamma^\mu \frac{\partial}{\partial x^\mu} - m\right)\psi(x) - e\gamma^\mu \Phi_{\mu\perp}(x)\psi(x) - e\gamma^\mu \Phi_{\mu//}(x)\psi(x)\right]$$

$$+ e^{-ie\chi(x)}e\gamma^\mu \left[\frac{\partial \chi(x)}{\partial x^\mu} - \partial_\mu(\Theta'(x) - \Theta(x))\right]\psi(x),$$

where (4-20) and (3-9) are used. We see that (3-17) is also invariable under the transformational relation (1-12) and (4-1) as long as we choose

$$\chi(x) = \Theta'(x) - \Theta(x)$$

$$= \frac{\partial}{\partial x^\nu}\int d^4 y D(x - (y + aA'(y)))\left(b_1 J_\lambda^{\prime\nu}(y) A^{\prime\lambda}(y) + b_2 J'(y) A^{\prime\nu}(y)\right) \qquad (4\text{-}22)$$

$$- \frac{\partial}{\partial x^\nu}\int d^4 y D(x - (y + aA(y)))\left(b_1 J_\lambda^\nu(y) A^\lambda(y) + b_2 J(y) A^\nu(y)\right).$$

In (4-22), $A'^\mu(y)$ is replaced by $A^\mu(y) + K^\mu(y)$, $K^\mu(x)$ is given by (4-12) ~ (4-19).

It is obvious that taking advantage of the above method we can prove that $L_D(x) + L_I(x)$ in (3-13) is also invariable under the transformational relation (1-12), (4-22) and (4-1), where $L_D(x)$ and $L_I(x)$ are given by (1-2) and (3-15), respectively.

We therefore have proved that all the action, the equations of motion of charged particle and electromagnetic field of the nonlocal interaction theory given by (3-13) ~ (3-15), (3-17) and (3-26) respectively are invariable under the transformational relation (1-12), (4-22) and (4-1).

The transformation (4-1) can guarantee that one can always choose a special component of $A^\mu(x)$ for which the temporal gauge condition

$$A^0(x) = 0 \qquad (4\text{-}23)$$

holds. In fact, according to the above discussion we have

$$A'^\mu(x) = A^\mu(x) + K^\mu(x) = A^\mu(x) + \partial^\mu \theta_{(0)}(x) + \sum_{n=1}^{\infty} a^n \partial^\mu \theta_{(n)}(x) + \sum_{n=1}^{\infty} a^n K^\mu_{\perp(n)}(x), \qquad (4\text{-}24)$$

especially, according to (4-15), (4-19) and $\dfrac{D^2 K^\mu_{(n)}(x)}{Dx^2} = \dfrac{D^2 K^\mu_{\perp(n)}(x)}{Dx^2}$ we have

$$K^\mu_{\perp(n)}(x) = \int d^4 y D(x - y)$$
$$\times R^\mu_{(n)}\left(y; A^\rho, A^\lambda_{,\tau}, \cdots, A^\alpha_{,\beta,\gamma,\cdots,\chi}; K^\mu_{\perp(1)}, K^\mu_{\perp(2)}, \cdots, K^\mu_{\perp(n-1)}; \theta_{(0)}, \theta_{(1)}, \cdots, \theta_{(n-1)}\right) \qquad (4\text{-}25)$$
$$\equiv \widetilde{R}^\mu_{(n)}\left(x; A^\rho, A^\lambda_{,\tau}, \cdots, A^\alpha_{,\beta,\gamma,\cdots,\chi}; K^\mu_{\perp(1)}, K^\mu_{\perp(2)}, \cdots, K^\mu_{\perp(n-1)}; \theta_{(0)}, \theta_{(1)}, \cdots, \theta_{(n-1)}\right),$$
$$n = 1, 2, 3, \cdots.$$

We see that $K^\mu_{\perp(n)}(x)$ is at most relevant to $\theta_{(n-1)}(x)$ but not to $\theta_{(n)}(x)$. Hence, if $A^0(x) \neq 0$

for a component of $A^\mu(x)$, then we can choose

$$\partial^0 \theta_{(0)}(x) = -A^0(x); \quad \partial^0 \theta_{(n)}(x) = -K^0_{\perp(n)}(x), \quad n = 1, 2, 3, \cdots, \tag{4-26}$$

and the component of $A'^\mu(x)$ thus satisfy $A'^0(x) = 0$.

The transformation (4-1) can also guarantee that one can always choose a special component of $A^\mu(x)$ for which the Lorentz gauge condition

$$\frac{\partial A^\lambda(y)}{\partial y^\lambda} = 0 \tag{4-27}$$

holds. In fact, from (4-24) and considering $K^\mu_{\perp(n),\mu}(x) = 0$, we have

$$A'^\mu{}_{,\mu}(x) = A^\mu{}_{,\mu}(x) + K^\mu{}_{,\mu}(x) = A^\mu{}_{,\mu}(x) + \partial_\mu \partial^\mu \theta_{(0)}(x) + \sum_{n=1}^{\infty} a^n \partial_\mu \partial^\mu \theta_{(n)}(x), \tag{4-28}$$

Hence, if $A^\mu{}_{,\mu}(x) \neq 0$ for a component of $A^\mu(x)$, then we can choose

$$\partial_\mu \partial^\mu \theta_{(0)}(x) = -A^\mu{}_{,\mu}(x); \quad \theta_{(n)}(x) = 0, \quad n = 1, 2, 3, \cdots, \tag{4-29}$$

the component of $A'^\mu(x)$ thus satisfy $A'^\mu{}_{,\mu}(x) = 0$.

Under the Lorentz gauge condition (4-27), the equation of motion of electromagnetic field (3-19) becomes

$$\partial_\nu \partial^\nu A^\mu(y) = eJ(y)\Omega^\mu_\nu(y)\int d^4x\, j^\nu(x)\delta^4(y + aA(y) - x), \tag{4-30}$$

after some calculation similar to (3-23) we have

$$\frac{\partial}{\partial y^\mu}\left(\partial_\nu \partial^\nu A^\mu(y)\right) = \partial_\nu \partial^\nu A^\mu{}_{,\mu}(y) = e\frac{\partial}{\partial y^\mu}\left(J(y)\Omega^\mu_\nu(y)\int d^4x\, j^\nu(x)\delta^4(x - (y + aA(y)))\right)$$

$$= eJ(y)\int d^4x \frac{\partial j^\lambda(x)}{\partial x^\lambda}\delta^4(x - (y + aA(y))),$$

according to the current conservation equation (1-10) we obtain

$$\partial_\nu \partial^\nu A^\mu{}_{,\mu}(y) = 0. \tag{4-31}$$

(4-31) means that $A^\mu{}_{,\mu}(y)$ is free field, this conclusion is the same as the current QED.

From (3-9) ~ (3-11) we see that if we choose $b_1 = 1$, $b_2 = 0$ and the Lorentz gauge condition $\frac{\partial A^\lambda(y)}{\partial y^\lambda} = 0$, then $\Phi^\mu_{//}(x) = \int d^4y \frac{\partial D(x - (y + aA(y)))}{\partial x_\mu} \frac{\partial A^\lambda(y)}{\partial y^\lambda} = 0$; and, further,

$$\frac{\partial \Phi^\mu(x)}{\partial x^\mu} = \frac{\partial \Phi^\mu_\perp(x)}{\partial x^\mu} = 0.$$

The transformational relation (1-12), (4-22) and (4-1) are a generalization of the gauge transformation (1-12) and (1-11), we call them *generalized gauge transformation*.

## 5  Quantization of the theory

### 5.1  Some notes about methods of quantization of nonlocal theory

If we want to use the method of canonical quantization to establish a corresponding quantum theory for the theory of quantum electrodynamics with nonlocal interaction given in this paper, then what we must do first is to obtain the momenta $\pi_\mu(y)$ conjugate to $A^\mu(y)$ from the action (3-13). However, if we use the form of the action given by (B-1), (B-2), (B-6) and (B-7) in the Appendix B of this paper, then we obtain $\pi_\mu(y) = 0$, the similar case for the current QED as well. The concrete calculation process of this conclusion can be found in the Appendix D of this paper.

Hence, if we want to use the method of canonical quantization, then it seems as if we should seek an appropriate action being equivalent to (3-13). However, in principle, the Hamiltonian representation of a nonlocal theory is full of uncertainties[7, 8].

A method in common use in quantization of nonlocal theory is that by introducing new field variables one changes the nonlocal theory to local form (See, for example, Ref [9 ~ 11]). In fact, the nonlocal theory given in this paper has already been written to local form via the variable $\Phi^\mu(x)$, because all (3-13) ~ (3-15) are local form. However, we can not use this method to realize quantization of the nonlocal theory, because via quantum theory about the field $\Phi^\mu(x)$ we can not obtain a correct quantum theory about the field $A^\mu(x)$. For example, if we choose the temporal gauge condition $A^0(x) = 0$, then $\Phi^0(x) \neq 0$; If we choose the Lorentz gauge condition $\frac{\partial A^\lambda(y)}{\partial y^\lambda} = 0$, and take $b_2 \neq 0$ in (3-9) ~ (3-11), then $\frac{\partial \Phi^\lambda(y)}{\partial y^\lambda} \neq 0$.

Besides the difference of choice of gauge condition, we investigate the case of free field. If we choose the Lorentz gauge condition $\frac{\partial A^\lambda(y)}{\partial y^\lambda} = 0$, then according to (3-22), $A^\mu(x)$ satisfy $\eta^{\mu\nu} \frac{\partial^2 A^\lambda(x)}{\partial x^\mu \partial x^\nu} = 0$. Hence, after both $A^\mu(t_0, \boldsymbol{x})$ and $\dot{A}^\mu(t_0, \boldsymbol{x})$ of initial time $t_0$ are given, we have [1, 6]

$$A^\mu(t, \boldsymbol{x}) = \int d^3 x' \Delta_1(t - t_0, \boldsymbol{x} - \boldsymbol{x}') A^\mu(t_0, \boldsymbol{x}') - \int d^3 x' \Delta(t - t_0, \boldsymbol{x} - \boldsymbol{x}') \dot{A}^\mu(t_0, \boldsymbol{x}'), \qquad (5\text{-}1)$$

where

$$\Delta(x - x') = -i \int \frac{d^3 k}{(2\pi)^{3/2} 2\omega} \left( e^{-ik \cdot (x - x')} - e^{ik \cdot (x - x')} \right), \quad \Delta_1(x - x') = \int \frac{d^3 k}{(2\pi)^{3/2} 2\omega} \left( e^{-ik \cdot (x - x')} + e^{ik \cdot (x - x')} \right).$$

After obtaining $A^\mu(x)$ by (5-1), we can obtain corresponding $\Phi^\mu(x)$ according to (A-15) in the Appendix A of this paper.

On the other hand, for this case, if we choose $b_1 = 1$, $b_2 = 0$ in (3-9) ~ (3-11), then according to the discussion below (4-31), for $\Phi^\mu(x)$ we have also $\frac{\partial \Phi^\mu(x)}{\partial x^\mu} = \frac{\partial \Phi^\mu_\perp(x)}{\partial x^\mu} = 0$ and $\eta^{\mu\nu}\frac{\partial^2 \Phi^\mu(x)}{\partial x^\mu \partial x^\nu} = 0$. Hence, if we obtain the solution of the equation $\eta^{\mu\nu}\frac{\partial^2 \Phi^\lambda(x)}{\partial x^\mu \partial x^\nu} = 0$ by the formula (5-1):

$$\widetilde{\Phi}^\mu(t, \boldsymbol{x}) = \int d^3 x' \Delta_1(t - t_0, \boldsymbol{x} - \boldsymbol{x}') \Phi^\mu(t_0, \boldsymbol{x}') - \int d^3 x' \Delta(t - t_0, \boldsymbol{x} - \boldsymbol{x}') \dot{\Phi}^\mu(t_0, \boldsymbol{x}'),$$

then we can prove that such $\widetilde{\Phi}^\mu(x) \neq \Phi^\mu(x)$ obtained by (A-15) in the Appendix A of this paper.

This result shows that we can not use methods of quantization of local theory for $\Phi^\mu(x)$ to realize quantization of the nonlocal theory for $A^\mu(x)$, the corresponding characteristic of this result in quantum theory is that we can not use the solutions of the equation $\eta^{\mu\nu}\frac{\partial^2 \Phi^\lambda(x)}{\partial x^\mu \partial x^\nu} = 0$ that the *field operators* $\Phi^\mu(x)$ satisfy in the Heisenberg picture to construct the Fock space of the field operators $A^\mu(x)$.

Another method in common use in quantization of nonlocal theory is changing nonlocal theory to higher-derivative theory[12~15]. However, from (A-16) ~ (A-20) in the Appendix A we see that what we now face is a theory including infinite higher time derivative terms; besides these infinite higher time derivative terms, there are still nonlocal spatial terms in the theory.

If we use the method of path integral and write out the vacuum-vacuum transition amplitude and the generating functional of the theory under certain gauge condition formally, the extra relative time variable arising from the retarded or advanced Green functions leads to trouble. (We don't discuss this question in detail here.)

In next discussion, we first give some characteristics of electromagnetic field under the temporal gauge condition, and then, state the Lehmann-Symanzik-Zimmermann formalism[1, 6, 16~21] of the current quantum electrodynamics under the temporal gauge condition, which is based on the equations of motion of operators but not on Lagrangian or Hamiltonian; finally, by generalizing this formalism we establish a quantum theory of the nonlocal theory given by this paper.

### 5.2 Some characteristics of electromagnetic field under the temporal gauge condition

At first, we discuss the classical case.

For the case of free electromagnetic field under the temporal gauge condition, the equation of motion of electromagnetic field reads

$$\ddot{\boldsymbol{A}}(T, \boldsymbol{X}) - \nabla \cdot \nabla \boldsymbol{A}(T, \boldsymbol{X}) + \nabla \nabla \cdot \boldsymbol{A}(T, \boldsymbol{X}) = 0. \tag{5-2}$$

Employing the standard method of Green functions[22], the odd and even solutions of (5-2) are

$$\overset{\rightarrow\rightarrow}{\Delta}_1(T, T_0; \boldsymbol{X}, \boldsymbol{X}')$$

$$= \int \frac{d^3 K}{(2\pi)^3} \left\{ \sum_{\lambda=1}^{2} \left[ \frac{\sin[|\boldsymbol{K}|(T-T_0)]}{|\boldsymbol{K}|} e^{i\boldsymbol{K}\cdot(\boldsymbol{X}-\boldsymbol{X}')} \boldsymbol{\varepsilon}_{(\lambda)}(\boldsymbol{K})\boldsymbol{\varepsilon}_{(\lambda)}(\boldsymbol{K}) \right] + (T-T_0) e^{i\boldsymbol{K}\cdot(\boldsymbol{X}-\boldsymbol{X}')} \boldsymbol{\varepsilon}_{(3)}(\boldsymbol{K})\boldsymbol{\varepsilon}_{(3)}(\boldsymbol{K}) \right\} \tag{5-3}$$

$$= \int \frac{d^3 K}{(2\pi)^3} \left\{ \frac{\sin[|\boldsymbol{K}|(T-T_0)]}{|\boldsymbol{K}|} \left( \overset{\rightarrow}{I} - \boldsymbol{\varepsilon}_{(3)}(\boldsymbol{K})\boldsymbol{\varepsilon}_{(3)}(\boldsymbol{K}) \right) + (T-T_0) \boldsymbol{\varepsilon}_{(3)}(\boldsymbol{K})\boldsymbol{\varepsilon}_{(3)}(\boldsymbol{K}) \right\} e^{i\boldsymbol{K}\cdot(\boldsymbol{X}-\boldsymbol{X}')},$$

$$\overrightarrow{\overrightarrow{\Delta}}(T,T_0;X,X') = \int \frac{d^3K}{(2\pi)^3} \left\{ \sum_{\lambda=1}^{2} \cos[|K|(T-T_0)] e^{iK\cdot(X-X')} \varepsilon_{(\lambda)}(K)\varepsilon_{(\lambda)}(K) + e^{iK\cdot(X-X')} \varepsilon_{(3)}(K)\varepsilon_{(3)}(K) \right\}$$

$$= \int \frac{d^3K}{(2\pi)^3} \left\{ \cos[|K|(T-T_0)] \left( \overrightarrow{\overrightarrow{I}} - \varepsilon_{(3)}(K)\varepsilon_{(3)}(K) \right) + \varepsilon_{(3)}(K)\varepsilon_{(3)}(K) \right\} e^{iK\cdot(X-X')}, \quad (5\text{-}4)$$

respectively; where 3-dimentional polarisation vectors $\varepsilon_{(\lambda)}(K)$ $(\lambda = 1, 2, 3)$ satisfy[6]

$$\varepsilon_{(3)}(K) = \frac{K}{|K|}, \varepsilon_{(\lambda)}(K)\cdot\varepsilon_{(\lambda')}(K) = \delta_{(\lambda)(\lambda')}, \sum_{\lambda=1}^{3} \varepsilon_{(\lambda)}(K)\varepsilon_{(\lambda)}(K) = \overrightarrow{\overrightarrow{I}}, \varepsilon_{(\lambda)}(-K) = (-1)^\lambda \varepsilon_{(\lambda)}(K). \quad (5\text{-}5)$$

We have

$$\overrightarrow{\overrightarrow{\Delta}}(T,T_0;X,X') = \frac{\partial \overrightarrow{\overrightarrow{\Delta}}_1(T,T_0;X,X')}{\partial T}. \quad (5\text{-}6)$$

After $A(T_0,X)$ and $\dot{A}(T_0,X)$ of initial time $T_0$ are given, the solution of (5-2) is

$$A(T,X) = \int d^3X' \overrightarrow{\overrightarrow{\Delta}}(T,T_0;X,X')\cdot A(T_0,X') + \int d^3X' \overrightarrow{\overrightarrow{\Delta}}_1(T,T_0;X,X')\cdot \dot{A}(T_0,X'). \quad (5\text{-}7)$$

In the calculation of (5-7), for instance,

$$\left( \overrightarrow{\overrightarrow{\Delta}}_1(T,T_0;X,X')\cdot \dot{A}(T_0,X') \right)^i$$

$$= \int \frac{d^3K}{(2\pi)^3} e^{iK\cdot(X-X')} \sum_{j=1}^{3} \left( \left\{ \frac{\sin[|K|(T-T_0)]}{|K|} \left( \delta^{ij} - \frac{K^i K^j}{|K|^2} \right) + (T-T_0)\frac{K^i K^j}{|K|^2} \right\} \dot{A}^j(T_0,X') \right).$$

Using $\overrightarrow{\overrightarrow{\Delta}}_1(T,T_0;X,X')$ we can introduce the corresponding retarded, advanced Green functions

$$\overrightarrow{\overrightarrow{\Delta}}_{\substack{\text{ret}\\\text{adv}}}(T,T';X,X') = \pm\theta(\pm(T-T_0))\overrightarrow{\overrightarrow{\Delta}}_1(T,T';X,X'), \quad (5\text{-}8)$$

which satisfy

$$\frac{\partial^2 \overrightarrow{\overrightarrow{\Delta}}_{\substack{\text{ret}\\\text{adv}}}(T,T';X,X')}{\partial T^2} - \nabla\cdot\nabla \overrightarrow{\overrightarrow{\Delta}}_{\substack{\text{ret}\\\text{adv}}}(T,T';X,X') + \nabla\nabla\cdot \overrightarrow{\overrightarrow{\Delta}}_{\substack{\text{ret}\\\text{adv}}}(T,T';X,X') \quad (5\text{-}9)$$

$$= \delta^4(X-X')\overrightarrow{\overrightarrow{I}}.$$

Taking advantage of $\overrightarrow{\overrightarrow{\Delta}}_{\substack{\text{ret}\\\text{adv}}}(T,T';X,X')$, a solution of the equation of motion of electromagnetic field with source

$$\ddot{A}(T,X) - \nabla\cdot\nabla A(T,X) + \nabla\nabla\cdot A(T,X) = e\zeta(T,X) \quad (5\text{-}10)$$

can be written to the form

$$A(T,X) = \overline{A}(T,X) + \int dT' \int d^3X' \overrightarrow{\overrightarrow{\Delta}}_{\substack{\text{ret}\\\text{adv}}}(T,T';X,X')\cdot e\zeta(T',X'), \quad (5\text{-}11)$$

where $\overline{A}(T,X)$ is a solution of the equation of motion of free electromagnetic field (5-2).

We only discuss quantum free electromagnetic field here. For this case, (5-2) is now an equation that the operators $A(T, X)$ satisfy in the Heisenberg picture, which can be obtained by the following commutation relations, the Hamiltonian and the equations of motion:

$$[A^i(T, X), \pi^j(T, X')] = i\delta^{ij}\delta^3(X - X'), \quad \text{others} = 0; \tag{5-12}$$

$$H_{EM} = \int d^3X \left( \frac{1}{2}\pi(T, X) \cdot \pi(T, X) + \frac{1}{2}(\nabla \times A(T, X)) \cdot (\nabla \times A(T, X)) \right), \tag{5-13}$$

$$\dot{A}(T, X) = -i[A(T, X), H_{EM}] = \pi(T, X), \tag{5-14}$$

$$\dot{\pi}(T, X) = -i[\pi(T, X), H_{EM}] = \nabla \cdot \nabla A(T, X) - \nabla \nabla \cdot A(T, X). \tag{5-15}$$

For seeking a separable solution of the eigenvalue equation of (5-13), we set

$$A(T, X) = \int \frac{d^3K}{(2\pi)^{3/2}} \sum_{\lambda=1}^{2} \left( v_{(\lambda)}^* \varepsilon_{(\lambda)}(K) a_{(\lambda)}(T, K) + v_{(\lambda)} \varepsilon_{(\lambda)}(-K) a_{(\lambda)}^*(T, -K) \right) e^{iK \cdot X}$$
$$- i \int d^3X' \int \frac{d^3K}{(2\pi)^3} \varepsilon_{(3)}(K) e^{iK \cdot (X-X')} Q(T, X'), \tag{5-16}$$

$$\pi(T, X) = \int \frac{d^3K}{(2\pi)^{3/2}} \sum_{\lambda=1}^{2} \left( u_{(\lambda)}^* \varepsilon_{(\lambda)}(-K) a_{(\lambda)}(T, -K) + u_{(\lambda)} \varepsilon_{(\lambda)}(K) a_{(\lambda)}^*(T, K) \right) e^{-iK \cdot X}$$
$$+ i \int d^3X' \int \frac{d^3K}{(2\pi)^3} \varepsilon_{(3)}(K) e^{-iK \cdot (X-X')} P(T, X'), \tag{5-17}$$

where

$$\sqrt{v_{(\lambda)}^* v_{(\lambda)}} = \frac{1}{\sqrt{2|K|}}, \quad u_{(\lambda)} = i|K|v_{(\lambda)}; \quad \lambda = 1, 2. \tag{5-18}$$

According to (5-16) ~ (5-18) and (5-5) we have

$$a_{(\lambda)}(T, K) = -i \int \frac{d^3X}{(2\pi)^{3/2}} \varepsilon_{(\lambda)}(K) \cdot \left( u_{(\lambda)} A(T, X) - v_{(\lambda)} \pi(T, X) \right) e^{-iK \cdot X},$$
$$a_{(\lambda)}^*(T, K) = i \int \frac{d^3X}{(2\pi)^{3/2}} \varepsilon_{(\lambda)}(K) \cdot \left( u_{(\lambda)}^* A(T, X) - v_{(\lambda)}^* \pi(T, X) \right) e^{iK \cdot X}; \quad \lambda = 1, 2; \tag{5-19}$$

$$Q(T, X) = i \int d^3X' \int \frac{d^3K}{(2\pi)^3} \varepsilon_{(3)}(K) \cdot A(T, X') e^{iK \cdot (X-X')},$$
$$P(T, X) = -i \int d^3X' \int \frac{d^3K}{(2\pi)^3} \varepsilon_{(3)}(K) \cdot \pi(T, X') e^{-iK \cdot (X-X')}. \tag{5-20}$$

From (5-12), (5-19) and (5-20) we obtain the following commutation relations:

$$[a_{(\lambda)}(T, K), a_{(\lambda')}^*(T, K')] = \delta_{(\lambda)(\lambda')}\delta^3(K - K'), \quad [Q(T, X), P(T, X')] = i\delta^3(X - X'), \text{ others=0}. \tag{5-21}$$

Substituting (5-16) and (5-17) to (5-13), and using (5-21), we obtain

$$H_{EM} = \int d^3K |K| \left( \sum_{\lambda=1}^{2} a_{(\lambda)}^*(T, K) a_{(\lambda)}(T, K) + \frac{1}{2} \right) + \int d^3X \frac{1}{2} P^2(T, X). \tag{5-22}$$

It is easy to obtain the set of eigenvectors of $H_{EM}$ given by (5-22).

## 5.3 The Lehmann-Symanzik-Zimmermann formalism of the current quantum electrodynamics

We state without explanation the Lehmann-Symanzik-Zimmermann formalism [1, 6, 16~21] of the current quantum electrodynamics under the Heisenberg picture, occupation number representation and the temporal gauge condition (4-23).

① The equations of motion of the field operators $\psi(x)$, $\bar{\psi}(x)$ and $A(X)$:

$$\left(i\gamma^\mu \partial_\mu - m\right)\psi(x) = e\gamma^\mu \Lambda_\mu(x)\psi(x), \tag{5-23}$$

$$\bar{\psi}(x)\left(i\gamma^\mu \bar{\partial}_\mu + m\right) = -e\bar{\psi}(x)\gamma^\mu \Lambda_\mu(x), \tag{5-24}$$

$$\frac{\partial}{\partial T}\dot{A}(T,X) - \nabla \cdot \nabla A(T,X) + \nabla \nabla \cdot A(T,X) = e\zeta(T,X). \tag{5-25}$$

In (5-23) ~ (5-25)

$$\Lambda^0(x) = 0, \quad \Lambda(x) = A(x), \tag{5-26}$$

$$\zeta(T,X) = j(T,X) = \bar{\psi}(T,X)\gamma\psi(T,X). \tag{5-27}$$

According to (5-23) and (5-24) we have

$$\frac{\partial j^\mu(x)}{\partial x^\mu} = \frac{\partial j^0(x)}{\partial x^0} + \nabla \cdot j(x) = \frac{\partial\left(\bar{\psi}(x)\gamma^0\psi(x)\right)}{\partial x^0} + \nabla \cdot \left(\bar{\psi}(x)\gamma\psi(x)\right) = 0. \tag{5-28}$$

② A constraint condition pressing on physical state vectors

A *physical* state vector $|\Psi_{phy}\rangle$ in the Heisenberg picture describing actual system of electrons and photons must satisfies

$$\langle\Psi_{phy}|\left(\nabla \cdot \dot{A}(T,X) + e\zeta^0(T,X)\right)|\Psi_{phy}\rangle = 0, \tag{5-29}$$

where

$$\zeta^0(T,X) = j^0(T,X) = \bar{\psi}(T,X)\gamma^0\psi(T,X). \tag{5-30}$$

In fact, a physical state vector $|\Psi_{phy}\rangle$ satisfies (5-29) if and only if, for the operators $\dot{A}(T_0,X)$ and $\zeta^0(T_0,X)$ of initial time $T_0$, it satisfies

$$\langle\Psi_{phy}|\left(\nabla \cdot \dot{A}(T_0,X) + e\zeta^0(T_0,X)\right)|\Psi_{phy}\rangle = 0. \tag{5-31}$$

Because, according to (5-25), (5-27), (5-28) and (5-30) we have

$$\frac{\partial}{\partial T}\nabla \cdot \dot{A}(T,X) = e\nabla \cdot \zeta(T,X) = e\nabla \cdot j(T,X) = -e\frac{\partial j^0(T,X)}{\partial T} = -e\frac{\partial \zeta^0(T,X)}{\partial T},$$

namely,

$$\frac{\partial}{\partial T}\left(\nabla \cdot \dot{A}(T,X) + e\zeta^0(T,X)\right) = 0; \tag{5-32}$$

And, further, notice that a state vector $|\Psi\rangle$ in the Heisenberg picture is independent of time, we

have

$$\frac{\partial}{\partial T}\left(\langle\Psi_{\text{phy}}|(\nabla\cdot\dot{A}(T,X)+e\zeta^0(T,X))|\Psi_{\text{phy}}\rangle\right)=0. \tag{5-33}$$

Hence, if $|\Psi_{\text{phy}}\rangle$ satisfies (5-31), then according to (5-33), for the operators $\dot{A}(T,X)$ and $\zeta^0(T,X)$ of arbitrary time $T$, it satisfies (5-29).

③ "In" and "Out" fields

By the following definitions we introduce the field operators $\psi_{\text{in/out}}(x)$, $\bar{\psi}_{\text{in/out}}(x)$ and $A_{\text{in/out}}(X)$:

$$\sqrt{Z_2}\psi_{\text{in/out}}(x)=\psi(x)-e\int d^4y\, S_{\text{ret/adv}}(x-y)\gamma^\mu \Lambda_\mu(y)\psi(y), \tag{5-34}$$

$$\sqrt{Z_2}\bar{\psi}_{\text{in/out}}(x)=\bar{\psi}(x)-e\int d^4y\,\bar{\psi}(y)\gamma^\mu \Lambda_\mu(y)S_{\text{adv/ret}}(y-x), \tag{5-35}$$

$$\sqrt{Z_3}A_{\text{in/out}}(X)=A(X)-e\int d^4Y\,\vec{\Delta}_{\text{ret/adv}}(X-Y)\cdot\zeta(Y). \tag{5-36}$$

In (5-34) and (5-35), $\vec{\Delta}_{\text{ret/adv}}(X-Y)$ are given by (5-8); $\Lambda_\mu(y)$ and $\zeta(Y)$ are given by (5-26) and (5-27), respectively; $S_{\text{ret/adv}}(x-y)$ read[6]

$$S_{\text{ret/adv}}(x-y)=-\left(i\gamma^\mu\frac{\partial}{\partial x^\mu}+m\right)\Delta_{\text{ret/adv}}(x-y),\quad \Delta_{\text{ret/adv}}(x-y)=\mp\theta(\pm(x^0-y^0))\Delta(x-y), \tag{5-37}$$

$$\Delta(x-y)=-i\int\frac{d^3p}{(2\pi)^3 2\sqrt{\mathbf{p}\cdot\mathbf{p}+m^2}}\left(e^{-ip\cdot(x-y)}-e^{ip\cdot(x-y)}\right),\quad p^0=\sqrt{\mathbf{p}\cdot\mathbf{p}+m^2}; \tag{5-38}$$

④ The asymptotic conditions

Using the idea of "weak convergence"[1, 6, 16~21], we have the asymptotic conditions:

$$\psi(x)\to\sqrt{Z_2}\psi_{\text{in/out}}(x),\quad t\to\mp\infty, \tag{5-39}$$

$$\bar{\psi}(x)\to\sqrt{Z_2}\bar{\psi}_{\text{in/out}}(x),\quad t\to\mp\infty, \tag{5-40}$$

$$A(x)\to\sqrt{Z_3}A_{\text{in/out}}(X),\quad t\to\mp\infty. \tag{5-41}$$

In (5-34) ~ (5-36) and (5-39) ~ (5-41), $\sqrt{Z_2}$ and $\sqrt{Z_3}$ are arisen from the adiabatic factor $e^{\varepsilon t}$ in the interaction terms[19].

⑤ The commutation relations of "In" fields

$$\{\psi_{\text{in}\alpha}(t,\mathbf{x}),\psi_{\text{in}\beta}^+(t,\mathbf{x}')\}=i\delta_{\alpha\beta}\delta^3(\mathbf{x}-\mathbf{x}'),\ [A_{\text{in}}^i(T,\mathbf{X}),\dot{A}_{\text{in}}^j(T,\mathbf{X}')]=i\delta^{ij}\delta^3(\mathbf{X}-\mathbf{X}'); \tag{5-42}$$
others $=0$.

⑥ The Fock space

The field operators $\psi_{\text{in}}(x)$ and $A_{\text{in}}(T,\mathbf{X})$ can be written to the following forms:

$$\psi_{\text{in}}(x) = \int d^3p \sum_{\pm s} \left( b_{\text{in}}(\boldsymbol{p},s) U_{ps}(x) + d_{\text{in}}^+(\boldsymbol{p},s) V_{ps}(x) \right), \tag{5-43}$$

$$\boldsymbol{A}_{\text{in}}(T,\boldsymbol{X}) = \int \frac{d^3K}{(2\pi)^{3/2}} \sum_{\lambda=1}^{2} \left( v_{(\lambda)}^* \boldsymbol{\varepsilon}_{(\lambda)}(\boldsymbol{K}) a_{(\lambda)\text{in}}(T,\boldsymbol{K}) + v_{(\lambda)} \boldsymbol{\varepsilon}_{(\lambda)}(-\boldsymbol{K}) a_{(\lambda)\text{in}}^*(T,-\boldsymbol{K}) \right) e^{i\boldsymbol{K}\cdot\boldsymbol{X}}$$
$$- i \int d^3X' \int \frac{d^3K}{(2\pi)^3} \boldsymbol{\varepsilon}_{(3)}(\boldsymbol{K}) e^{i\boldsymbol{K}\cdot(\boldsymbol{X}-\boldsymbol{X}')} Q_{\text{in}}(T,\boldsymbol{X}'); \tag{5-44}$$

The commutation relations of the operators $b_{\text{in}}(\boldsymbol{p},s)$, $b_{\text{in}}^+(\boldsymbol{p},s)$, $d_{\text{in}}(\boldsymbol{p},s)$, $d_{\text{in}}^+(\boldsymbol{p},s)$, $a_{(\lambda)\text{in}}(T,\boldsymbol{K})$, $a_{(\lambda)\text{in}}^*(T,\boldsymbol{K})$, $Q_{\text{in}}(T,\boldsymbol{X})$ and $P_{\text{in}}(T,\boldsymbol{X})$ read

$$\begin{aligned}
&\{b_{\text{in}}(\boldsymbol{p},s), b_{\text{in}}^+(\boldsymbol{p}',s')\} = \delta_{ss'}\delta^3(\boldsymbol{p}-\boldsymbol{p}'), \quad \{d_{\text{in}}(\boldsymbol{p},s), d_{\text{in}}^+(\boldsymbol{p}',s')\} = \delta_{ss'}\delta^3(\boldsymbol{p}-\boldsymbol{p}'), \\
&[a_{(\lambda)\text{in}}(T,\boldsymbol{K}), a_{(\lambda')\text{in}}^*(T,\boldsymbol{K}')] = \delta_{(\lambda)(\lambda')}\delta^3(\boldsymbol{K}-\boldsymbol{K}'), \quad \lambda = 1,2, \\
&[Q_{\text{in}}(T,\boldsymbol{X}), P_{\text{in}}(T,\boldsymbol{X}')] = i\delta^3(\boldsymbol{X}-\boldsymbol{X}'); \text{ others} = 0.
\end{aligned} \tag{5-45}$$

The Fock space of the theory is constructed by the set of eigenvectors of the following two operators:

$$H_e = \int d^3p \sqrt{\boldsymbol{p}\cdot\boldsymbol{p}+m^2} \sum_{\pm s} \left( b_{\text{in}}^+(p,s)b_{\text{in}}(p,s) + d_{\text{in}}^+(p,s)d_{\text{in}}(p,s) - 1 \right), \tag{5-46}$$

$$H_{\text{EM}} = \int d^3K |\boldsymbol{K}| \left( \sum_{\lambda=1}^{2} a_{(\lambda)\text{in}}^*(T,\boldsymbol{K}) a_{(\lambda)\text{in}}(T,\boldsymbol{K}) + \frac{1}{2} \right) + \int d^3X \frac{1}{2} P_{\text{in}}^2(T,\boldsymbol{X}). \tag{5-47}$$

The characteristics of (5-43) and (5-46) are well-known, (5-44) and (5-47) correspond to (5-16) and (5-22), respectively.

Using such Fock space and according to (5-36), the constraint condition (5-31) becomes

$$\langle \Psi_{\text{phy}} | (\nabla \cdot \dot{\boldsymbol{A}}_{\text{in}}(T,\boldsymbol{X}) + e\zeta_{\text{in}}^0(T,\boldsymbol{X})) | \Psi_{\text{phy}} \rangle = 0. \tag{5-48}$$

In (5-48), $\zeta_{\text{in}}^0(T,\boldsymbol{X})$ is given by (5-30) in which $\psi(x)$ and $\overline{\psi}(x)$ are substituted by $\psi_{\text{in}}(x)$ and $\overline{\psi}_{\text{in}}(x)$, respectively.

⑦ The S-matrix

By the following definition we introduce the S-matrix of the theory[6]:

$$\langle \beta_{\text{in}} | S = \langle \beta_{\text{out}} |, \quad \langle \beta_{\text{out}} | S^{-1} = \langle \beta_{\text{in}} |, \quad \langle \beta_{\text{out}} | \beta_{\text{in}} \rangle = \langle \beta_{\text{in}} | S | \beta_{\text{in}} \rangle, \tag{5-49}$$

And, using $\hat{M}$ to denote the field operators $\psi(x)$, $\overline{\psi}(x)$ and $A(x)$, we have

$$\hat{M}_{\text{out}}(x) = S^{-1} \hat{M}_{\text{in}}(x) S. \tag{5-50}$$

Hence, according to (5-23) ~ (5-25), (5-34) ~ (5-36) and (5-39) ~ (5-41), we can calculate the S-matrix in principle.

⑧ The Lehmann-Symanzik-Zimmermann reduction formulas

The method writing down the corresponding Lehmann-Symanzik-Zimmermann reduction formula for a S-matrix element is well-known[1, 6, 16~21], taking advantage of the reduction formula we can calculate a *S*-matrix element by the field operators $\psi(x)$, $\bar{\psi}(x)$ and $A(X)$, which can be obtained from (5-34) ~ (5-36) via perturbation method.

We therefore have a self-sufficient system of the current quantum electrodynamics; in principle, we can obtain solution for arbitrary question in the current quantum electrodynamics. The main characteristic of this Lehmann-Symanzik-Zimmermann formalism is that it rests with the equations of motion of operators, and avoids Lagrangian or Hamiltonian. Thus, we can generalize this formalism to the case of the nonlocal theory given by this paper.

**5.4 A quantum theory of the nonlocal theory given by this paper**

Under the temporal gauge condition (4-23), for the functions $J(x)$, $J^\alpha_\beta(x)$ and $\Omega^\mu_\nu(x)$ given in §2.1 we have

$$J^0_\mu = \delta^0_\mu, \quad \Omega^0_\mu = \delta^0_\mu, \quad \Omega^i_0 = -aA^j{}_{,0}\Omega^i_j, \quad J^i_k\Omega^k_j = \delta^i_j, \quad \Omega^i_k J^k_j = \delta^i_j; \tag{5-51}$$

$$J = J(A^l{}_{,m}(x)) = \left\|\delta^l_m + a\frac{\partial A^l(x^0, \boldsymbol{x})}{\partial x^m}\right\|, \quad J^i_j = J^i_j(A^l{}_{,m}(x)), \quad \Omega^i_j = \Omega^i_j(A^l{}_{,m}(x)). \tag{5-52}$$

Notice that (5-52) predicates that all the functions $J(x)$, $J^i_j(x)$ and $\Omega^i_j(x)$ are independent of time derivative term $A^i{}_{,0}$. And we can obtain some characteristics about 3-dimentional quantities $J$, $J^i_j$ and $\Omega^i_j$ by using methods shown in §2. For example, corresponding to (2-34), (2-23) and (2-36), we have

$$\frac{\partial \delta^3(\boldsymbol{x} - (\boldsymbol{X} + a\boldsymbol{A}(T, \boldsymbol{X})))}{\partial X^l} = -J^n_l(A^i{}_{,j}(T, \boldsymbol{X}))\frac{\partial \delta^3(\boldsymbol{x} - (\boldsymbol{X} + a\boldsymbol{A}(T, \boldsymbol{X})))}{\partial x^n}, \tag{5-53}$$

$$\frac{\partial J(A^i{}_{,j}(T, \boldsymbol{X}))}{\partial A^u{}_{,v}(T, \boldsymbol{X})} = aJ(A^i{}_{,j}(T, \boldsymbol{X}))\Omega^v_u(A^i{}_{,j}(T, \boldsymbol{X})), \tag{5-54}$$

$$\frac{\partial \delta^3(\boldsymbol{x} - (\boldsymbol{X} + a\boldsymbol{A}(T, \boldsymbol{X})))}{\partial A^l(T, \boldsymbol{X})} = a\Omega^m_l(A^i{}_{,j}(T, \boldsymbol{X}))\frac{\partial \delta^3(\boldsymbol{x} - (\boldsymbol{X} + a\boldsymbol{A}(T, \boldsymbol{X})))}{\partial X^m}. \tag{5-55}$$

Under the temporal gauge condition (4-23), (3-8), (A-9), (A-10) and (A-14) becomes

$$\Phi^0_{(0)}(x) = 0, \quad \Phi^i_{(0)}(t, \boldsymbol{x}) = \int d^3y\, \delta^3(\boldsymbol{x} - (\boldsymbol{y} + a\boldsymbol{A}(t, \boldsymbol{y})))A^i(t, \boldsymbol{y}); \tag{5-56}$$

$$\Phi^0_{\perp(1)}(x) = -\frac{\partial^2}{\partial x^i \partial x^j}\int d^4y\, D(x^0 - y^0, \boldsymbol{x} - (\boldsymbol{y} + a\boldsymbol{A}(y)))A^i(y)A^{j,0}(y), \tag{5-57}$$

$$\Phi^i_{\perp(1)}(x) = \frac{\partial^2}{\partial x^0 \partial x^j}\int d^4y\, D(x^0 - y^0, \boldsymbol{x} - (\boldsymbol{y} + a\boldsymbol{A}(y)))A^i(y)A^{j,0}(y)$$
$$+ \frac{\partial^2}{\partial x^j \partial x^k}\int d^4y\, D(x^0 - y^0, \boldsymbol{x} - (\boldsymbol{y} + a\boldsymbol{A}(y)))(A^i(y)A^{j,k}(y) - A^k(y)A^{j,i}(y)); \tag{5-58}$$

$$\Phi^0_{\perp(2)}(x) = -\frac{\partial}{\partial x^j}\int d^4y\, D(x^0 - y^0, \boldsymbol{x} - (\boldsymbol{y} + a\boldsymbol{A}(y)))A^j{}_{,k}(y)A^{k,0}(y), \tag{5-59}$$

$$\Phi^{i}_{\perp(2)}(x) = \frac{\partial}{\partial x^0}\int d^4y D\big(x^0-y^0, \boldsymbol{x}-(\boldsymbol{y}+a\boldsymbol{A}(y))\big)A^{i}{}_{,j}(y)A^{j,0}(y)$$
$$+ \frac{\partial}{\partial x^j}\int d^4y D\big(x^0-y^0, \boldsymbol{x}-(\boldsymbol{y}+a\boldsymbol{A}(y))\big)\Big(\big(A^{i}{}_{,k}(y)A^{k,j}(y) - A^{j}{}_{,k}(y)A^{k,i}(y)\big)$$
$$- aA^{i}{}_{,0}(y)A^{j}{}_{,k}(y)A^{k,0}(y) + aA^{i}{}_{,k}(y)A^{j}{}_{,0}(y)A^{k,0}(y)$$
$$- aA^{i}{}_{,l}(y)A^{j}{}_{,m}(y)A^{m,l}(y) + aA^{i}{}_{,l}(y)A^{j}{}_{,m}(y)A^{l,m}(y)\Big); \quad (5\text{-}60)$$

$$\Phi^{\mu}_{/\!/(1)}(x) = \frac{\partial^2}{\partial x_\mu \partial x^j}\int d^4y D\big(x^0-y^0, \boldsymbol{x}-(\boldsymbol{y}+a\boldsymbol{A}(y))\big)\left(b_1 A^{j}{}_{,k}(y)A^{k}(z) + b_2 \frac{J(z)-1}{a}A^{j}(z)\right). \quad (5\text{-}61)$$

For establishing a quantum theory of the nonlocal theory given by this paper via the Lehmann-Symanzik-Zimmermann approach, based on the discussion in §5.3, what we must do are only the following three steps:

① Instead of (5-26) and (5-27), in (5-23) ~ (5-25), (5-34) ~ (5-36) and (5-39) ~ (5-41), we take

$$\Lambda^0(x) = a\big(\Phi^0_{\perp(1)}(x)\big)_{\overline{W}} + a\big(\Phi^0_{\perp(2)}(x)\big)_{\overline{W}} + a\Phi^0_{/\!/(1)}(x), \quad (5\text{-}62)$$

$$\Lambda^i(x) = \Phi^i_{(0)}(x) + a\big(\Phi^i_{\perp(1)}(x)\big)_{\overline{W}} + a\big(\Phi^i_{\perp(2)}(x)\big)_{\overline{W}} + a\Phi^i_{/\!/(1)}(x); \quad (5\text{-}63)$$

$$\zeta^l(T, \boldsymbol{X}) = J\big(A^i{}_{,j}(T,\boldsymbol{X})\big)\Omega^l_m\big(A^i{}_{,j}(T,\boldsymbol{X})\big)\int d^3x j^m(T,\boldsymbol{x})\delta^3\big(\boldsymbol{x}-(\boldsymbol{X}+a\boldsymbol{A}(T,\boldsymbol{X}))\big)$$
$$- a\big(\dot{A}^m(T,\boldsymbol{X})J\big(A^i{}_{,j}(T,\boldsymbol{X})\big)\Omega^l_m\big(A^i{}_{,j}(T,\boldsymbol{X})\big)\int d^3x j^0(T,\boldsymbol{x})\delta^3\big(\boldsymbol{x}-(\boldsymbol{X}+a\boldsymbol{A}(T,\boldsymbol{X}))\big)\big)_{\overline{W}}. \quad (5\text{-}64)$$

where $\Phi^{\mu}_{(0)}(x)$, $\Phi^{\mu}_{\perp(1)}(x)$, $\Phi^{\mu}_{\perp(2)}(x)$ and $\Phi^{\mu}_{/\!/(1)}(x)$ in (5-62) and (5-63) are given by (5-54) ~ (5-59); $(\cdots)_{\overline{W}}$ means an appropriate operators ordering, of which the exact definition will be given in below discussion. The form (5-64) of $\zeta^l(T, \boldsymbol{X})$ is obtained according to the equation of motion (3-21), the temporal gauge condition $A^0 = 0$, (5-51) and (5-52), in which the appropriate operators ordering has been considered.

② Instead of (5-31), in the constraint condition (5-30),

$$\zeta^0(T, \boldsymbol{X}) = J\big(A^i{}_{,j}(T,\boldsymbol{X})\big)\int d^3x j^0(T,\boldsymbol{x})\delta^3\big(\boldsymbol{x}-(\boldsymbol{X}+a\boldsymbol{A}(T,\boldsymbol{X}))\big). \quad (5\text{-}65)$$

The form (5-65) of $\zeta^0(T, \boldsymbol{X})$ is also obtained according to (3-20), $A^0 = 0$, (5-51) and (5-52).

Similar to the case of the current quantum electrodynamics, a physical state vector $|\Psi_{\text{phy}}\rangle$ satisfies (5-30), in which $\zeta^0(T, \boldsymbol{X})$ is given by (5-65), if and only if, for the operators $\dot{A}(T_0, \boldsymbol{X})$ and $\zeta^0(T_0, \boldsymbol{X})$ of initial time $T_0$, it satisfies (5-32). Because, according to (5-26) and (5-64) we have

$$\frac{\partial}{\partial T}\nabla\cdot\dot{A}(T,X)=e\nabla\cdot\zeta(T,X)=e\zeta^{l}{}_{,l}(T,X)$$

$$=eJ\!\left(A^{i}{}_{,j}(T,X)\right)\Omega_{m}^{l}\!\left(A^{i}{}_{,j}(T,X)\right)\!\int d^{3}x\, j^{m}(T,x)\frac{\partial\delta^{3}(x-(X+aA(T,X)))}{\partial X^{l}}$$

$$-ea\!\left(\!\left(\dot{A}^{m}{}_{,l}(T,X)J\!\left(A^{i}{}_{,j}(T,X)\right)\Omega_{m}^{l}\!\left(A^{i}{}_{,j}(T,X)\right)\!\int d^{3}x\, j^{0}(T,x)\delta^{3}(x-(X+aA(T,X)))\right)\!\bigg|_{\overline{W}}\right.\quad(5\text{-}66)$$

$$\left.+\!\left(\dot{A}^{m}(T,X)J\!\left(A^{i}{}_{,j}(T,X)\right)\Omega_{m}^{l}\!\left(A^{i}{}_{,j}(T,X)\right)\!\int d^{3}x\, j^{0}(T,x)\frac{\partial\delta^{3}(x-(X+aA(T,X)))}{\partial X^{l}}\right)\!\bigg|_{\overline{W}}\right),$$

for the first term in (5-66), using (5-53), integration by parts for $\int d^{3}x$ and (5-28), for the rest terms using (5-54) and (5-55), we obtain

$$\frac{\partial}{\partial T}\nabla\cdot\dot{A}(T,X)=-eJ\!\left(A^{i}{}_{,j}(T,X)\right)\Omega_{m}^{l}\!\left(A^{i}{}_{,j}(T,X)\right)\!\int d^{3}x\,\frac{\partial j^{0}(T,x)}{\partial T}\delta^{3}(x-(X+aA(T,X)))$$

$$-e\!\left(\!\left(\dot{A}^{m}{}_{,l}(T,X)\frac{\partial J\!\left(A^{i}{}_{,j}(T,X)\right)}{\partial A^{m}{}_{,l}(T,X)}\!\int d^{3}x\, j^{0}(T,x)\delta^{3}(x-(X+aA(T,X)))\right)\!\bigg|_{\overline{W}}\right.\quad(5\text{-}67)$$

$$\left.+\!\left(\dot{A}^{m}(T,X)J\!\left(A^{i}{}_{,j}(T,X)\right)\!\int d^{3}x\, j^{0}(T,x)\frac{\partial\delta^{3}(x-(X+aA(T,X)))}{\partial A^{m}(T,X)}\right)\!\bigg|_{\overline{W}}\right).$$

On the other hand, for arbitrary operator $\Pi\!\left(A^{i}(T,X),A^{l}{}_{,m}(T,X)\right)$ as a function of $A^{i}(T,X)$ and $A^{l}{}_{,m}(T,X)$, we define

$$\frac{\partial\Pi\!\left(A^{i}(T,X),A^{l}{}_{,m}(T,X)\right)}{\partial T}$$

$$=\!\left(\dot{A}^{w}(T,X)\frac{\partial\Pi\!\left(A^{i}(T,X),A^{l}{}_{,m}(T,X)\right)}{\partial A^{w}(T,X)}\right)\!\bigg|_{\overline{W}}+\!\left(\dot{A}^{u}{}_{,v}(T,X)\frac{\partial\Pi\!\left(A^{i}(T,X),A^{l}{}_{,m}(T,X)\right)}{\partial A^{u}{}_{,v}(T,X)}\right)\!\bigg|_{\overline{W}}.\quad(5\text{-}68)$$

Thus, for $\zeta^{0}(T,X)$ given by (5-65) we have

$$\frac{\partial\zeta^{0}(T,X)}{\partial T}=J\!\left(A^{i}{}_{,j}(T,X)\right)\!\int d^{3}x\,\frac{\partial j^{0}(T,X)}{\partial T}\delta^{3}(x-(X+aA(T,X)))$$

$$+\!\left(\dot{A}^{u}{}_{,v}(T,X)\frac{\partial J\!\left(A^{i}{}_{,j}(T,X)\right)}{\partial A^{u}{}_{,v}(T,X)}\!\int d^{3}x\, j^{0}(T,X)\delta^{3}(x-(X+aA(T,X)))\right)\!\bigg|_{\overline{W}}\quad(5\text{-}69)$$

$$+\!\left(\dot{A}^{w}(T,X)J\!\left(A^{i}{}_{,j}(T,X)\right)\!\int d^{3}x\, j^{0}(T,X)\frac{\partial\delta^{3}(x-(X+aA(T,X)))}{\partial A^{w}(T,X)}\right)\!\bigg|_{\overline{W}}.$$

Comparing (5-67) and (5-69), we have

$$\frac{\partial}{\partial T}\nabla\cdot\dot{A}(T,X)=-e\frac{\partial\zeta^{0}(T,X)}{\partial T},$$

namely, (5-32) still holds. Hence, through similar discussion, we obtain the conclusion that a physical state vector $|\Psi_{\text{phy}}\rangle$ satisfies (5-29) if and only if it satisfies (5-31), and, further, (5-48).

Of course, in (5-29) ~ (5-33) and (5-48), $\zeta^{0}(T,X)$ is now given by (5-65).

③ Definition of the appropriate operators ordering

Notice that, in the current QED, according to (5-25), (5-27) and the commutation relations (5-42) of "In" and "Out" fields we can *prove* the commutation relations of the field operators $\psi(x)$, $\bar{\psi}(x)$ and $A(X)$:

$$\{\psi_\alpha(t,\boldsymbol{x}),\psi_\beta^+(t,\boldsymbol{x}')\}=i\delta_{\alpha\beta}\delta^3(\boldsymbol{x}-\boldsymbol{x}'),\quad [A^i(T,\boldsymbol{X}),\dot{A}^j(T,\boldsymbol{X}')]=i\delta^{ij}\delta^3(\boldsymbol{X}-\boldsymbol{X}');\ \text{others}=0. \quad (5\text{-}70)$$

Namely, (5-70) are not axioms but *educible conclusions*.

On the other hand, for the nonlocal theory given in this paper, (5-70) does not hold. For instance, according to (5-25) and (5-64) we have

$$\frac{\partial}{\partial T}[A^i(T,\boldsymbol{X}),\dot{A}^j(T,\boldsymbol{X}')]=[\dot{A}^i(T,\boldsymbol{X}),\dot{A}^j(T,\boldsymbol{X}')]-ae[A^i(T,\boldsymbol{X}),\ddot{A}^j(T,\boldsymbol{X}')]$$
$$=-ae[A^i(T,\boldsymbol{X}),(\dot{A}^k(T,\boldsymbol{X}')f_k^j(A^i_{,j}(T,\boldsymbol{X}')))_{\overline{W}}], \quad (5\text{-}71)$$

where

$$f_m^l(A^i_{,j}(T,\boldsymbol{X}))=J(A^i_{,j}(T,\boldsymbol{X}))\Omega_m^l(A^i_{,j}(T,\boldsymbol{X}))\int d^3x j^0(T,\boldsymbol{x})\delta^3(\boldsymbol{x}-(\boldsymbol{X}+a\boldsymbol{A}(T,\boldsymbol{X}))). \quad (5\text{-}72)$$

In the calculation of (5-71), we have used $[A^i(T,\boldsymbol{X}),A^j(T,\boldsymbol{X}')]=0$ and $[\dot{A}^i(T,\boldsymbol{X}),\dot{A}^j(T,\boldsymbol{X}')]=0$, we can prove that these two formulas hold for the nonlocal theory given in this section.

From (5-71) we see that $\frac{\partial}{\partial T}[A^i(T,\boldsymbol{X}),\dot{A}^j(T,\boldsymbol{X}')]\neq 0$. Hence, even if

$$[A^i(T,\boldsymbol{X}),\dot{A}^j(T,\boldsymbol{X}')]\Big|_{T\to-\infty}=[A^i_{\text{in}}(T,\boldsymbol{X}),\dot{A}^j_{\text{in}}(T,\boldsymbol{X}')]=i\delta^{ij}\delta^3(\boldsymbol{X}-\boldsymbol{X}'), \quad (5\text{-}73)$$

we still have $[A^i(T,\boldsymbol{X}),\dot{A}^j(T,\boldsymbol{X}')]\neq i\delta^{ij}\delta^3(\boldsymbol{X}-\boldsymbol{X}')$ for arbitrary time *T*.

We therefore can not define $(\cdots)_{\overline{W}}$ in (5-62) ~ (5-64) and (5-68) as the Weyl ordering, because the Weyl ordering is arisen from the commutation relations (5-70).

On the other hand, if we use perturbation method for (5-34) ~ (5-36) to obtain series solutions of the field operators $\psi(x)$, $\bar{\psi}(x)$, $A(X)$ and $\dot{A}(X)$, then all $\psi(x)$, $\bar{\psi}(x)$, $A(X)$ and $\dot{A}(X)$ are functions (series) of $\psi_{\text{in}}(x)$, $\bar{\psi}_{\text{in}}(x)$, $A_{\text{in}}(X)$ and $\dot{A}_{\text{in}}(X)$, whose the commutation relations are given by (5-42). Thus, we can try to define the generalized Weyl ordering $(V(\psi,\bar{\psi},A,\dot{A}))_{\overline{W}}$ as such manner: after $\psi(x)$, $\bar{\psi}(x)$, $A(X)$ and $\dot{A}(X)$ are expressed by $\psi_{\text{in}}(x)$, $\bar{\psi}_{\text{in}}(x)$, $A_{\text{in}}(X)$ and $\dot{A}_{\text{in}}(X)$, in $V(\psi,\bar{\psi},A,\dot{A})=\tilde{V}(\psi_{\text{in}},\bar{\psi}_{\text{in}},A_{\text{in}},\dot{A}_{\text{in}})$, all $\psi_{\text{in}}(x)$, $\bar{\psi}_{\text{in}}(x)$, $A_{\text{in}}(X)$ and $\dot{A}_{\text{in}}(X)$ obey the Weyl ordering.

Another definition of the appropriate operators ordering is that, after $\psi_{\text{in}}(x)$, $\bar{\psi}_{\text{in}}(x)$, $A_{\text{in}}(X)$ and $\dot{A}_{\text{in}}(X)$ are expressed by $b_{\text{in}}(\boldsymbol{p},s)$, $b_{\text{in}}^+(\boldsymbol{p},s)$, $d_{\text{in}}(\boldsymbol{p},s)$, $d_{\text{in}}^+(\boldsymbol{p},s)$, $a_{(\lambda)\text{in}}(T,\boldsymbol{K})$, $a_{(\lambda)\text{in}}^*(T,\boldsymbol{K})$, $Q_{\text{in}}(T,\boldsymbol{X})$ and $P_{\text{in}}(T,\boldsymbol{X})$, in the expression

$$V(\psi, \overline{\psi}, A, \dot{A}) = \widetilde{V}(\psi_{in}, \overline{\psi}_{in}, A_{in}, \dot{A}_{in})$$
$$= \overline{V}(b_{in}(\boldsymbol{p},s), b_{in}^+(\boldsymbol{p},s), d_{in}(\boldsymbol{p},s), d_{in}^+(\boldsymbol{p},s), a_{(\lambda)in}(T,\boldsymbol{K}), a_{(\lambda)in}^*(T,\boldsymbol{K}), Q_{in}(T,\boldsymbol{X}), P_{in}(T,\boldsymbol{X})),$$

all the operators $b_{in}(\boldsymbol{p},s)$, $b_{in}^+(\boldsymbol{p},s)$, $d_{in}(\boldsymbol{p},s)$, $d_{in}^+(\boldsymbol{p},s)$, $a_{(\lambda)in}(T,\boldsymbol{K})$, $a_{(\lambda)in}^*(T,\boldsymbol{K})$, $Q_{in}(T,\boldsymbol{X})$ and $P_{in}(T,\boldsymbol{X})$ obey so called "normal product" ordering.

In any case, the definition $(\cdots)_{\overline{W}}$ of the appropriate operators ordering in (5-62) ~ (5-64) and (5-68) is after all determined by comparing theoretic calculation results with experimental data.

Via the above three steps and remaining the all rest formulas in §5.3, we have established a self-sufficient quantum theory of the nonlocal theory given by this paper. Similar to the current quantum electrodynamics, in principle, we can obtain solution for arbitrary question in such quantum theory of the nonlocal theory.

Some questions, for example, gauge invariance of the *S*-matrix and microscopic causality, i.e., the condition that the commutator of two Heisenberg field operators taken at two spacelike separated points strictly vanish, of such quantum theory of the nonlocal theory given in this section will be studied further.

## 6 Summary

In this paper, we present a theory of quantum electrodynamics with nonlocal interaction, all the action, the equations of motion of charged particle and electromagnetic field are given by (3-13) ~ (3-15), (3-17) and (3-19), respectively; and all these qualities are invariable under the generalized gauge transformation (1-12), (4-22) and (4-1). The generalized gauge transformation can guarantee that the temporal gauge condition $A^0(x) = 0$ holds or the Lorentz gauge condition $\frac{\partial A^\mu(x)}{\partial x^\mu} = 0$ holds, respectively. It is important that, for the case of free fields, charged particle and electromagnetic field obey the Dirac equation and the Maxwell equation of free fields, respectively; for the case with interaction, both the equations of motion of charged particle and electromagnetic field lead to current conservation $j^\mu{}_{,\mu}(x) = 0$ naturally. Besides these characteristics, the Lorentz invariance of the theory is obvious, and we can verify easily that the theory returns to the current QED when *a* = 0.

Hence, the theory of quantum electrodynamics with nonlocal interaction satisfies all conditions that a generalized theory must satisfy.

Besides these characteristics, we have established the corresponding quantum theory under the temporal gauge condition by taking advantage of the Lehmann-Symanzik-Zimmermann formalism. Of course, for quantization of the theory, one can establish the corresponding quantum theory under the Lorentz gauge condition by the same approach (see the Appendix E of this paper), or uses other methods, e.g., functional method. However, in my opinion, the quantum theory under the temporal gauge condition should be as the foundation of other quantum theory.

On the other hand, for the quantum theory given in §5.4, besides the problem of gauge invariance of the *S*-matrix and microscopic causality, it is obvious that it is very complicated when we use it to calculate actual physical process; for which we therefore must develop convenient calculation technology.

## Appendix A  The forms of $\Phi_\perp^\mu(x)$ and $\Phi^\mu(x)$

Via integration by parts, $\Phi_\perp^\mu(x)$ given by (3-6) can be expressed to the following forms:

$$\begin{aligned}\Phi_\perp^\mu(x) &= \int d^4y \widetilde{F}^{\mu\nu}(y)\frac{\partial}{\partial y^\nu}D(x-y) \\ &= \int d^4y \left(\int d^4z \delta^4(z+aA(z)-y)J_\rho^\mu(z)J_\sigma^\nu(z)F^{\rho\sigma}(z)\right)\frac{d^4k}{(2\pi)^4}\frac{-ik_\nu}{k^2}e^{-ik\cdot(x-y)} \\ &= \int d^4z J_\rho^\mu(z)J_\sigma^\nu(z)F^{\rho\sigma}(z)\frac{d^4k}{(2\pi)^4}\frac{-ik_\nu}{k^2}e^{-ik\cdot(x-(z+aA(z)))},\end{aligned} \qquad (A\text{-}1)$$

using (2-58), $\Phi_\perp^\mu(x)$ can also be written to the form

$$\Phi_\perp^\mu(x) = -\frac{\partial}{\partial x^\nu}\int d^4y D(x-(y+aA(y)))J_\rho^\mu(y)J_\sigma^\nu(y)F^{\rho\sigma}(y). \qquad (A\text{-}2)$$

Introducing

$$\begin{aligned}\varphi_{(2)}^{\mu\nu}(y) &= \frac{1}{a}\left(J_\rho^\mu(y)J_\sigma^\nu(y)F^{\rho\sigma}(y)-F^{\mu\nu}(y)\right) \\ &= -\left(A^\mu_{,\lambda}(y)A^{\lambda,\nu}(y)-A^\nu_{,\lambda}(y)A^{\lambda,\mu}(y)\right)+aA^\mu_{,\rho}(y)A^\nu_{,\sigma}(y)F^{\rho\sigma}(y);\end{aligned} \qquad (A\text{-}3)$$

we have

$$\Phi_\perp^\mu(x) = -\frac{\partial}{\partial x^\nu}\int d^4y D(x-(y+aA(y)))F^{\mu\nu}(y) - a\frac{\partial}{\partial x^\nu}\int d^4y D(x-(y+aA(y)))\varphi_{(2)}^{\mu\nu}(y). \qquad (A\text{-}4)$$

The first term in (A-4) can be written to the form

$$\begin{aligned}&-\frac{\partial}{\partial x^\nu}\int d^4y D(x-(y+aA(y)))F^{\mu\nu}(y) \\ &= -\frac{\partial}{\partial x^\nu}\int d^4y D(x-(y+aA(y)))\left(\eta^{\mu\sigma}\frac{\partial A^\nu(y)}{\partial y^\sigma}-\eta^{\nu\sigma}\frac{\partial A^\mu(y)}{\partial y^\sigma}\right) \\ &= \frac{\partial}{\partial x^\nu}\int d^4y \frac{\partial D(x-(y+aA(y)))}{\partial y^\sigma}\left(\eta^{\mu\sigma}A^\nu(y)-\eta^{\nu\sigma}A^\mu(y)\right) \\ &= -\frac{\partial^2}{\partial x^\nu \partial x^\rho}\int d^4y D(x-(y+aA(y)))J_\sigma^\rho(y)\left(\eta^{\mu\sigma}A^\nu(y)-\eta^{\nu\sigma}A^\mu(y)\right) \\ &= -\frac{\partial^2}{\partial x^\nu \partial x^\rho}\int d^4y D(x-(y+aA(y)))\left(\eta^{\mu\sigma}A^\nu(y)-\eta^{\nu\sigma}A^\mu(y)\right)\delta_\sigma^\rho \\ &\quad -a\frac{\partial^2}{\partial x^\nu \partial x^\rho}\int d^4y D(x-(y+aA(y)))A^\rho_{,\sigma}(y)\left(\eta^{\mu\sigma}A^\nu(y)-\eta^{\nu\sigma}A^\mu(y)\right),\end{aligned} \qquad (A\text{-}5)$$

in the above calculation, (2-35) is used. The first term in the last expression of (A-5) reads

$$-\frac{\partial^2}{\partial x^\nu \partial x^\rho} \int d^4 y D(x-(y+aA(y)))(\eta^{\mu\sigma} A^\nu(y) - \eta^{\nu\sigma} A^\mu(y))\delta_\sigma^\rho$$

$$= \int d^4 y (\eta^{\mu\sigma} A^\nu(y) - \eta^{\nu\sigma} A^\mu(y))\delta_\sigma^\rho \frac{d^4 k}{(2\pi)^4} \frac{-k_\rho k_\nu}{k^2} e^{-ik\cdot(x-(y+aA(y)))}$$

$$= \int d^4 y A^\nu(y) \frac{d^4 k}{(2\pi)^4} \frac{k^2 \delta_\nu^\mu - k^\mu k_\nu}{k^2} e^{-ik\cdot(x-(y+aA(y)))} \quad \text{(A-6)}$$

$$= \int d^4 y \left( \int d^4 z \delta^4(z+aA(z)-y) A^\nu(z) \right) \frac{d^4 k}{(2\pi)^4} \frac{k^2 \delta_\nu^\mu - k^\mu k_\nu}{k^2} e^{ik\cdot(x-y)},$$

substituting (A-6), (A-5) to (A-4), and defining

$$\varphi_{(1)\lambda}{}^{\mu\nu}(y) = A^\mu(y) A_\lambda{}^{,\nu}(y) - A^\nu(y) A_\lambda{}^{,\mu}(y), \quad \text{(A-7)}$$

we obtain

$$\Phi_\perp^\mu(x) = \int d^4 y \left( \int d^4 z \delta^4(y-(z+aA(z))) A^\nu(z) \right) \frac{d^4 k}{(2\pi)^4} \frac{k^2 \delta_\nu^\mu - k^\mu k_\nu}{k^2} e^{ik\cdot(x-y)} \quad \text{(A-8)}$$
$$+ a\Phi_{\perp(1)}^\mu(x) + a\Phi_{\perp(2)}^\mu(x);$$

$$\Phi_{\perp(1)}^\mu(x) = \frac{\partial^2}{\partial x^\nu \partial x_\lambda} \int d^4 y D(x-(y+aA(y))) \varphi_{(1)\lambda}{}^{\mu\nu}(y), \quad \text{(A-9)}$$

$$\Phi_{\perp(2)}^\mu(x) = -\frac{\partial}{\partial x^\nu} \int d^4 y D(x-(y+aA(y))) \varphi_{(2)}^{\mu\nu}(y); \quad \text{(A-10)}$$

The above functions have the following characteristics:

$$\varphi_{(1)\lambda}{}^{\mu\nu}(y) = -\varphi_{(1)\lambda}{}^{\nu\mu}(y), \quad \varphi_{(2)}^{\mu\nu}(y) = -\varphi_{(2)}^{\nu\mu}(y); \quad \text{(A-11)}$$

$$\Phi_{\perp(1),\mu}^\mu(x) = 0, \quad \Phi_{\perp(2),\mu}^\mu(x) = 0. \quad \text{(A-12)}$$

On the other hand, from (3-9) and (3-11) we have

$$\Phi_{//}^\mu(x) = \int \frac{d^4 k}{(2\pi)^4} \frac{k^\mu k_\nu}{k^2} e^{ik\cdot(x-y)} \int d^4 y \left( \int d^4 z \delta^4(y-(z+aA(z))) A^\nu(z) \right) + a\Phi_{//(1)}^\mu(x), \quad \text{(A-13)}$$

$$\Phi_{//(1)}^\mu(x) = \frac{\partial^2}{\partial x_\mu \partial x^\nu} \int d^4 y D(x-(y+aA(y))) \left( b_1 A^\nu{}_{,\lambda}(y) A^\lambda(z) + b_2 \frac{J(z)-1}{a} A^\nu(z) \right). \quad \text{(A-14)}$$

Finally, we obtain

$$\Phi^\mu(x) = \Phi_\perp^\mu(x) + \Phi_{//}^\mu(x) = \Phi_{(0)}^\mu(x) + a\Phi_{\perp(1)}^\mu(x) + a\Phi_{\perp(2)}^\mu(x) + a\Phi_{//(1)}^\mu(x). \quad \text{(A-15)}$$

where $\Phi_{(0)}^\mu(x)$ is given by (3-8).

Using the method obtaining (4-7) from (4-6), we have

$$\Phi_{(0)}^\mu(x) \equiv A^\mu(x) + \sum_{n=1}^\infty \frac{(-1)^n a^n}{n!} \Phi_{(0)(n)}^\mu(x); \quad \text{(A-16)}$$

Similar to (4-6) we have

$$D(x-(y+aA(y))) = D(y+aA(y)-x) = D(y-x) + aA^\lambda(y)\frac{\partial D(y-x)}{\partial y^\lambda}$$
$$+\frac{1}{2!}a^2 A^\rho(y)A^\sigma(y)\frac{\partial^2 D(y-x)}{\partial y^\rho \partial y^\sigma} + \cdots + \frac{1}{n!}a^n A^n(y)\frac{\partial^n D(y-x)}{\partial y^n} + \cdots, \qquad (A-17)$$

from (A-9), (A-10), (A-14) and (A-17) we have

$$\Phi^\mu_{\perp(1)}(x) + \Phi^\mu_{\perp(2)}(x) + \Phi^\mu_{//(1)}(x) = \int d^4 y D(x-(y+aA(y)))\frac{\partial}{\partial y^\rho}\left\{\Omega^\rho_\nu(y)\right.$$
$$\times\left[\frac{\partial}{\partial y^\sigma}\left(\eta^{\lambda\tau}\Omega^\sigma_\tau(y)\varphi_{(1)\lambda}{}^{\mu\nu}(y)\right) - \varphi^{\mu\nu}_{(2)}(y) + \frac{\partial}{\partial y^\sigma}\left(\eta^{\mu\tau}\Omega^\sigma_\tau(y)\left(b_1 A^\nu{}_{,\lambda}(y)A^\lambda(z) + b_2\frac{J(z)-1}{a}A^\nu(z)\right)\right)\right]\right\}$$
$$\equiv \int d^4 y D(x-y)\sum_{n=0}^\infty a^n \widetilde{\Phi}^\mu_{(n)}(y); \qquad (A-18)$$

If we choose

$$D(x-y) = D_{\text{ret}}(x-y) = \frac{1}{4\pi|\mathbf{x}-\mathbf{y}|}\theta(x^0-y^0)\delta\left((x^0-y^0)-|\mathbf{x}-\mathbf{y}|\right)$$
$$= \frac{1}{4\pi|\mathbf{x}-\mathbf{y}|}\theta(x^0-y^0)\int\frac{dk}{2\pi}e^{ik((x^0-y^0)-|\mathbf{x}-\mathbf{y}|)} = \frac{1}{4\pi|\mathbf{x}-\mathbf{y}|}\theta(x^0-y^0)\int\frac{dk}{2\pi}e^{-ik|\mathbf{x}-\mathbf{y}|}e^{ik(x^0-y^0)}$$
$$\equiv \frac{1}{4\pi|\mathbf{x}-\mathbf{y}|}\theta(x^0-y^0)\exp\left(|\mathbf{x}-\mathbf{y}|\frac{\partial}{\partial y^0}\right)\int\frac{dk}{2\pi}e^{ik(x^0-y^0)}$$
$$= \frac{1}{4\pi|\mathbf{x}-\mathbf{y}|}\theta(x^0-y^0)\exp\left(|\mathbf{x}-\mathbf{y}|\frac{\partial}{\partial y^0}\right)\delta(x^0-y^0),$$

then substituting the above formula to (A-18), after integration by parts and ignoring surface terms, we have

$$\Phi^\mu_{\perp(1)}(x) + \Phi^\mu_{\perp(2)}(x) + \Phi^\mu_{//(1)}(x) = \int d^4 y \frac{1}{4\pi|\mathbf{x}-\mathbf{y}|}\theta(x^0-y^0)\exp\left(|\mathbf{x}-\mathbf{y}|\frac{\partial}{\partial y^0}\right)\delta(x^0-y^0)\sum_{n=0}^\infty a^n \widetilde{\Phi}^\mu_{(n)}(y)$$
$$= \int d^4 y \frac{1}{4\pi|\mathbf{x}-\mathbf{y}|}\delta(x^0-y^0)\exp\left(-|\mathbf{x}-\mathbf{y}|\frac{\partial}{\partial y^0}\right)\left(\theta(x^0-y^0)\sum_{n=0}^\infty a^n \widetilde{\Phi}^\mu_{(n)}(y)\right)$$
$$= \int d^3 y \frac{1}{4\pi|\mathbf{x}-\mathbf{y}|}\left[\exp\left(-|\mathbf{x}-\mathbf{y}|\frac{\partial}{\partial y^0}\right)\left(\theta(x^0-y^0)\sum_{n=0}^\infty a^n \widetilde{\Phi}^\mu_{(n)}(y^0,\mathbf{y})\right)\right]_{y^0\to x^0}. \qquad (A-19)$$

Substituting (A-16) and (A-19) to (A-15), we obtain

$$\Phi^\mu(x) = A^\mu(x) + \sum_{n=1}^\infty \frac{(-1)^n a^n}{n!}\Phi^\mu_{(0)(n)}(x)$$
$$+ \int d^3 y \frac{1}{4\pi|\mathbf{x}-\mathbf{y}|}\left[\exp\left(-|\mathbf{x}-\mathbf{y}|\frac{\partial}{\partial y^0}\right)\left(\theta(x^0-y^0)\sum_{n=0}^\infty a^n \widetilde{\Phi}^\mu_{(n)}(y^0,\mathbf{y})\right)\right]_{y^0\to x^0}. \qquad (A-20)$$

## Appendix B  The derivation of the equation of motion of electromagnetic field according to the action principle

We write the action (3-13) to the form:

$$S = \int d^4x L_D(x) + S_{EM} + S_{I\perp} + S_{I//},  \tag{B-1}$$

$$S_{EM} = \int d^4x L_{EM}(x) = -\frac{1}{2}\int d^4x \eta_{\alpha\mu}\eta_{\beta\nu}\Phi_\perp^{\alpha,\beta}(x)\left(\Phi_\perp^{\mu,\nu}(x) - \Phi_\perp^{\nu,\mu}(x)\right)$$

$$= -\frac{1}{2}\int d^4x \Phi_\perp^{\alpha,\beta}(x)\Phi_{\perp\alpha,\beta}(x) + \frac{1}{2}\int d^4x \Phi_{\perp,\beta}^\alpha(x)\Phi_{\perp,\alpha}^\beta(x);$$

Considering $\int d^4x \Phi_{\perp,\beta}^\alpha(x)\Phi_{\perp,\alpha}^\beta(x) = -\int d^4x \Phi_\perp^\alpha(x)\Phi_{\perp,\alpha,\beta}^\beta(x) = -\int d^4x \Phi_\perp^\alpha(x)\left(\Phi_{\perp,\beta}^\beta(x)\right)_{,\alpha} = 0$, and using (A-2), we have

$$S_{EM} = -\frac{1}{2}\eta_{\alpha\beta}\int d^4x \Phi_\perp^{\alpha,\gamma}(x)\Phi_{\perp,\gamma}^\beta(x)$$

$$= -\frac{1}{2}\eta_{\alpha\beta}\int d^4x \left(\int d^4u \frac{\partial^2 D(x-(u+aA(u)))}{\partial x_\gamma \partial x^\tau} J_\omega^\alpha(u)J_\chi^\tau(u)F^{\omega\chi}(u)\right) \tag{B-2}$$

$$\times \left(\int d^4v \frac{\partial^2 D(x-(v+aA(v)))}{\partial x^\gamma \partial x^\lambda} J_\rho^\beta(v)J_\sigma^\lambda(v)F^{\rho\sigma}(v)\right);$$

Calculating integration by parts for $x^\gamma$ and $x^\tau$, and notice $\frac{\partial^2 D(x-(z+aA(z)))}{\partial x^\gamma \partial x_\gamma} = \delta^4(x-(z+aA(z)))$, we have

$$S_{EM} = -\frac{1}{2}\eta_{\alpha\beta}\int d^4x \left(\int d^4u \delta^4(x-(u+aA(u)))J_\omega^\alpha(u)J_\chi^\tau(u)F^{\omega\chi}(u)\right)$$

$$\times \left(\int d^4v \frac{\partial^2 D(x-(v+aA(v)))}{\partial x^\tau \partial x^\lambda} J_\rho^\beta(v)J_\sigma^\lambda(v)F^{\rho\sigma}(v)\right), \tag{B-3}$$

we can verify that (B-3) returns to (1-23) when $a = 0$.

After long-winded calculation, we can obtain another form of $S_{EM}$:

$$S_{EM} = -\frac{1}{2}\int \frac{d^4z}{J(z)} F_{\alpha\beta}(z)F^{\alpha\beta}(z)$$

$$+ \frac{1}{4}\int \frac{d^4z}{J(z)}\left(\Omega_\rho^\alpha(z)A^{\rho,\beta}(z) - \Omega_\rho^\beta(z)A^{\rho,\alpha}(z)\right)\left(\Omega_\beta^\sigma(z)A_{\alpha,\sigma}(z) - \Omega_\alpha^\sigma(z)A_{\beta,\sigma}(z)\right)$$

$$- \frac{1}{2}\int d^4z \frac{\partial J^{-1}(z)}{\partial z^\alpha} A_\beta(z)\left\{2J_\gamma^\beta(z)F^{\alpha\gamma}(z) + \Omega_\sigma^\alpha(z)\left(\Omega_\rho^\sigma(z)A^{\rho,\beta}(z) - \Omega_\rho^\beta(z)A^{\rho,\sigma}(z)\right)\right.$$

$$\left. + \left[\eta^{\beta\rho}\frac{\partial(A^\alpha(z)\Omega_\rho^\sigma(z))}{\partial z^\sigma} - \eta^{\alpha\rho}\frac{\partial(A^\beta(z)\Omega_\rho^\sigma(z))}{\partial z^\sigma}\right]\right\}$$

$$- \frac{1}{2}a^2\int d^4z \eta_{\alpha\beta}\varphi_{(1)\lambda}^{\alpha\rho}(z)\frac{d^4k}{(2\pi)^4}\frac{k_\rho k_\sigma k^\lambda k^\tau}{k^2} e^{ik\cdot((z+aA(z))-(x+aA(x)))}d^4x \varphi_{(1)\tau}^{\beta\sigma}(x)$$

$$- a^2\int d^4z \eta_{\alpha\beta}\varphi_{(1)\lambda}^{\alpha\rho}(z)\frac{d^4k}{(2\pi)^4}\frac{ik_\rho k_\sigma k^\lambda}{k^2} e^{ik\cdot((z+aA(z))-(x+aA(x)))}d^4x \varphi_{(2)}^{\beta\sigma}(x)$$

$$- \frac{1}{2}a^2\int d^4z \eta_{\alpha\beta}\varphi_{(2)}^{\alpha\rho}(z)\frac{d^4k}{(2\pi)^4}\frac{k_\rho k_\sigma}{k^2} e^{ik\cdot((z+aA(z))-(x+aA(x)))}d^4x \varphi_{(2)}^{\beta\sigma}(x). \tag{B-4}$$

where $\varphi_{(1)\lambda}{}^{\mu\nu}$ and $\varphi_{(2)}^{\mu\nu}$ are given by (A-7) and (A-3), respectively. Notice that the second derivative term $A^{\gamma}{}_{,\chi,\kappa}(y)$ appears in (B-4) due to $\dfrac{\partial J^{-1}(z)}{\partial z^{\alpha}} = -\dfrac{a}{J(z)}\Omega_{\mu}^{\nu}(z)A^{\mu}{}_{,\nu,\alpha}(z)$ and

$$\frac{\partial \Omega_{\rho}^{\sigma}(z)}{\partial z^{\sigma}} = -a\Omega_{\mu}^{\sigma}(z)\Omega_{\rho}^{\nu}(z)A^{\mu}{}_{,\nu,\sigma}(z).$$

If we use (3-16), then

$$S_{\text{EM}}(x) = \sum_{i=1}^{2} d_i\, S_{\text{EM}(i)}(x), \qquad (B-5)$$

the form of $S_{\text{EM}(i)}(x)$ is (B-4), but in which the function $D(x-(y+aA(y)))$ is replaced by $D_{(i)}(x-(y+aA(y)))$.

Using (A-2) and (3-9), we have

$$S_{\text{I}\perp} = -e\int d^4 x\, j_{\alpha}(x)\Phi_{\perp}^{\alpha}(x) = e\int d^4 x\, j_{\alpha}(x)\left(\int d^4 y\, \frac{\partial D(x-(y+aA(y)))}{\partial x^{\nu}} J_{\rho}^{\alpha}(y)J_{\sigma}^{\nu}(y)F^{\rho\sigma}(y)\right), \qquad (B-6)$$

$$\begin{aligned}S_{\text{I}/\!/} &= -e\int d^4 x\, j_{\alpha}(x)\Phi_{/\!/}^{\alpha}(x) \\ &= -e\int d^4 x\, j^{\alpha}(x)\left(\int d^4 y\, \frac{\partial^2 D(x-(y+aA(y)))}{\partial x^{\alpha}\partial x^{\nu}}\left(b_1 J_{\lambda}^{\nu}(y)A^{\lambda}(y) + b_2 J(y)A^{\nu}(y)\right)\right).\end{aligned} \qquad (B-7)$$

The variational equation for $A^{\mu}(y)$ becomes

$$\frac{\delta S}{\delta A^{\mu}(y)} = \frac{\delta S_{\text{EM}}}{\delta A^{\mu}(y)} + \frac{\delta S_{\text{I}\perp}}{\delta A^{\mu}(y)} + \frac{\delta S_{\text{I}/\!/}}{\delta A^{\mu}(y)} = 0. \qquad (B-8)$$

We employ (B-2) to calculate $\dfrac{\delta S_{\text{EM}}}{\delta A^{\mu}(y)}$ and obtain

$$\frac{\delta S_{\text{EM}}}{\delta A^\mu(y)}$$

$$= -\frac{1}{2}\eta_{\alpha\beta}\int d^4x \left( \int d^4u \left( \frac{\partial}{\partial A^\mu(u)} \frac{\partial^2 D(x-(u+aA(u)))}{\partial x_\gamma \partial x^\tau} \right) \delta^4(u-y) J_\omega^\alpha(u) J_\chi^\tau(u) F^{\omega\chi}(u) \right)$$

$$\times \left( \int d^4v \frac{\partial^2 D(x-(v+aA(v)))}{\partial x^\gamma \partial x^\lambda} J_\rho^\beta(v) J_\sigma^\lambda(v) F^{\rho\sigma}(v) \right)$$

$$-\frac{1}{2}\eta_{\alpha\beta}\int d^4x \left( \int d^4u \frac{\partial^2 D(x-(u+aA(u)))}{\partial x_\gamma \partial x^\tau} \frac{\partial \left( J_\omega^\alpha(u) J_\chi^\tau(u) F^{\omega\chi}(u) \right)}{\partial A^\mu_{,\nu}(u)} \frac{\partial \delta^4(u-y)}{\partial u^\nu} \right)$$

$$\times \left( \int d^4v \frac{\partial^2 D(x-(v+aA(v)))}{\partial x^\gamma \partial x^\lambda} J_\rho^\beta(v) J_\sigma^\lambda(v) F^{\rho\sigma}(v) \right)$$

$$-\frac{1}{2}\eta_{\alpha\beta}\int d^4x \left( \int d^4u \frac{\partial^2 D(x-(u+aA(u)))}{\partial x_\gamma \partial x^\tau} J_\omega^\alpha(u) J_\chi^\tau(u) F^{\omega\chi}(u) \right)$$

$$\times \left( \int d^4v \left( \frac{\partial}{\partial A^\mu(v)} \frac{\partial^2 D(x-(v+aA(v)))}{\partial x^\gamma \partial x^\lambda} \right) \delta^4(v-y) J_\rho^\beta(v) J_\sigma^\lambda(v) F^{\rho\sigma}(v) \right)$$

$$-\frac{1}{2}\eta_{\alpha\beta}\int d^4x \left( \int d^4u \frac{\partial^2 D(x-(u+aA(u)))}{\partial x_\gamma \partial x^\tau} J_\omega^\alpha(u) J_\chi^\tau(u) F^{\omega\chi}(u) \right)$$

$$\times \left( \int d^4v \frac{\partial^2 D(x-(v+aA(v)))}{\partial x^\gamma \partial x^\lambda} \frac{\partial \left( J_\rho^\beta(v) J_\sigma^\lambda(v) F^{\rho\sigma}(v) \right)}{\partial A^\mu_{,\nu}(v)} \frac{\partial \delta^4(v-y)}{\partial v^\nu} \right)$$

$$= \eta_{\alpha\beta} \int d^4x \left( \int d^4z \frac{\partial \delta^4(x-(z+aA(z)))}{\partial x^\tau} J_\omega^\alpha(z) J_\chi^\tau(z) F^{\omega\chi}(z) \right)$$

$$\times \left( \left( \frac{\partial}{\partial A^\mu(y)} \frac{\partial D(x-(y+aA(y)))}{\partial x^\lambda} \right) J_\rho^\beta(y) J_\sigma^\lambda(y) F^{\rho\sigma}(y) \right.$$

$$\left. - \frac{\partial}{\partial y^\nu} \left( \frac{\partial D(x-(y+aA(y)))}{\partial x^\lambda} \frac{\partial \left( J_\rho^\beta(y) J_\sigma^\lambda(y) F^{\rho\sigma}(y) \right)}{\partial A^\mu_{,\nu}(y)} \right) \right)$$

$$= \eta_{\alpha\beta} \int d^4x \left( \int d^4z \delta^4(x-(z+aA(z))) J_\omega^\alpha(z) F^{\omega\chi}_{,\chi}(z) \right)$$

$$\times \left( \left( \frac{\partial}{\partial A^\mu(y)} \frac{\partial D(x-(y+aA(y)))}{\partial x^\lambda} \right) J_\rho^\beta(y) J_\sigma^\lambda(y) F^{\rho\sigma}(y) \right.$$

$$\left. - \frac{\partial}{\partial y^\nu} \left( \frac{\partial D(x-(y+aA(y)))}{\partial x^\lambda} \frac{\partial \left( J_\rho^\beta(y) J_\sigma^\lambda(y) F^{\rho\sigma}(y) \right)}{\partial A^\mu_{,\nu}(y)} \right) \right); \quad \text{(B-9)}$$

In the above calculation process, we have used integration by parts for $x^\gamma$ and the formula $\frac{\partial^2 D(x-(z+aA(z)))}{\partial x^\gamma \partial x_\gamma} = \delta^4(x-(z+aA(z)))$ according to (2-59).

According to (B-6) we obtain

$$\frac{\delta S_{I\perp}}{\delta A^{\mu}(y)} = e\int d^{4}x j_{\alpha}(x)\left(\int d^{4}z \left(\frac{\partial}{\partial A^{\mu}(z)}\frac{\partial D(x-(z+aA(z)))}{\partial x^{\beta}}\right)\delta^{4}(z-y)J_{\rho}^{\alpha}(z)J_{\sigma}^{\beta}(z)F^{\rho\sigma}(z)\right)$$

$$+e\int d^{4}x j_{\alpha}(x)\left(\int d^{4}y \frac{\partial D(x-(z+aA(z)))}{\partial x^{\beta}}\frac{\partial\left(J_{\rho}^{\alpha}(z)J_{\sigma}^{\beta}(z)F^{\rho\sigma}(z)\right)}{\partial A^{\mu}{}_{,\nu}(z)}\frac{\partial\delta^{4}(z-y)}{\partial z^{\nu}}\right)$$

$$=\eta_{\alpha\beta}e\int d^{4}x j^{\alpha}(x)\left(\left(\frac{\partial}{\partial A^{\mu}(y)}\frac{\partial D(x-(y+aA(y)))}{\partial x^{\lambda}}\right)J_{\rho}^{\beta}(y)J_{\sigma}^{\lambda}(y)F^{\rho\sigma}(y)\right.$$

$$\left.-\frac{\partial}{\partial v^{\nu}}\left(\frac{\partial D(x-(y+aA(y)))}{\partial x^{\lambda}}\frac{\partial\left(J_{\rho}^{\beta}(y)J_{\sigma}^{\lambda}(y)F^{\rho\sigma}(y)\right)}{\partial A^{\mu}{}_{,\nu}(y)}\right)\right);$$

(B-10)

According to (B-7) we have

$$\frac{\delta S_{I//}}{\delta A^{\mu}(y)}$$

$$=-e\int d^{4}x j^{\alpha}(x)\left(\int d^{4}z \left(\frac{\partial}{\partial A^{\mu}(z)}\frac{\partial^{2}D(x-(z+aA(z)))}{\partial x^{\alpha}\partial x^{\beta}}\right)\delta^{4}(z-y)\left(b_{1}J_{\lambda}^{\beta}(z)A^{\lambda}(z)+b_{2}J(z)A^{\beta}(z)\right)\right)$$

$$-e\int d^{4}x j^{\alpha}(x)\left(\int d^{4}z \frac{\partial^{2}D(x-(z+aA(z)))}{\partial x^{\alpha}\partial x^{\beta}}\left(b_{1}J_{\lambda}^{\beta}(z)\frac{\partial A^{\lambda}(z)}{\partial A^{\mu}(z)}+b_{2}J(z)\frac{\partial A^{\beta}(z)}{\partial A^{\mu}(z)}\right)\delta^{4}(z-y)\right)$$

$$-e\int d^{4}x j^{\alpha}(x)\left(\int d^{4}z \frac{\partial^{2}D(x-(z+aA(z)))}{\partial x^{\alpha}\partial x^{\beta}}\left(b_{1}\frac{\partial J_{\lambda}^{\beta}(z)}{\partial A^{\mu}{}_{,\nu}(z)}A^{\lambda}(z)+b_{2}\frac{\partial J(z)}{\partial A^{\mu}{}_{,\nu}(z)}A^{\beta}(z)\right)\frac{\partial\delta^{4}(z-y)}{\partial z^{\nu}}\right)$$

$$=-e\int d^{4}x j^{\alpha}(x)\left(\frac{\partial}{\partial A^{\mu}(y)}\frac{\partial^{2}D(x-(y+aA(y)))}{\partial x^{\alpha}\partial x^{\beta}}\right)\left(b_{1}J_{\lambda}^{\beta}(y)A^{\lambda}(y)+b_{2}J(y)A^{\beta}(y)\right)$$

$$-e\int d^{4}x j^{\alpha}(x)\frac{\partial^{2}D(x-(y+aA(y)))}{\partial x^{\alpha}\partial x^{\beta}}\left(b_{1}J_{\mu}^{\beta}(y)+b_{2}J(y)\delta_{\mu}^{\beta}\right)$$

$$+ae\int d^{4}x j^{\alpha}(x)\left(\frac{\partial}{\partial y^{\nu}}\frac{\partial^{2}D(x-(y+aA(y)))}{\partial x^{\alpha}\partial x^{\beta}}\right)\left(b_{1}\delta_{\mu}^{\beta}A^{\nu}(y)+b_{2}J(y)\Omega_{\mu}^{\nu}(y)A^{\beta}(y)\right)$$

$$+ae\int d^{4}x j^{\alpha}(x)\frac{\partial^{2}D(x-(y+aA(y)))}{\partial x^{\alpha}\partial x^{\beta}}\frac{\partial}{\partial y^{\nu}}\left(b_{1}\delta_{\mu}^{\beta}A^{\nu}(y)+b_{2}J(y)\Omega_{\mu}^{\nu}(y)A^{\beta}(y)\right),$$

using (2-29), (2-30) and (2-58) we can prove

$$\left(\frac{\partial}{\partial A^{\mu}(y)}\frac{\partial^{2}D(x-(y+aA(y)))}{\partial x^{\alpha}\partial x^{\beta}}\right)J_{\lambda}^{\beta}(y)A^{\lambda}(y)=a\left(\frac{\partial}{\partial y^{\nu}}\frac{\partial^{2}D(x-(y+aA(y)))}{\partial x^{\alpha}\partial x^{\beta}}\right)\delta_{\mu}^{\beta}A^{\nu}(z),$$

$$\left(\frac{\partial}{\partial A^{\mu}(y)}\frac{\partial^{2}D(x-(y+aA(y)))}{\partial x^{\alpha}\partial x^{\beta}}\right)A^{\beta}(y)=a\left(\frac{\partial}{\partial y^{\nu}}\frac{\partial^{2}D(x-(y+aA(y)))}{\partial x^{\alpha}\partial x^{\beta}}\right)\Omega_{\mu}^{\nu}(y)A^{\beta}(y),$$

we therefore obtain

$$\frac{\delta S_{I//}}{\delta A^{\mu}(y)}=-e\int d^{4}x j^{\alpha}(x)\frac{\partial^{2}D(x-(y+aA(y)))}{\partial x^{\alpha}\partial x^{\beta}}\left(b_{1}J_{\mu}^{\beta}(y)+b_{2}J(y)\Omega_{\mu}^{\beta}(y)-ab_{1}\delta_{\mu}^{\beta}A^{\nu}{}_{,\nu}(y)\right).$$ (B-11)

Substituting (B-9), (B-10) and (B-11) to (B-8), we obtain *the equation of motion of electromagnetic field*:

$$\eta_{\alpha\beta}\int d^4x \left(\int d^4z \delta^4(x-(z+aA(z)))J^\alpha_\omega(z)F^{\omega\chi}{}_{,\chi}(z)+ej^\alpha(x)\right)$$

$$\times\left(\left(\frac{\partial}{\partial A^\mu(y)}\frac{\partial D(x-(y+aA(y)))}{\partial x^\lambda}\right)J^\beta_\rho(y)J^\lambda_\sigma(y)F^{\rho\sigma}(y)\right.$$

$$\left.-\frac{\partial}{\partial y^\nu}\left(\frac{\partial D(x-(y+aA(y)))}{\partial x^\lambda}\frac{\partial(J^\beta_\rho(y)J^\lambda_\sigma(y)F^{\rho\sigma}(y))}{\partial A^\mu{}_{,\nu}(y)}\right)\right) \quad \text{(B-12)}$$

$$-e\int d^4x j^\alpha(x)\frac{\partial^2 D(x-(y+aA(y)))}{\partial x^\alpha \partial x^\beta}\left(b_1 J^\beta_\mu(y)+b_2 J(y)\Omega^\beta_\mu(y)-ab_1\delta^\beta_\mu A^\nu{}_{,\nu}(y)\right)=0.$$

The equation (B-12) is so complicated that we have to try to predigest it.

At first, according to $F^{\omega\chi}{}_{,\chi,\omega}=0$, we have

$$-\int d^4x\left(\int d^4z \delta^4(x-(z+aA(z)))J^\alpha_\omega(z)F^{\omega\chi}{}_{,\chi}(z)\right)\frac{\partial^2 D(x-(y+aA(y)))}{\partial x^\alpha \partial x^\beta}$$

$$=\int d^4x\left(\int d^4z \frac{\partial \delta^4(x-(z+aA(z)))}{\partial x^\alpha}J^\alpha_\omega(z)F^{\omega\chi}{}_{,\chi}(z)\right)\frac{\partial D(x-(y+aA(y)))}{\partial x^\beta} \quad \text{(B-13)}$$

$$=\int d^4x\left(\int d^4z \delta^4(x-(z+aA(z)))F^{\omega\chi}{}_{,\chi,\omega}(z)\right)\frac{\partial D(x-(y+aA(y)))}{\partial x^\beta}=0,$$

we therefore can add the term

$$-\int d^4x\left(\int d^4z \delta^4(x-(z+aA(z)))J^\alpha_\omega(z)F^{\omega\chi}{}_{,\chi}(z)\right)\frac{\partial^2 D(x-(y+aA(y)))}{\partial x^\alpha \partial x^\beta}$$

$$\times\left(b_1 J^\beta_\mu(y)+b_2 J(y)\Omega^\beta_\mu(y)-ab_1\delta^\beta_\mu A^\nu{}_{,\nu}(y)\right)$$

to (B-12), which thus becomes

$$\eta_{\alpha\beta}\int d^4x \left(\int d^4z \delta^4(x-(z+aA(z)))J^\alpha_\omega(z)F^{\omega\chi}{}_{,\chi}(z)+ej^\alpha(x)\right)\Delta^\beta_\mu(x,y)=0, \quad \text{(B-14)}$$

where

$$\Delta^\alpha_\mu(x,y)=\left(\frac{\partial}{\partial A^\mu(y)}\frac{\partial D(x-(y+aA(y)))}{\partial x^\lambda}\right)J^\alpha_\rho(y)J^\lambda_\sigma(y)F^{\rho\sigma}(y)$$

$$-\frac{\partial}{\partial y^\nu}\left(\frac{\partial D(x-(y+aA(y)))}{\partial x^\lambda}\frac{\partial(J^\alpha_\rho(y)J^\lambda_\sigma(y)F^{\rho\sigma}(y))}{\partial A^\mu{}_{,\nu}(y)}\right) \quad \text{(B-15)}$$

$$-\frac{\partial^2 D(x-(y+aA(y)))}{\partial x_\alpha \partial x^\beta}\left(b_1 J^\beta_\mu(y)+b_2 J(y)\Omega^\beta_\mu(y)-ab_1\delta^\beta_\mu A^\nu{}_{,\nu}(y)\right).$$

Using some formulas in §2, we can prove

$$\Delta^\alpha_\mu(x,y)=\frac{\partial^2 D(x-(y+aA(y)))}{\partial x^\beta \partial x^\gamma}\Xi^{\alpha\beta\gamma}_{\mu(1)}(y)-a\frac{\partial D(x-(y+aA(y)))}{\partial x^\beta}\frac{\partial \Xi^{\alpha\beta\gamma}_{\mu(2)}(y)}{\partial y^\gamma}, \quad \text{(B-16)}$$

where

$$\Xi^{\alpha\beta\gamma}_{\mu(1)}(y)=\eta^{\rho\sigma}J^\alpha_\rho(y)J^\beta_\sigma(y)J^\gamma_\mu(y)-\eta^{\rho\sigma}J^\beta_\rho(y)J^\gamma_\sigma(y)J^\alpha_\mu(y)$$

$$-\eta^{\alpha\beta}\left(b_1 J^\gamma_\mu(y)+b_2 J(y)\Omega^\gamma_\mu(y)-ab_1\delta^\gamma_\mu A^\nu{}_{,\nu}(z)\right), \quad \text{(B-17)}$$

$$\begin{aligned}
\varXi^{\alpha\beta\gamma}_{\mu(2)}(y) &= \eta^{\lambda\gamma}\left(J^{\alpha}_{\lambda}(y)A^{\beta}_{\ ,\mu}(y) - J^{\beta}_{\lambda}(y)A^{\alpha}_{\ ,\mu}(y)\right) \\
&\quad + \delta^{\beta}_{\mu}A^{\gamma,\alpha}(y) - \delta^{\alpha}_{\mu}A^{\gamma,\beta}(y) + aF^{\lambda\gamma}(y)\left(\delta^{\beta}_{\mu}A^{\alpha}_{\ ,\lambda}(y) - \delta^{\alpha}_{\mu}A^{\beta}_{\ ,\lambda}(y)\right).
\end{aligned} \quad \text{(B-18)}$$

Assuming the inverse of $\varDelta^{\alpha}_{\mu}(x,y)$ reads $M^{\mu}_{\nu}(x,y)$:

$$\int d^4 y \varDelta^{\alpha}_{\mu}(x,y) M^{\mu}_{\nu}(y,z) = \delta^{\alpha}_{\nu}\delta^4(x-z), \quad \text{(B-19)}$$

according to (2-44) we have

$$\begin{aligned}
\delta^4(x-z) &= \int J(y)d^4 y \delta^4(x-(y+aA(y)))\delta^4(y+aA(y)-z) \\
&= \int J(y)d^4 y \eta^{\beta\gamma}\frac{\partial^2 D(x-(y+aA(y)))}{\partial x^{\beta}\partial x^{\gamma}}\delta^4(y+aA(y)-z) \\
&= -\int d^4 y \eta^{\beta\gamma}\frac{\partial D(x-(y+aA(y)))}{\partial x^{\beta}}\frac{\partial}{\partial y^{\lambda}}\left(J(y)\Omega^{\lambda}_{\gamma}(y)\delta^4(y+aA(y)-z)\right) \\
&= -\int d^4 y \eta^{\beta\gamma}\frac{\partial D(x-(y+aA(y)))}{\partial x^{\beta}}J(y)\Omega^{\lambda}_{\gamma}(y)\frac{\partial \delta^4(y+aA(y)-z)}{\partial y^{\lambda}} \\
&= \int J(y)d^4 y \eta^{\beta\gamma}\frac{\partial D(x-(y+aA(y)))}{\partial x^{\beta}}\frac{\partial \delta^4(y+aA(y)-z)}{\partial z^{\gamma}};
\end{aligned} \quad \text{(B-20)}$$

Substituting both (B-16) and (B-20) to (B-19), we have

$$\begin{aligned}
&\delta^{\alpha}_{\nu}\int J(y)d^4 y \eta^{\beta\gamma}\frac{\partial D(x-(y+aA(y)))}{\partial x^{\beta}}\frac{\partial \delta^4(y+aA(y)-z)}{\partial z^{\gamma}} \\
&= \int d^4 y \left(\frac{\partial^2 D(x-(y+aA(y)))}{\partial x^{\beta}\partial x^{\gamma}}\varXi^{\alpha\beta\gamma}_{\mu(1)}(y) - a\frac{\partial D(x-(y+aA(y)))}{\partial x^{\beta}}\frac{\partial \varXi^{\alpha\beta\gamma}_{\mu(2)}(y)}{\partial y^{\gamma}}\right)M^{\mu}_{\nu}(y,z) \\
&= \int d^4 y \frac{\partial D(x-(y+aA(y)))}{\partial x^{\beta}}\left(\frac{\partial}{\partial y^{\lambda}}\left(\Omega^{\lambda}_{\gamma}(y)\overline{\varXi}^{\alpha\beta\gamma}_{\mu(1)}(y)M^{\mu}_{\nu}(y,z)\right) - a\frac{\partial \varXi^{\alpha\beta\gamma}_{\mu(2)}(y)}{\partial y^{\gamma}}M^{\mu}_{\nu}(y,z)\right),
\end{aligned} \quad \text{(B-21)}$$

where

$$\overline{\varXi}^{\alpha\beta\gamma}_{\mu(1)}(y) = \frac{1}{2}\left(\varXi^{\alpha\beta\gamma}_{\mu(1)}(y) + \varXi^{\alpha\gamma\beta}_{\mu(1)}(y)\right), \quad \text{(B-22)}$$

we see that $M^{\mu}_{\nu}(y,z)$ satisfies:

$$\begin{aligned}
&\frac{\partial}{\partial y^{\lambda}}\left(\Omega^{\lambda}_{\gamma}(y)\overline{\varXi}^{\alpha\beta\gamma}_{\mu(1)}(y)M^{\mu}_{\nu}(y,z)\right) - a\frac{\partial \varXi^{\alpha\beta\gamma}_{\mu(2)}(y)}{\partial y^{\gamma}}M^{\mu}_{\nu}(y,z) \\
&= \delta^{\alpha}_{\nu}\eta^{\beta\gamma}J(y)\frac{\partial \delta^4(y+aA(y)-z)}{\partial z^{\gamma}} = -\delta^{\alpha}_{\nu}\eta^{\beta\gamma}J(y)\int \frac{d^4 k}{(2\pi)^4}ik_{\gamma}e^{ik\cdot(y+aA(y)-z)}.
\end{aligned} \quad \text{(B-23)}$$

We can try to seek the solution of $M^{\mu}_{\nu}(y,z)$ having the form:

$$M^{\mu}_{\nu}(y,z) = J(y)\int \frac{d^4 k}{(2\pi)^4}m^{\mu}_{\nu}(k,y)e^{ik\cdot(y+aA(y)-z)}. \quad \text{(B-24)}$$

Substituting (B-24) to (B-23), we obtain an equation that $m^{\mu}_{\nu}(k,y)$ must satisfy:

$$\overline{\Xi}_{\mu(1)}^{\alpha\beta\gamma}(y)\left(\Omega_{\gamma}^{\lambda}(y)\frac{\partial m_{\nu}^{\mu}(k,y)}{\partial y^{\lambda}}+ik_{\gamma}m_{\nu}^{\mu}(k,y)\right)$$
$$+\left(\Omega_{\gamma}^{\lambda}(y)\frac{\partial \overline{\Xi}_{\mu(1)}^{\alpha\beta\gamma}(y)}{\partial y^{\lambda}}-a\frac{\partial \Xi_{\mu(2)}^{\alpha\beta\gamma}(y)}{\partial y^{\gamma}}\right)m_{\nu}^{\mu}(k,y)=-\delta_{\nu}^{\alpha}\eta^{\beta\gamma}ik_{\gamma}\,. \quad \text{(B-25)}$$

And, further, we can try to seek the solution of $m_{\nu}^{\mu}(k,y)$ having the following form:

$$m_{\nu}^{\mu}(k,y)=\sum_{n=0}^{\infty}a^{n}m_{\nu(n)}^{\mu}(k,y)\,, \quad \text{(B-26)}$$

substituting the above expression to (B-25) and comparing the power of $a$ one by one, we have

$$\frac{\partial m_{\nu(0)}^{\mu}(k,y)}{\partial y^{\lambda}}+ik_{\lambda}m_{\nu(0)}^{\mu}(k,y)=i\delta_{\nu}^{\mu}k_{\lambda}\,, \quad \text{(B-27)}$$

$$\frac{\partial m_{\nu(n)}^{\mu}(k,y)}{\partial y^{\lambda}}+ik_{\lambda}m_{\nu(n)}^{\mu}(k,y)$$
$$=\widetilde{m}_{\nu\lambda(n)}^{\mu}(m_{\sigma(0)}^{\rho},m_{\sigma(1)}^{\rho},\cdots,m_{\sigma(n-1)}^{\rho};k,A^{\alpha},A^{\alpha}{}_{,\beta},A^{\alpha}{}_{,\beta,\gamma})\,, \quad n=1,2,\cdots. \quad \text{(B-28)}$$

In (B-28), the function $\widetilde{m}_{\nu\lambda(n)}^{\mu}$ is known after $m_{\sigma(0)}^{\rho},m_{\sigma(1)}^{\rho},\cdots,m_{\sigma(n-1)}^{\rho}$ are determined. It is obvious that the both equations (B-27) and (B-28) have solutions. We therefore have proved that the inverse $M_{\nu}^{\mu}(x,y)$ of $\mathit{\Lambda}_{\mu}^{\alpha}(x,y)$ exists, and can be determined formally by (B-24), (B-26), (B-27) and (B-28).

Calculating $\int d^{4}yM_{\nu}^{\mu}(y,w)$ for (B-14), and considering (B-19), we obtain

$$\int d^{4}z\delta^{4}\!\left(w-(z+aA(z))\right)J_{\omega}^{\alpha}(z)F^{\omega\chi}{}_{,\chi}(z)+ej^{\alpha}(w)=0\,. \quad \text{(B-29)}$$

(B-29) is just (3-24), we thus have proved that if the equation (B-12) of motion of electromagnetic field that is obtained by the variational equation for $A^{\mu}(y)$ (B-8) directly holds, then (3-24) holds. On the other hand, it is obvious that if (B-29) holds, then (B-14), namely, (B-12) holds. Hence, we have proved that the equation of motion of electromagnetic field (3-19).

## Appendix C  A method discussing the gauge invariance of the equation of motion of electromagnetic field

Notice

$$F^{\mu\nu}(x)=\eta^{\mu\lambda}\frac{\partial A^{\nu}(x)}{\partial x^{\lambda}}-\eta^{\nu\lambda}\frac{\partial A^{\mu}(x)}{\partial x^{\lambda}}=\frac{1}{a}\!\left(\eta^{\mu\lambda}J_{\lambda}^{\nu}(x)-\eta^{\nu\lambda}J_{\lambda}^{\mu}(x)\right), \quad \text{(C-1)}$$

from (3-25) for $A'^{\mu}(x)$ and using (C-1) we have

$$\frac{\partial \widetilde{F}'^{\mu\nu}(x)}{\partial x^\nu} = \int d^4 y \delta^4\left(x-(y+aA'(y))\right) J'^\mu_\rho(y) \frac{\partial F'^{\rho\sigma}(y)}{\partial y^\sigma}$$

$$= \int J'(y') d^4 y' \delta^4\left(x-(y'+aA'(y))\right) \frac{1}{J'(y')} J'^\mu_\rho(y') \frac{\partial}{\partial y'^\sigma} \left[\frac{1}{a}\left(\eta^{\rho\lambda} J'^\sigma_\lambda(y') - \eta^{\sigma\lambda} J'^\rho_\lambda(y')\right)\right]$$

$$= \int J'(y') d^4 y' \delta^4\left(x-(y'+aA'(y))\right) \frac{1}{J'(y')} \frac{\partial(y'^\mu + aA'^\mu(y'))}{\partial y'^\rho}$$

$$\times \frac{1}{a} \frac{\partial}{\partial y'^\sigma}\left[\eta^{\rho\lambda} \frac{\partial(y'^\sigma + aA'^\sigma(y'))}{\partial y'^\lambda} - \eta^{\sigma\lambda} \frac{\partial(y'^\mu + aA'^\mu(y'))}{\partial y'^\lambda}\right]. \qquad (C-2)$$

If we set $\xi^\mu = y'^\mu + aA'^\mu(y')$, then $y'^\mu = y'^\mu(\xi)$, notice

$$d^4\xi = \left\|\frac{\partial(y'^\alpha + aA'^\alpha(y'))}{\partial y'^\beta}\right\| d^4 y' = \left\|J'^\alpha_\beta(y')\right\| d^4 y' = J'(y') d^4 y',$$

(C-2) becomes

$$\frac{\partial \widetilde{F}'^{\mu\nu}(x)}{\partial x^\nu} = \int d^4\xi \delta^4(x-\xi) \frac{1}{J'(y'(\xi))} \frac{\partial \xi^\mu}{\partial y'^\rho} \frac{1}{a} \frac{\partial}{\partial y'^\sigma}\left(\eta^{\rho\lambda} \frac{\partial \xi^\sigma}{\partial y'^\lambda} - \eta^{\sigma\lambda} \frac{\partial \xi^\rho}{\partial y'^\lambda}\right). \qquad (C-3)$$

We now set $\xi^\mu = y^\mu + aA^\mu(y)$, the relation between $A'^\mu(x)$ and $A^\mu(x)$ is given by (4-1), (C-3) becomes

$$\frac{\partial \widetilde{F}'^{\mu\nu}(x)}{\partial x^\nu} = \int J(y) d^4 y \delta^4\left(x-(y+aA(y))\right) \frac{1}{J'(y'(\xi(y)))} \frac{\partial(y^\mu + aA^\mu(y))}{\partial y'^\rho}$$

$$\times \frac{1}{a} \frac{\partial}{\partial y'^\sigma}\left[\eta^{\rho\lambda} \frac{\partial(y^\sigma + aA^\sigma(y))}{\partial y'^\lambda} - \eta^{\sigma\lambda} \frac{\partial(y^\mu + aA^\mu(y))}{\partial y'^\lambda}\right]$$

$$= \int J(y) d^4 y \delta^4\left(x-(y+aA(y))\right) \frac{1}{J'(y'(y))} \frac{\partial y^\alpha}{\partial y'^\rho} \frac{\partial(y^\mu + aA^\mu(y))}{\partial y^\alpha}$$

$$\times \frac{1}{a} \frac{\partial y^\beta}{\partial y'^\sigma} \frac{\partial}{\partial y^\beta}\left\{\frac{\partial y^\tau}{\partial y'^\lambda}\left[\eta^{\rho\lambda} \frac{\partial(y^\sigma + aA^\sigma(y))}{\partial y^\tau} - \eta^{\sigma\lambda} \frac{\partial(y^\rho + aA^\rho(y))}{\partial y^\tau}\right]\right\} \qquad (C-4)$$

$$= \int J(y) d^4 y \delta^4\left(x-(y+aA(y))\right) \frac{1}{J'(y'(y))} \frac{\partial y^\alpha}{\partial y'^\rho} J^\mu_\alpha(y)$$

$$\times \frac{1}{a} \frac{\partial y^\beta}{\partial y'^\sigma} \frac{\partial}{\partial y^\beta}\left[\frac{\partial y^\tau}{\partial y'^\lambda}\left(\eta^{\rho\lambda} J^\sigma_\tau(y) - \eta^{\sigma\lambda} J^\rho_\tau(y)\right)\right],$$

in the above expression, the relation $y^\mu = y^\mu(y')$ or $y'^\mu = y'^\mu(y)$ is determined by

$$y^\mu + aA^\mu(y) = \xi^\mu = y'^\mu + aA'^\mu(y') = y'^\mu + aA^\mu(y') + aK^\mu(y'). \qquad (C-5)$$

In (C-5), (4-1) is used.

Comparing (C-4) and $\dfrac{\partial \widetilde{F}^{\mu\nu}(x)}{\partial x^\nu}$ given by (3-25), we obtain

$$\frac{1}{J'(y'(y))} \frac{\partial y^\mu}{\partial y'^\rho} \frac{\partial y^\nu}{\partial y'^\sigma} \frac{\partial}{\partial y^\nu}\left[\frac{\partial y^\tau}{\partial y'^\lambda}\left(\eta^{\rho\lambda} J^\sigma_\tau(y) - \eta^{\sigma\lambda} J^\rho_\tau(y)\right)\right] = \frac{a}{J(y)} \frac{\partial F^{\mu\nu}(y)}{\partial y^\nu}. \qquad (C-6)$$

In principle, after obtaining $y^\mu = y^\mu(y')$ and $y'^\mu = y'^\mu(y)$ from (C-5), we can calculate $\dfrac{\partial y^\alpha}{\partial y'^\rho}$ and substitute these terms to (C-6); we therefore obtain an equation that the function $K^\mu(x)$ must satisfy.

As an example, we derive the equation (4-5) for the case that $K^\mu(x)$ is infinitesimal.

At first, assuming

$$y'^\mu = y'^\mu(y) = y'^\mu_{(0)}(y) + aY^\mu(y) \tag{C-7}$$

and substituting it to (C-5), comparing the coefficients of the power of $a$ one by one, we can obtain easily

$$y'^\mu_{(0)}(y) = y^\mu, \tag{C-8}$$

$$\begin{aligned}
& Y^\mu(y) - A^\mu(y) + A^\mu(y + aY(y)) + K^\mu(y + aY(y)) \\
&= Y^\mu(y) + aY^\lambda(y)\frac{\partial A^\mu(y)}{\partial y^\lambda} + \frac{1}{2!}a^2 Y^\rho(y)Y^\sigma(y)\frac{\partial^2 A^\mu(y)}{\partial y^\rho \partial y^\sigma} + \cdots + \frac{1}{n!}a^n Y^n(y)\frac{\partial^n A^\mu(y)}{\partial y^n} + \cdots \\
& + K^\mu(y) + aY^\lambda(y)\frac{\partial K^\mu(y)}{\partial y^\lambda} + \frac{1}{2!}a^2 Y^\rho(y)Y^\sigma(y)\frac{\partial^2 K^\mu(y)}{\partial y^\rho \partial y^\sigma} + \cdots + \frac{1}{n!}a^n Y^n(y)\frac{\partial^n K^\mu(y)}{\partial y^n} + \cdots \\
&= 0;
\end{aligned} \tag{C-9}$$

And then, setting

$$Y^\mu(y) = Y^\mu_{(0)}(y) + aY^\mu_{(1)}(y) + a^2 Y^\mu_{(2)}(y) + \cdots + a^n Y^\mu_{(n)}(y) + \cdots, \tag{C-10}$$

and substituting (C-10) to (C-9), we first obtain

$$Y^\mu_{(0)}(y) = -K^\mu(y); \tag{C-11}$$

and, further, comparing the coefficients of the power of $a$ one by one and ignoring all the qualities of higher than first infinitesimal, we obtain

$$\begin{aligned}
Y^\mu_{(n)}(y) &= -Y^\nu_{(n-1)}(y) A^\mu{}_{,\nu}(y) \\
&= (-1)^{n-1} A^\mu{}_{,\lambda_1}(y) A^{\lambda_1}{}_{,\lambda_2}(y) \cdots A^{\lambda_{n-1}}{}_{,\lambda_n}(y) K^{\lambda_n}(y) \quad (n=1,2,3,\cdots).
\end{aligned} \tag{C-12}$$

Substituting (C-11) and (C-12) to (C-10), we obtain an expression of $Y^\mu(y)$:

$$\begin{aligned}
Y^\mu(y) &= -K^\mu(y) + aA^\mu{}_{,\nu}(y)K^\nu(y) - a^2 A^\mu{}_{,\rho}(y)A^\rho{}_{,\sigma}(y)K^\sigma(y) + \cdots \\
&\quad + (-1)^{n-1} a^n A^\mu{}_{,\lambda_1}(y) A^{\lambda_1}{}_{,\lambda_2}(y) \cdots A^{\lambda_{n-1}}{}_{,\lambda_n}(y) K^{\lambda_n}(y) + \cdots,
\end{aligned} \tag{C-13}$$

for which we can prove

$$\left(\delta^\mu_\nu + aA^\mu{}_{,\nu}(y)\right) Y^\nu(y) = J^\mu_\nu(y) Y^\nu(y) = -K^\mu(y),$$

we therefore obtain

$$Y^\mu(y) = -\Omega^\mu_\nu(y) K^\nu(y). \tag{C-14}$$

For the sake of brevity, we define

$$\widetilde{K}^\mu(y) = \Omega^\mu_\nu(y) K^\nu(y), \tag{C-15}$$

substituting (C-8) and (C-14) to (C-7), and considering (C-15), we obtain

$$y'^\mu = y^\mu - a\Omega^\mu_\nu(y) K^\nu(y) = y^\mu - a\widetilde{K}^\mu(y). \tag{C-16}$$

Notice that $J'(y')$ in (C-6) is a function of $A'^\rho{}_{,\sigma}(y') = \dfrac{\partial A'^\rho(y')}{\partial y'^\sigma}$:

$$\begin{aligned}
J'(y') &= J\big(A'^\rho{}_{,\sigma}(y')\big) = J\big(A^\rho{}_{,\sigma}(y') + K^\rho{}_{,\sigma}(y')\big) \\
&= J\big(A^\rho{}_{,\sigma}(y')\big) + K^\alpha{}_{,\beta}(y') \frac{\partial J\big(A^\rho{}_{,\sigma}(y')\big)}{\partial A^\alpha{}_{,\beta}(y')} = J(y') + aJ(y')\Omega^\beta_\alpha(y')K^\alpha{}_{,\beta}(y') \\
&= J\big(y - a\widetilde{K}(y)\big) + aJ\big(y - a\widetilde{K}(y)\big)\Omega^\beta_\alpha\big(y - a\widetilde{K}(y)\big)K^\alpha{}_{,\beta}\big(y - a\widetilde{K}(y)\big) \\
&= J(y) - a\widetilde{K}^\lambda(y)\frac{\partial J(y)}{\partial y^\lambda} + aJ(y)\Omega^\beta_\alpha(y)K^\alpha{}_{,\beta}(y) \\
&= J(y) - a^2\Omega^\lambda_\tau(y)K^\tau(y)J(y)\Omega^\beta_\alpha(y)A^\alpha{}_{,\beta,\lambda}(y) + aJ(y)\Omega^\beta_\alpha(y)K^\alpha{}_{,\beta}(y) \\
&= J(y)\left[1 + a\frac{\partial\big(\Omega^\beta_\alpha(y)K^\alpha(y)\big)}{\partial y^\beta}\right] = J(y)\big(1 + a\widetilde{K}^\gamma{}_{,\gamma}(y)\big),
\end{aligned} \tag{C-17}$$

hence,

$$\frac{1}{J'(y'(y))} = \frac{1}{J(y)}\frac{1}{1 + a\widetilde{K}^\gamma{}_{,\gamma}(y)} = \frac{1 - a\widetilde{K}^\gamma{}_{,\gamma}(y)}{J(y)}. \tag{C-18}$$

According to (C-16) we have

$$\frac{\partial y'^\mu}{\partial y^\nu} = \delta^\mu_\nu - a\widetilde{K}^\mu{}_{,\nu}(y), \tag{C-19}$$

from (C-19) and $\dfrac{\partial y'^\mu}{\partial y^\lambda}\dfrac{\partial y^\lambda}{\partial y'^\nu} = \delta^\mu_\nu$ we obtain

$$\frac{\partial y^\mu}{\partial y'^\nu} = \delta^\mu_\nu + a\widetilde{K}^\mu{}_{,\nu}(y). \tag{C-20}$$

Substituting (C-18) and (C-20) to (C-6), we have

$$\frac{1 - a\widetilde{K}^\gamma{}_{,\gamma}(y)}{J(y)}\big(\delta^\mu_\rho + a\widetilde{K}^\mu{}_{,\rho}(y)\big)\big(\delta^\nu_\sigma + a\widetilde{K}^\nu{}_{,\sigma}(y)\big)\frac{\partial}{\partial y^\nu}\left[\big(\delta^\tau_\lambda + a\widetilde{K}^\tau{}_{,\lambda}(y)\big)\big(\eta^{\rho\lambda}J^\sigma_\tau(y) - \eta^{\sigma\lambda}J^\rho_\tau(y)\big)\right]$$
$$= \frac{a}{J(y)}\frac{\partial F^{\mu\nu}(y)}{\partial y^\nu}, \tag{C-21}$$

ignoring all the qualities of higher than first infinitesimal, we obtain

$$\frac{\partial}{\partial y^{\nu}}\left[\left(\eta^{\mu\rho}J_{\sigma}^{\nu}(y)-\eta^{\nu\rho}J_{\sigma}^{\mu}(y)\right)\widetilde{K}^{\sigma}{}_{,\rho}(y)\right]$$
$$+a\left[-\delta_{\rho}^{\mu}\delta_{\sigma}^{\nu}\widetilde{K}^{\gamma}{}_{,\gamma}(y)+\delta_{\rho}^{\mu}\widetilde{K}^{\nu}{}_{,\sigma}(y)+\delta_{\sigma}^{\nu}\widetilde{K}^{\mu}{}_{,\rho}(y)\right]\frac{\partial F^{\rho\sigma}(y)}{\partial y^{\nu}}=0,$$
(C-22)

we can prove that (C-22) is just (4-5).

## Appendix D  The calculation of the momenta $\pi_{\mu}(y)$ conjugate to $A^{\mu}(y)$ for the form of the action given by (B-1), (B-2), (B-6) and (B-7)

According to the definition of momenta $\pi_{\mu}(y)$ conjugate to $A^{\mu}(y)$ and the form of the action given by (B-1), we have

$$\pi_{\mu}(y)=\frac{\delta S}{\delta A^{\mu}{}_{,0}(y)}=\frac{\delta S_{\text{EM}}}{\delta A^{\mu}{}_{,0}(y)}+\frac{\delta S_{\text{I}\perp}}{\delta A^{\mu}{}_{,0}(y)}+\frac{\delta S_{\text{I}//}}{\delta A^{\mu}{}_{,0}(y)}.$$
(D-1)

We employ (B-2) to calculate $\dfrac{\delta S_{\text{EM}}}{\delta A^{\mu}{}_{,0}(y)}$ and obtain

$$\frac{\delta S_{\text{EM}}}{\delta A^{\mu}{}_{,0}(y)}=-\frac{1}{2}\eta_{\alpha\beta}\int d^{4}x\left(\int d^{4}u\,\frac{\partial^{2}D(x-(u+aA(u)))}{\partial x_{\gamma}\partial x^{\tau}}\frac{\partial\left(J_{\omega}^{\alpha}(u)J_{\chi}^{\tau}(u)F^{\omega\chi}(u)\right)}{\partial A^{\mu}{}_{,0}(u)}\delta^{4}(u-y)\right)$$
$$\times\left(\int d^{4}v\,\frac{\partial^{2}D(x-(v+aA(v)))}{\partial x^{\gamma}\partial x^{\lambda}}J_{\rho}^{\beta}(v)J_{\sigma}^{\lambda}(v)F^{\rho\sigma}(v)\right)$$
$$-\frac{1}{2}\eta_{\alpha\beta}\int d^{4}x\left(\int d^{4}u\,\frac{\partial^{2}D(x-(u+aA(u)))}{\partial x_{\gamma}\partial x^{\tau}}J_{\omega}^{\alpha}(u)J_{\chi}^{\tau}(u)F^{\omega\chi}(u)\right)$$
$$\times\left(\int d^{4}v\,\frac{\partial^{2}D(x-(v+aA(v)))}{\partial x^{\gamma}\partial x^{\lambda}}\frac{\partial\left(J_{\rho}^{\beta}(v)J_{\sigma}^{\lambda}(v)F^{\rho\sigma}(v)\right)}{\partial A^{\mu}{}_{,0}(v)}\delta^{4}(v-y)\right)$$
$$=\frac{\partial\left(J_{\omega}^{\alpha}(y)J_{\chi}^{\tau}(y)F^{\omega\chi}(y)\right)}{\partial A^{\mu}{}_{,0}(y)}\eta_{\alpha\beta}\int d^{4}x\,\frac{\partial D(x-(y+aA(y)))}{\partial x^{\tau}}$$
$$\times\left(\int d^{4}v\,\frac{\partial\delta^{4}(x-(v+aA(v)))}{\partial x^{\lambda}}J_{\rho}^{\beta}(v)J_{\sigma}^{\lambda}(v)F^{\rho\sigma}(v)\right)$$
$$=\frac{\partial\left(J_{\omega}^{\alpha}(y)J_{\chi}^{\tau}(y)F^{\omega\chi}(y)\right)}{\partial A^{\mu}{}_{,0}(y)}\eta_{\alpha\beta}\int d^{4}x\,\frac{\partial D(x-(y+aA(y)))}{\partial x^{\tau}}$$
$$\times\left(\int d^{4}v\,\delta^{4}(x-(v+aA(v)))J_{\rho}^{\beta}(v)F^{\rho\sigma}{}_{,\sigma}(v)\right),$$
(D-2)

in the above calculation process, we have used integration by parts for $x_{\gamma}$.

According to (B-6) we have

$$\frac{\delta S_{I\perp}}{\delta A^\mu_{,0}(y)} = e\int d^4x j_\alpha(x)\left(\int d^4z \frac{\partial D(x-(z+aA(z)))}{\partial x^\beta}\frac{\partial\left(J^\alpha_\rho(z)J^\beta_\sigma(z)F^{\rho\sigma}(z)\right)}{\partial A^\mu_{,0}(z)}\delta^4(z-y)\right)$$

$$= \frac{\partial\left(J^\alpha_\rho(y)J^\beta_\sigma(y)F^{\rho\sigma}(y)\right)}{\partial A^\mu_{,0}(y)} e\int d^4x j_\alpha(x)\frac{\partial D(x-(y+aA(y)))}{\partial x^\beta}.$$

(D-3)

From (D-2) and (D-3) we have

$$\frac{\delta S_{EM}}{\delta A^\mu_{,0}(y)} + \frac{\delta S_{I\perp}}{\delta A^\mu_{,0}(y)}$$
$$= e\eta_{\alpha\gamma}\frac{\partial\left(J^\alpha_\rho(y)J^\beta_\sigma(y)F^{\rho\sigma}(y)\right)}{\partial A^\mu_{,0}(y)}$$

(D-4)

$$\times \int d^4x\left(\int d^4z\delta^4(x-(z+aA(z)))J^\gamma_\omega(z)F^{\omega\chi}_{,\chi}(z)+ej^\gamma(x)\right)\frac{\partial D(x-(y+aA(y)))}{\partial x^\beta}=0.$$

According to (B-7) we obtain

$$\frac{\delta S_{I//}}{\delta A^\mu_{,0}(y)}$$

$$= -e\int d^4x j^\alpha(x)\left(\int d^4z \frac{\partial^2 D(x-(z+aA(z)))}{\partial x^\alpha \partial x^\nu}\left(b_1\frac{\partial J^\nu_\lambda(z)}{\partial A^\mu_{,0}(z)}A^\lambda(z)+b_2\frac{\partial J(z)}{\partial A^\mu_{,0}(z)}A^\nu(z)\right)\delta^4(z-y)\right)$$

$$= -ae\int d^4x j^\alpha(x)\frac{\partial^2 D(x-(y+aA(y)))}{\partial x^\alpha \partial x^\nu}\left(b_1\delta^\nu_\mu A^0(y)+b_2 J(y)\Omega^0_\mu(y)A^\nu(y)\right)$$

(D-5)

$$= -a\int d^4x\left(\int d^4z\delta^4(x-(z+aA(z)))J^\alpha_\omega(z)F^{\omega\chi}_{,\chi}(z)+ej^\alpha(x)\right)$$

$$\times \frac{\partial^2 D(x-(y+aA(y)))}{\partial x^\alpha \partial x^\nu}\left(b_1\delta^\nu_\mu A^0(y)+b_2 J(y)\Omega^0_\mu(y)A^\nu(y)\right)=0.$$

In the above calculation, (B-13) and (B-29) are used.

Substituting (D-4) and (D-5) to (D-1), we obtain

$$\pi_\mu(y)=0.$$  (D-6)

For the current QED, if we employ (1-24), (1-23), (1-19) and (1-20) to calculate the momenta $\pi_\mu(y)$ conjugate to, then we have obtain the same result. In fact, we have

$$\frac{\delta S^{(0)}_{EM}}{\delta A^\mu_{,0}(y)} = -\frac{1}{2}\eta_{\alpha\beta}\int d^4z \frac{\partial F^{\alpha\tau}(z)}{\partial A^\mu_{,0}(z)}\delta^4(z-y)\frac{d^4k}{(2\pi)^4}\frac{k_\tau k_\lambda}{k^2}e^{ik\cdot(z-x)}d^4x F^{\beta\lambda}(x)$$

$$-\frac{1}{2}\eta_{\alpha\beta}\int d^4z F^{\alpha\tau}(z)\frac{d^4k}{(2\pi)^4}\frac{k_\tau k_\lambda}{k^2}e^{ik\cdot(z-x)}d^4x \frac{\partial F^{\beta\lambda}(x)}{\partial A^\mu_{,0}(x)}\delta^4(x-y)$$

$$= -\eta_{\alpha\beta}\left(\delta^\tau_\mu \eta^{0\alpha}-\delta^\alpha_\mu \eta^{0\tau}\right)\int d^4z F^{\beta\lambda}(z)\frac{d^4k}{(2\pi)^4}\frac{k_\tau k_\lambda}{k^2}e^{-ik\cdot(y-z)}$$

(D-7)

$$= -\eta_{\alpha\beta}\left(\delta^\tau_\mu \eta^{0\alpha}-\delta^\alpha_\mu \eta^{0\tau}\right)\int d^4z F^{\beta\lambda}(z)\frac{d^4k}{(2\pi)^4}\frac{-ik_\tau}{k^2}\frac{\partial}{\partial z^\lambda}e^{-ik\cdot(y-z)}$$

$$= -\eta_{\alpha\beta}\left(\delta^\tau_\mu \eta^{0\alpha}-\delta^\alpha_\mu \eta^{0\tau}\right)\int d^4z \frac{\partial F^{\beta\lambda}(z)}{\partial z^\lambda}\frac{d^4k}{(2\pi)^4}\frac{ik_\tau}{k^2}e^{-ik\cdot(y-z)},$$

$$\frac{\delta S_{\mathrm{I}\perp}^{(0)}}{\delta A^\mu{}_{,0}(y)} = -e\int \mathrm{d}^4 x j_\alpha(x)\mathrm{d}^4 z \frac{\partial F^{\alpha\tau}(z)}{\partial A^\mu{}_{,0}(z)} \delta^4(z-y)\frac{\mathrm{d}^4 k}{(2\pi)^4}\frac{-\mathrm{i}k_\tau}{k^2}\mathrm{e}^{-\mathrm{i}k\cdot(x-z)}$$

$$= -e\left(\delta_\mu^\tau \eta^{0\alpha} - \delta_\mu^\alpha \eta^{0\tau}\right)\int \mathrm{d}^4 x j_\alpha(x)\frac{\mathrm{d}^4 k}{(2\pi)^4}\frac{-\mathrm{i}k_\tau}{k^2}\mathrm{e}^{-\mathrm{i}k\cdot(x-y)} \quad \text{(D-8)}$$

$$= -e\left(\delta_\mu^\tau \eta^{0\alpha} - \delta_\mu^\alpha \eta^{0\tau}\right)\int \mathrm{d}^4 z j_\alpha(z)\frac{\mathrm{d}^4 k}{(2\pi)^4}\frac{\mathrm{i}k_\tau}{k^2}\mathrm{e}^{-\mathrm{i}k\cdot(y-z)},$$

$$\frac{\delta S_{\mathrm{I}/\!/}^{(0)}}{\delta A^\mu{}_{,0}(y)} = 0 ; \quad \text{(D-9)}$$

and then, according to (D-7) ~ (D-9) we obtain

$$\pi_\mu(y) = \frac{\delta S}{\delta A^\mu{}_{,0}(y)} = \frac{\delta S_{\mathrm{EM}}^{(0)}}{\delta A^\mu{}_{,0}(y)} + \frac{\delta S_{\mathrm{I}\perp}^{(0)}}{\delta A^\mu{}_{,0}(y)} + \frac{\delta S_{\mathrm{I}/\!/}^{(0)}}{\delta A^\mu{}_{,0}(y)}$$

$$= -\left(\delta_\mu^\tau \eta^{0\alpha} - \delta_\mu^\alpha \eta^{0\tau}\right)\int \mathrm{d}^4 z\left(\eta_{\alpha\beta}\frac{\partial F^{\beta\lambda}(z)}{\partial z^\lambda} + ej_\alpha(z)\right)\frac{\mathrm{d}^4 k}{(2\pi)^4}\frac{\mathrm{i}k_\tau}{k^2}\mathrm{e}^{-\mathrm{i}k\cdot(y-z)} = 0. \quad \text{(D-10)}$$

In the above calculation, the equation of motion of electromagnetic field (1-8) is used.

# Appendix E  The Lehmann-Symanzik-Zimmermann formalism of quantum electrodynamics under the Lorentz gauge condition

## E.1  The current quantum electrodynamics

For obtaining the Lehmann-Symanzik-Zimmermann formalism of the current quantum electrodynamics under the Heisenberg picture, occupation number representation and the Lorentz gauge condition $\frac{\partial A^\mu(x)}{\partial x^\mu} = 0$, based on the discussion in §5.3, what we do are only the following two steps:

① Using the following nine equations instead of (5-25), (5-26), (5-27), (5-36), (5-41), (5-42), (5-44), (5-45) and (5-47), respectively:

$$\eta^{\rho\sigma}\frac{\partial^2 A^\mu(X)}{\partial X^\rho \partial X^\sigma} = e\zeta^\mu(X). \quad \text{(E-1)}$$

$$\Lambda_\mu(x) = A_\mu(x), \quad \text{(E-2)}$$

$$\zeta^\mu(X) = j^\mu(X) = \overline{\psi}(X)\gamma^\mu\psi(X). \quad \text{(E-3)}$$

$$\sqrt{Z_3}\,A^\mu_{\substack{\mathrm{in}\\\mathrm{out}}}(X) = A^\mu(X) - e\int \mathrm{d}^4 Y D_{\substack{\mathrm{ret}\\\mathrm{adv}}}(X-Y)\zeta^\mu(Y). \quad \text{(E-4)}$$

$$A^\mu(x) \to \sqrt{Z_3}\,A^\mu_{\substack{\mathrm{in}\\\mathrm{out}}}(X),\quad t \to \mp\infty. \quad \text{(E-5)}$$

$$\left\{\psi_{\mathrm{in}\alpha}(t,\boldsymbol{x}),\psi_{\mathrm{in}\beta}^+(t,\boldsymbol{x}')\right\} = \mathrm{i}\delta_{\alpha\beta}\delta^3(\boldsymbol{x}-\boldsymbol{x}'),\ \left[A^\mu_{\mathrm{in}}(T,\boldsymbol{X}),\dot{A}^\nu_{\mathrm{in}}(T,\boldsymbol{X}')\right] = -\mathrm{i}\eta^{\mu\nu}\delta^3(\boldsymbol{X}-\boldsymbol{X}'); \quad \text{(E-6)}$$
others $= 0$.

$$A_{\text{in}}^{\mu}(X) = \int \frac{d^3K}{(2\pi)^{3/2}\sqrt{2|\mathbf{K}|}} \sum_{\lambda=0}^{3} \varepsilon_{(\lambda)}^{\mu}(\mathbf{K}) \left( a_{(\lambda)\text{in}}(\mathbf{K}) e^{-iK\cdot X} + a_{(\lambda)\text{in}}^{*}(\mathbf{K}) e^{iK\cdot X} \right) \quad \text{(E-7)}$$

$$\left\{ b_{\text{in}}(\mathbf{p},s), b_{\text{in}}^{+}(\mathbf{p}',s') \right\} = \delta_{ss'}\delta^3(\mathbf{p}-\mathbf{p}'), \quad \left\{ d_{\text{in}}(\mathbf{p},s), d_{\text{in}}^{+}(\mathbf{p}',s') \right\} = \delta_{ss'}\delta^3(\mathbf{p}-\mathbf{p}'),$$
$$\left[ a_{(\lambda)\text{in}}(\mathbf{K}), a_{(\lambda')\text{in}}^{*}(\mathbf{K}') \right] = -\eta_{(\lambda)(\lambda')}\delta^3(\mathbf{K}-\mathbf{K}'); \text{ others} = 0. \quad \text{(E-8)}$$

$$H_{\text{EM}} = \int d^3K |\mathbf{K}| \left( \sum_{i=0}^{3} a_{(i)\text{in}}^{*}(\mathbf{K}) a_{(i)\text{in}}(\mathbf{K}) - a_{(0)\text{in}}^{*}(\mathbf{K}) a_{(0)\text{in}}(\mathbf{K}) \right). \quad \text{(E-9)}$$

Now, in (5-23), (5-24), (5-34) and (5-35), $\Lambda_{\mu}(x)$ is given by (E-2); $D_{\text{ret}\atop\text{adv}}(X-Y)$ in (E-4) is given by (2-60).

② A constraint condition pressing on physical state vectors

Instead of the discussion (5-29) ~ (5-33) about constraint condition pressing on physical state vectors, under the Lorentz gauge condition $\frac{\partial A^{\mu}(x)}{\partial x^{\mu}} = 0$, a physical state vector $|\Psi_{\text{phy}}\rangle$ in the Heisenberg picture describing actual system of electrons and photons must satisfies

$$\langle \Psi_{\text{phy}} | \frac{\partial A^{\lambda}(X)}{\partial X^{\lambda}} | \Psi_{\text{phy}} \rangle = \langle \Psi_{\text{phy}} | \frac{\partial A^{0}(T,\mathbf{X})}{\partial T} | \Psi_{\text{phy}} \rangle + \langle \Psi_{\text{phy}} | \nabla \cdot \mathbf{A}(T,\mathbf{X}) | \Psi_{\text{phy}} \rangle = 0, \quad \text{(E-10)}$$

We have proved that $\frac{\partial A^{\lambda}(X)}{\partial X^{\lambda}}$ is free field in the nonlocal theory given by this paper (See (4-31)), this conclude of course holds for the current QED; hence, a physical state vector $|\Psi_{\text{phy}}\rangle$ satisfies (E-10) if and only if, for the operators $A^{\mu}(T_0,\mathbf{X})$ of initial time $T_0$, it satisfies

$$\langle \Psi_{\text{phy}} | \frac{\partial A^{0}(T_0,\mathbf{X})}{\partial T_0} | \Psi_{\text{phy}} \rangle + \langle \Psi_{\text{phy}} | \nabla \cdot \mathbf{A}(T_0,\mathbf{X}) | \Psi_{\text{phy}} \rangle = 0. \quad \text{(E-11)}$$

And, further, for the Fock space constructed by the set of eigenvectors of the two operators given by (5-46) and (E-9), according to (E-4), the constraint condition (E-11) becomes

$$\langle \Psi_{\text{phy}} | \frac{\partial A_{\text{in}}^{\lambda}(X)}{\partial X^{\lambda}} | \Psi_{\text{phy}} \rangle = \langle \Psi_{\text{phy}} | \frac{\partial A_{\text{in}}^{0}(T,\mathbf{X})}{\partial T} | \Psi_{\text{phy}} \rangle + \langle \Psi_{\text{phy}} | \nabla \cdot \mathbf{A}_{\text{in}}(T,\mathbf{X}) | \Psi_{\text{phy}} \rangle = 0. \quad \text{(E-12)}$$

For this Fock space, indefinite metric and the work of the constraint condition (E-12) on the eigenvectors of the operator given by (E-9) are necessary, we no longer repeat these well-known contents.

Via the above two steps and remaining the all rest formulas in §5.3, we have established the Lehmann-Symanzik-Zimmermann formalism of the current quantum electrodynamics under the Lorentz gauge condition $\frac{\partial A^{\mu}(x)}{\partial x^{\mu}} = 0$.

### E.2 A quantum theory of the nonlocal theory given by this paper

For establishing a quantum theory of the nonlocal theory given by this paper via the Lehmann-Symanzik-Zimmermann approach, based on the above discussion, what we must do is only, instead of (E-2) and (E-3), to take

$$\Lambda_\mu(x) = \Phi_\mu(x), \qquad (\text{E-13})$$

$$\zeta^\mu(X) = \left( J(X)\Omega_\nu^\mu(X) \int d^4x\, j^\nu(x)\delta^4(x-(X+aA(X))) \right)\Big|_{\overline{W}}. \qquad (\text{E-14})$$

In (E-13), $\Phi_\mu(x)$ is given by (A-15). Under the Lorentz gauge condition $\dfrac{\partial A^\mu(x)}{\partial x^\mu} = 0$, $J(x)\Omega_\nu^\mu(x)$ in (E-14) becomes

$$\begin{aligned}
J(x)\Omega_\nu^\mu(x) = {} & \delta_\nu^\mu - aA^\mu{}_{,\nu}(x) + \frac{a^2}{2}\left(-\delta_\nu^\mu A^\alpha{}_{,\beta}(x)A^\beta{}_{,\alpha}(x) + 2A^\mu{}_{,\lambda}(x)A^\lambda{}_{,\nu}(x)\right) \\
& + \frac{a^3}{6}\Big(2\delta_\nu^\mu A^\alpha{}_{,\beta}(x)A^\beta{}_{,\gamma}(x)A^\gamma{}_{,\alpha}(x) + 3A^\mu{}_{,\nu}(x)A^\alpha{}_{,\beta}(x)A^\beta{}_{,\alpha}(x) \\
& - 6A^\mu{}_{,\alpha}(x)A^\alpha{}_{,\beta}(x)A^\beta{}_{,\nu}(x)\Big),
\end{aligned} \qquad (\text{E-15})$$

We can try to definite that the appropriate operators ordering $(\cdots)_{\overline{W}}$ in (E-14) is that, after $\psi(x)$, $\overline{\psi}(x)$ and $A^\mu(X)$ are expressed by $\psi_{\text{in}}(x)$, $\overline{\psi}_{\text{in}}(x)$ and $A_{\text{in}}^\mu(X)$ via perturbation method to solving the equations (5-34), (5-35) and (E-4) of the operators $\psi(x)$, $\overline{\psi}(x)$ and $A^\mu(X)$, and, further, after $\psi_{\text{in}}(x)$, $\overline{\psi}_{\text{in}}(x)$ and $A_{\text{in}}^\mu(X)$ are expressed by $b_{\text{in}}(p,s)$, $b_{\text{in}}^+(p,s)$, $d_{\text{in}}(p,s)$, $d_{\text{in}}^+(p,s)$, $a_{(\lambda)\text{in}}(K)$ and $a_{(\lambda)\text{in}}^*(K)$, in an expression

$$V(\psi,\overline{\psi},A,\dot{A}) = \widetilde{V}(\psi_{\text{in}},\overline{\psi}_{\text{in}},A_{\text{in}},\dot{A}_{\text{in}}) = \overline{V}(b_{\text{in}}(p,s),b_{\text{in}}^+(p,s),d_{\text{in}}(p,s),d_{\text{in}}^+(p,s),a_{(\lambda)\text{in}}(K),a_{(\lambda)\text{in}}^*(K)),$$

all the operators $b_{\text{in}}(p,s)$, $b_{\text{in}}^+(p,s)$, $d_{\text{in}}(p,s)$, $d_{\text{in}}^+(p,s)$, $a_{(\lambda)\text{in}}(K)$ and $a_{(\lambda)\text{in}}^*(K)$ obey so called "normal product" ordering.

We therefore have established a self-sufficient quantum theory under the Lorentz gauge condition $\dfrac{\partial A^\mu(x)}{\partial x^\mu} = 0$ of the nonlocal theory given by this paper. Similar to the current quantum electrodynamics, in principle, we can obtain solution for arbitrary question in such quantum theory of the nonlocal theory.

Notice that according (E-4) and (E-14), $\dfrac{\partial}{\partial T}\left[A^\mu(T,X), \dot{A}^\nu(T,X')\right] \neq 0$; hence, even if

$$\left[A^\mu(T,X), \dot{A}^\nu(T,X')\right]\Big|_{T\to-\infty} = \left[A_{\text{in}}^\mu(T,X), \dot{A}_{\text{in}}^\nu(T,X')\right] = -i\eta^{\mu\nu}\delta^3(X-X'),$$

we still have $\left[A^\mu(T,X), \dot{A}^\nu(T,X')\right] \neq -i\eta^{\mu\nu}\delta^3(X-X')$ for arbitrary time $T$. This is different from the current quantum electrodynamics. For the current QED, according (E-4), (E-3) and (E-6), we can prove $\left[A^\mu(T,X), \dot{A}^\nu(T,X')\right] = -i\eta^{\mu\nu}\delta^3(X-X')$ for arbitrary time $T$.

Gauge invariance of the *S*-matrix and microscopic causality of such quantum theory of the nonlocal theory given in this section will be studied further.